\documentclass[journal]{IEEEtran}
\usepackage{epsfig,amsmath,amssymb,epsf,cite,scalefnt,graphicx,subfig,array} 
\usepackage{pifont,epstopdf,grffile}
\usepackage{multirow}

\def\bu{{\mathbf{u}}}   \def\bx{{\mathbf{x}}}

   \def\bI{{\mathbf{I}}}

   \def\bX{{\mathbf{X}}}

 \def\ibb{{\pmb{b}}}   
  \def\ibh{{\pmb{h}}}

\def\ibu{{\pmb{u}}}   \def\ibx{{\pmb{x}}} 
\def\ibz{{\pmb{z}}}

   \def\ibI{{\pmb{I}}}

   \def\ibX{{\pmb{X}}}

     \def\d4{\!\!\!\!}              
                  \def\gam{\gamma}

 \def\bPhi{\mathbf{\Phi}} \def\bTheta{\mathbf{\Theta}}   \def\lam{\lambda} 
\def\bal{\boldsymbol{\alpha}} \def\blam{\boldsymbol{\lambda}}
\def\bzeta{\boldsymbol{\zeta}}
\def\bLambda{\mathbf{\Lambda}}

  \def\-{\! - \!}  \def\+{\! + \!}  \def\={\! = \!}  \def\>{\! > \!}

\newcommand{\bef}{\begin{figure}}
\newcommand{\eef}{\end{figure}}
\newcommand{\beq}{\begin{eqnarray}}
\newcommand{\eeq}{\end{eqnarray}}

\newcommand{\qed}{\nobreak \ifvmode \relax \else
\ifdim\lastskip<1.5em \hskip-\lastskip \hskip1.5em plus0em
minus0.5em \fi \nobreak \vrule height0.5em width0.5em
depth0.25em\fi}

\ifCLASSINFOpdf
\else
\fi
\begin{document}
%
\title{Robust Real-time Ellipse Fitting Based on Lagrange Programming Neural Network and Locally Competitive Algorithm}
%
%
%

\author{Hao~Wang,
        Chi-Sing~Leung,~\IEEEmembership{Senior Member,~IEEE,}
        Hing Cheung So,~\IEEEmembership{Fellow,~IEEE,}
        Junli~Liang,~\IEEEmembership{Senior Member,~IEEE,}
        Ruibin~Feng,
        and~Zifa~Han.
\thanks{Hao~Wang, Chi-Sing~Leung, Hing Cheung So, Ruibin~Feng, and~Zifa~Han are with the Department of Electronic Engineering, City University of Hong Kong, Hong Kong.}
\thanks{Junli~Liang is with Northwestern Polytechnical University, Xi'an 710072, China.}}

%
%

\markboth{IEEE TRANSACTIONS ON ,~Vol.~, No.~, ~2018}%
{Shell \MakeLowercase{\textit{et al.}}: Robust Ellipse Fitting based on Lagrange Programming Neural Network and Locally Competitive Algorithm}
%



\maketitle

\begin{abstract}
Given a set of 2-dimensional (2-D) scattering points, which are usually obtained from the edge detection process, the aim of ellipse fitting is to construct an elliptic equation that best fits the collected observations.
However, some of the scattering points may contain outliers due to imperfect edge detection. To address this issue, we devise a robust real-time ellipse fitting approach based on two kinds of analog neural network, Lagrange programming neural network (LPNN) and locally competitive algorithm (LCA).
First, to alleviate the influence of these outliers, the fitting task is formulated as a nonsmooth constrained optimization problem in which the objective function is either an $l_1$-norm or $l_0$-norm term.
It is because compared with the $l_2$-norm in some traditional ellipse fitting models, the $l_p$-norm with $p<2$ is less sensitive to outliers.
Then, to calculate a real-time solution of this optimization problem, LPNN is applied. As the LPNN model cannot handle the non-differentiable term in its objective, the concept of LCA is introduced and combined with the LPNN framework.
Simulation and experimental results show that the proposed ellipse fitting approach is superior to several state-of-the-art algorithms.
 \end{abstract}

\begin{IEEEkeywords}
Ellipse fitting, Outlier, Real-time solution, Lagrange programming neural network (LPNN), Locally competitive algorithm (LCA).
\end{IEEEkeywords}

%
\IEEEpeerreviewmaketitle

\section{Introduction}\label{section1}

%
%
%
Fitting of geometric primitives with given coplanar points is required in many research areas such as physics~\cite{chernov1984effective}, biology~\cite{paton1970conic}, industrial inspection, automatic manufacture, and computer vision. In particular, an ellipse, which generalizes a circle, is a common geometric primitive in image processing. However, ellipse fitting is much more difficult than circle fitting due to the reason that the curvature of the ellipse is not uniform, that is, low-curvature points contribute more to fitting than those at high curvature. Furthermore, the equation of ellipse is more complicated than that of circle where the former and latter are functions of 5 and 3 parameters, respectively.

Numerous ellipse fitting algorithms have been developed in the literature.
In general, they can be roughly classified into two main types. One is clustering which includes Hough transform (HT) and its variants~\cite{duda1972use,ballard1981generalizing}.
The basic idea is to search the 5 parameters of the ellipse in a 5-dimensional (5-D) space.
Apparently, costly computations are needed.
Another one is the least squares (LS) method where the key idea is to calculate the elliptical parameters by minimizing an error metric between the geometric primitives and collected data points~\cite{barwick_very_2009}.
Obviously, the LS techniques is more computationally efficient than the clustering approach. And it can be further divided into geometric based and algebraic based methods.
For the former, the error metric is the sum of the orthogonal distances between the 2-D measurements and corresponding points in the constructed ellipse~\cite{nakagawa1979note}, and the fitting problem can be formulated as a nonlinear program~\cite{rosin1995nonparametric}.
On the other hand, the algebraic based methods are extensively studied because they are generally simple and computationally attractive~\cite{ahn_least-squares_2001,rosin_note_1993,maini_enhanced_2006}.
Nevertheless, constraint needs to be introduced here in order to guarantee that the solution is valid.
There are many different algebraic methods~\cite{bookstein1979fitting,kanatani1994statistical,sampson1982fitting,gander1994least} because the choices of the constraints are not unique.
Among them, the constrained least squares (CLS) method \cite{gander1994least} is a representative example, which introduces a unit-norm constraint on the elliptical parameter vector. Even though the algebraic based methods work very well in many cases, their sensitivities to outliers limit their applications.
It is because the 2-D scattering points are usually acquired from edge detection where it is difficult to avoid disturbances including outliers. As a result, there is a need to devise robust algebraic solutions.
It is worth noting that a few efforts on robust ellipse fitting have already been made recently. These include the sparsity based method (SBM) \cite{liang_robust_2013} and robust CLS (RCLS) algorithm \cite{liang_robust_2015}. The former utilizes the $l_1$-norm to resist outliers and calculates the elliptical parameters by solving a second-order cone programming (SOCP) problem. While the latter introduces the maximum correntropy criterion and quadratic constraint to handle the problem.

In this paper, we develop a novel robust ellipse fitting approach based on the Lagrange programming neural network (LPNN)~\cite{zhang1992lagrange,nagamatu1996stability,paper:lpnn1,paper:lpnn2,lpnn3,lpnn4} and locally competitive algorithm (LCA) \cite{rozell2008sparse},\cite{balavoine2012convergence}.
It is also an algebraic based method.
First, the problem is formulated as a constrained optimization problem.
Analogous to \cite{liang_robust_2013}, the $l_p$-norm ($p=1$ or $p=0$) is used as its objective to achieve robustness against outliers.
Then, the LPNN framework is applied to solve the problem.
Since the LPNN framework requires that its objective function and constraints are twice differentiable, the internal state concept of the LCA is utilized to convert the non-differentiable components due to the $l_1$-norm and $l_0$-norm as differentiable expressions. 

The rest of this paper is organized as follows. The background of ellipse fitting, LPNN and LCA are described in Section~\ref{section2}.
In Section~\ref{section3}, the proposed ellipse fitting algorithms are developed and their digital realization are presented.
The local stability of the LPNN approach is proved in Section~\ref{section4}. Numerical results for algorithm evaluation and comparison are provided in Section~\ref{section5}. Finally, conclusions are drawn in Section~\ref{section6}.

\section{Background}\label{section2}
\subsection{Notation}
We use a lower-case or upper-case letter to represent a scalar while vectors and matrices are denoted by bold lower-case and upper-case letters, respectively. The transpose operator is denoted as $(\centerdot)^ \mathrm{T}$, and $\ibI$ and $\mathbf{0}$ represent the identity matrix and zero matrix of appropriate dimensions, respectively. Other mathematical symbols are defined in their first appearance.

\subsection{Rudimentary Knowledge of Ellipse Fitting}
An axis-aligned ellipse with center at $(c_x,c_y)$, axes parallelled to the $x$-axis and $y$-axis of lengths $a$ and $b$ can be expressed as:
\beq\label{eq-1.1}
\frac{(x-c_x)^2}{a^2}+\frac{(y-c_y)^2}{b^2}=1
\eeq
This particular parametric model is frequently used in the diameter control system of silicon single crystal growth~\cite{liu_bayesian_2011}.
For the more general case, a non-axis aligned ellipse centered at $(c_x,c_y)$ with a counter-clockwise rotation of $\theta$ can be described as
\beq\label{eq-1.2}
\frac{((x-c_x)\cos\theta+(y-c_y)\sin\theta)^2}{a^2}+ \nonumber \\ \frac{(-(x-c_x)\sin\theta+(y-c_y)\cos\theta)^2}{b^2}=1.
\eeq
The task of ellipse fitting is to find the five parameters $\{a,b,c_x,c_y,\theta\}$. However, it is very difficult to estimate them directly because equation~\eqref{eq-1.2} is highly nonlinear. Instead, many ellipse fitting algorithms~\cite{ahn_least-squares_2001,maini_enhanced_2006,fitzgibbon_direct_1999} consider the second-order polynomial model:
\beq\label{eq-1.3}
Ax^2+Bxy+Cy^2+Dx+Ey+F=0,
\eeq
where the six parameters $\{A,B,C,D,E,F\}$ are related to
$\{a,b,c_x,c_y,\theta\}$ as:
\begin{eqnarray}
A&=&\frac{\cos^2\theta}{a^2}+\frac{\sin^2\theta}{b^2}\\
B&=&2\cos\theta \sin\theta\left(\frac{1}{a^2}-\frac{1}{b^2}\right)\\
C&=&\frac{\sin^2\theta}{a^2}+\frac{\cos^2\theta}{b^2}\\
D&=&\frac{-2c_x \cos^2\theta-2c_y \sin\theta \cos\theta}{a^2} \nonumber \\
& &+\frac{-2c_x \sin^2\theta+2c_y \sin\theta \cos\theta}{b^2}
\end{eqnarray}
\begin{eqnarray}
E&=&\frac{-2c_y \sin^2\theta-2c_x \sin\theta \cos\theta}{a^2} \nonumber \\
&&+\frac{-2c_y \cos^2\theta+2c_x \sin\theta \cos\theta}{b^2}\\
F&=&\frac{(c_x \cos \theta+c_y \sin\theta)^2}{a^2} \nonumber \\
 & &+\frac{(c_x \sin\theta-c_y \cos\theta)^2}{b^2}-1.
\end{eqnarray}

Let $\mathcal{D}=\{(x_i,y_i): i=1,\cdots,N \}$  be a set of 2-D scattering points of an ellipse.
Denote
\beq
\bal&=&[A,B,C,D,E,F]^ \mathrm{T}, \\
\ibx_i&=&[x_i^2,x_iy_i,y_i^2,x_i,y_i,1]^\mathrm{T},\\
\ibX & = & [\ibx_1,\cdots,\ibx_N].
\eeq
In the absence of measurement errors, \eqref{eq-1.3} can be rewritten as:
\beq\label{eq-1.4}
\ibX^\mathrm{T} \bal ={[\ibx_1^\mathrm{T}\bal,\cdots,\ibx_N^\mathrm{T}\bal]}^\mathrm{T}=\mathbf{0}.
\eeq
Where $\ibx_i^\mathrm{T}\bal$ is called the "algebraic distance" which can be used to measure the fitting error of point $(x_i,y_i)$~\cite{fitzgibbon_direct_1999}. Hence, in a noisy environment, the traditional CLS algorithm considers the following constrained optimization problem:
\begin{subequations}\label{eq-1.5}
\beq
\min\limits_{\bal} \,\,& \left\| \ibX^ \mathrm{T} \bal \right\|^2_2 \\
\mbox{s.t.}\,\, &\bal^\mathrm{T}\bal=1.
\eeq
\end{subequations}
Where the objective function in~(\ref{eq-1.5}a) is the sum of squared algebraic distances.
In~(\ref{eq-1.5}b), the unit-norm constraint is used to avoid the redundant solutions (the solutions with linear correlation), and the trivial solution ($\bal=\mathbf{0}$).
The CLS approach is efficient for ellipse fitting, providing that the noise in data obeys a Gaussian distribution.
When the data set contains impulsive disturbances or even outliers, the CLS solution may have a large deviation from the actual ellipse.

It is worth pointing out that the CLS solution may also correspond to a hyperbola or parabola~\cite{gander1994least} because these two geometric primitives can be expressed by~\eqref{eq-1.4} as well.
To eliminate these possibilities, an additional constraint is introduced:
\beq
B^2-4AC<0
\eeq
which aims to guarantee the solution corresponding to an ellipse only~\cite{liang_robust_2015}. On the other hand, $B^2-4AC>0$ and $B^2-4AC=0$ result in a hyperbola and parabola, respectively.

\subsection{Lagrange Programming Neural Network}
The LPNN is an analog neural network computational approach, which can be implemented by hardware circuit. It is very effective when real-time solutions are required. Generally, it can be used to solve a general nonlinear constrained optimization problem~\cite{zhang1992lagrange}, given by
\begin{subequations}\label{eq-1.6}
\beq
\min\limits_{\ibz} &\,\,f(\ibz) \\
\mbox{s.t.}&\,\, \ibh(\ibz)=0,
\eeq
\end{subequations}
where $\ibz=[z_1,\cdots,z_n]^\mathrm{T}$ is the variable vector being optimized,
 $f:\mathbb{R}^n \to \mathbb{R}$ is the objective function, $\ibh:\mathbb{R}^n \to \mathbb{R}^m$ with $m<n$ represents $m$ equality constraints, and $f$ and $\ibh$
should be twice differentiable. The first step in the LPNN approach is to define the Lagrangian:
\beq
\label{eq-1.7}
L(\ibz,\bzeta)=f(\ibz)+\bzeta^\mathrm{T}\ibh(\ibz)
\eeq
where $\bzeta=[\zeta_1,\cdots,\zeta_m]^\mathrm{T}$ is the Lagrange multiplier vector.
There are two kinds of neurons in LPNN, namely, variable neurons and Lagrangian neurons.
The $n$ variable neurons are used to hold the decision variable vector $\ibz$ while
the $m$ Lagrangian neurons deal with the Lagrange multiplier vector
$\bzeta$.
In the LPNN framework, the dynamics of the neurons are defined as
\begin{subequations}\label{eq-1.8}
\beq
\frac{d\ibz}{dt}=& \displaystyle -\frac{\partial L(\ibz,\bzeta)}{\partial \ibz} \\
\frac{d\bzeta}{dt}=& \displaystyle \frac{\partial L(\ibz,\bzeta)}{\partial \bzeta}.
\eeq
\end{subequations}
The differential equations in (\ref{eq-1.8})
govern the state transition of the neurons.
After the neurons settle down at an equilibria, the solution is obtained by measuring the neuron outputs at this stable equilibrium point.
The purpose of (\ref{eq-1.8}a) is to seek for a state with the minimum objective value while
(\ref{eq-1.8}b) aims to constrain the system state such that it falls into the feasible region.
From (\ref{eq-1.8}), the network will settle down at a stable state if several mild conditions are satisfied~\cite{zhang1992lagrange,lpnn3,lpnn4}.
It is also clear that $f$ and $\ibh$ should be differentiable, otherwise the dynamics cannot be defined.

\subsection{Locally Competitive Algorithm}
The LCA, introduced by~\cite{rozell2008sparse}, is also an analog neural network which can be used to handle the following unconstrained optimization problem
\beq\label{eq-1.9}
\min L_{\rm lca}= \frac{1}{2} \| \ibb -\bPhi \ibz \|_2^2  + \lambda \| \ibz \|_1,
\eeq
where $\ibz \in \mathbb{R}^n$, $\ibb \in \mathbb{R}^m$ and $\bPhi \in \mathbb{R}^{m \times n}$ ($m < n$).
For this optimization problem, LCA uses $n$ neurons to hold the variable vector $\ibz$. To minimize the cost function $L_{\rm lca}$, its gradient with respect to $\ibz$ needs to be calculated. Because the term $\lambda \| \ibz \|_1$ is non-differentiable at zero point, we can anticipate the problem in computing the gradient of $L_{\rm lca}$ is achieve by setting up a approximate differential equation of \eqref{eq-1.9}.
In mathematics, the sub-differential denoted as $\partial \| \ibz \|_1$ can be used to describe the gradient of  $\| \ibz \|_1$. And the sub-differential at a non-differentiable point is equal to a set\footnote{For the absolute function $|z|$, the sub-differential $\partial |z|$ at $z=0$ is equal to $[-1,1]$.}.

The LCA introduces an internal state vector $\ibu=[u_1,\cdots,u_n]^\mathrm{T} $ for the neuron output vector $\ibz$. The mapping between $\ibz$ and $\ibu$ is given by
\begin{equation}
z_i = T_{\lambda}(u_i) = \left\{ \begin{array}{lcl}
0, &  |u_i| \leq \lambda, \\
u_i - \lambda \mbox{sign} (u_i),  & |u_i| > \lambda.
\end{array}\right.
\label{internal2}
\end{equation}
In the LCA, $\ibz$ and $\ibu$ are the output state variable and internal state variable vectors, respectively. $\lambda$ is a scalar which denotes the threshold of the function.

Furthermore, according to the proof in the appendices of~\cite{rozell2008sparse}, we have
\begin{equation}
\label{internal1x}
\lambda \partial \|\ibz \|_1 \owns \ibu - \ibz.
\end{equation}
At a non-differentiable point, $\ibu - \ibz$ can be seen as a gradient selection method.
The LCA defines its dynamics with respect to $\ibu$ rather than of $\ibz$.
Because for the dynamics of $\ibu$, the sub-differentiable term can be replaced according to the relationship given in (\ref{internal1x}), then we have
\begin{equation}
\frac{d \ibu}{dt}=-\partial_{\ibz} {L}_{\rm lca}= \bPhi^\mathrm{T} \ibb - (\bPhi^T \bPhi  - \ibI) \ibz - \ibu.
\label{eqn:dyna}
\end{equation}
It should be noticed that if the dynamics of $\ibz$ is used, we need to implement $\partial \|\ibz \|_1$ which is equal to a set for $\forall z_i=0,\,\, i=1, \dots, n$. While for $d \ibu/dt$, the term $\partial \|\ibz \|_1$ can be replaced by $\ibu - \ibz$.

In~\cite{rozell2008sparse}, a more general threshold function has been proposed which is given by
\beq\label{eq-1.14}
z_i=T_{(\eta,\delta,\lambda)}(u_i)=\mbox{sign} (u_i)\frac{ |u_i|-\delta\lambda}{1 + e^{-\eta(|u_i|-\lambda)}}.
\eeq
Where the value of $\lambda$ still denotes the threshold, the $\eta$ is a parameter to control the speed of the threshold transition, and $\delta \in [0, 1]$ indicates what fraction of an additive adjustment is made for values above threshold. Some examples of this general threshold function are provided in Fig.1.
With this threshold function, a more general objective function can be solved, which is given by
\beq\label{eq-general}
\tilde{L}_{\rm lca}=  \frac{1}{2} \| \ibb -\bPhi \ibz \|_2^2  +  \lambda  \sum_{i=1}^n \psi_{(\eta,\delta,\lambda)}(z_i).
\eeq
Furthermore, for any $z_i=T_{(\eta,\delta,\lambda)}(u_i)$, the relationship between $u_i$, $z_i$ and $\partial \psi_{(\eta,\delta,\lambda)}(z_i)/ \partial z_i$ is
\begin{equation}
\label{internal1x2016}
\lambda \frac{\partial \psi_{(\eta,\delta,\lambda)}(z_i)}{\partial z_i} \equiv u_i - z_i.
\end{equation}

Someone may want to construct the exact form of $\psi_{(\eta,\delta,\lambda)}(\cdot)$ but note that its analytical expression cannot be obtained generally. Nevertheless, this does not limit the application of the LCA because the neural dynamics are expressed in terms of the threshold function $T_{(\eta,\delta,\lambda)}(u_i)$ rather than the exact penalty term.
Setting $\eta \rightarrow \infty$, $\delta=0$ and $\lambda=1$, we obtain
an ideal hard threshold function~\cite{rozell2008sparse}:
\beq
\label{eq-thresholdl0a}
z_i=T_{(\infty,0,1)}(u_i)=
\left\{ \begin{array}{lcl}
0,  & |u_i| \leq 1, \\
u_i, & |u_i| > 1.
\end{array}\right.
\eeq
Also, the corresponding penalty term is close to the  $l_0$-norm component:
\beq\label{eq-generalL0}
\lambda  \sum_{i=1}^n \psi_{(\infty,0,1)} (z_i) = \frac{1}{2} \sum_{i=1}^n \mathcal{I}(|z_i|>1),
\eeq
where $\mathcal{I}(\cdot)$ is an indicator function.
Note that according to (\ref{eq-thresholdl0a}),
the variables $z_i$ produced
by the ideal threshold function cannot take values in the range
of $[-1,0)$ and $(0,1]$.
The details of (\ref{eq-thresholdl0a}) and (\ref{eq-generalL0}) are provided in \cite{rozell2008sparse}.

If we set $\eta \rightarrow \infty$ and $\delta=1$, then the
general threshold function is reduced to
the soft threshold function~\cite{rozell2008sparse}, given by
\beq
\label{eq-thresholdl0aa}
z_i=T_{(\infty,1,\lambda)}(u_i)=T_{\lambda}(u_i)
\eeq
and the penalty term becomes the  $l_1$-norm function:
\beq\label{eq-generalL1aa}
\lambda  \sum_{i=1}^n \psi_{(\infty,1,\lambda)} (z_i) = \lambda \|\ibz\|_1.
\eeq
The behavior of the dynamics under different settings has been studied in~\cite{rozell2008sparse,balavoine2012convergence,balavoine2011global}. However, the limitation of LCA is that it can handle the unconstrained optimization problem only.

\begin{figure}[htb]
\centering
\centerline{\includegraphics[width=2.5in]{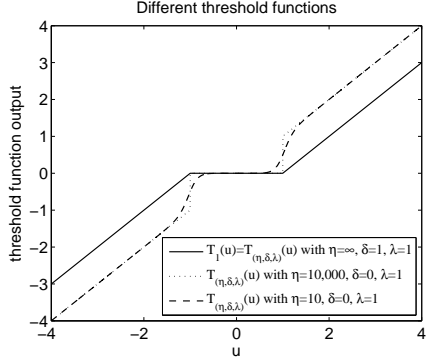}}
\caption{Examples of general threshold function.}
\label{threshold}
\end{figure}

\section{Development of Proposed Algorithm} \label{section3}
\subsection{Problem Formulation}
In the CLS method, the $l_2$-norm is used as its objective function, i.e., $\|\bal^\mathrm{ T }\ibX\|^2_2$.
It is well known that the $l_2$-norm works well in Gaussian noise environments, but is sensitive to outliers. 
In the presence of impulsive noise or outliers, the performance of model with $l_p$-norm ($p<2$) will outperform that with the $l_2$-norm. 
In this study, we focus on two particular $l_p$-norms, namely, $l_1$-norm and $l_0$-norm.
The corresponding formulations are
\begin{subequations}\label{eq-1.15kk1}
\beq
\min\limits_{\bal} \,\,& \left\| \ibX^\mathrm{T} \bal \right\|_1 \\
\mbox{s.t.}\,\, &     \bal^\mathrm{T}\bal=1,\\
\, \, \, \, & B^2-4AC<0,
\eeq
\end{subequations}
and
\begin{subequations}\label{eq-1.15kk2}
\beq
\min\limits_{\bal} \,\,& \left\| \ibX^\mathrm{T} \bal \right\|_0 \\
\mbox{s.t.}\,\, &     \bal^\mathrm{T}\bal=1,\\
\, \, \, \, & B^2-4AC<0.
\eeq
\end{subequations}

To achieve the real-time solution, we use LPNN to solve the optimization problem \eqref{eq-1.15kk1} and \eqref{eq-1.15kk2}. 
Prior to applying the LPNN framework, we need to resolve two issues.
First, the inequality constraint in (\ref{eq-1.15kk1}) and (\ref{eq-1.15kk2}) should be convert to an equality, because the LPNN framework can only handle problems with equality constraints. 
Another issue is that the objective function in (\ref{eq-1.15kk1}) and (\ref{eq-1.15kk2}) are non-differentiable, while LPNN can only solve the problem with differentiable objective and constraints.

To deal with the first issue, we introduce a new variable $G$ into the inequality constraint to change it into an equality one, thus $B^2-4AC+G^2=\epsilon$ where $\epsilon$ is a small negative scalar ($\epsilon=-10^{-12}$ in our experiments).
The formulations of (\ref{eq-1.15kk1}) and (\ref{eq-1.15kk2}) are then modified as
\begin{subequations}\label{eq-1.15kk3}
\beq
\min\limits_{\tilde{\bal}} \,\,& \left\| \tilde{\ibX}^\mathrm{T} \tilde{\bal} \right\|_1 \\
\mbox{s.t.}\,\, &\tilde{\bal}^\mathrm{ T }\bPhi \tilde{\bal}=1,\\
&\tilde{\bal}^\mathrm{ T }\bTheta \tilde{\bal}=\epsilon,
\eeq
\end{subequations}
and
\begin{subequations}\label{eq-1.15kk4}
\beq
\min\limits_{\tilde{\bal}} \,\,& \left\| \tilde{\ibX}^\mathrm{T} \tilde{\bal} \right\|_0 \\
\mbox{s.t.}\,\, &\tilde{\bal}^\mathrm{ T }\bPhi \tilde{\bal}=1,\\
&\tilde{\bal}^\mathrm{ T }\bTheta \tilde{\bal}=\epsilon,
\eeq
\end{subequations}
where
\beq
\tilde{\bal}&=&[A, B, C, D, E, F, G]^ \mathrm{ T }, \nonumber \\
\tilde{\ibX}&=&[\tilde{\bx}_1, \tilde{\bx}_2,\cdots,\tilde{\bx}_N], \nonumber \\
\tilde{\ibx_i}&=&[x_i^2, x_i y_i,y_i^2, x_i,y_i,1,0]^\mathrm{T},\nonumber\\
\bPhi&=&\left[
\begin{matrix}
\bI_{6\times6}&\mathbf{0}_{6\times1}&\\
\mathbf{0}_{1\times6}&0&
\end{matrix}
\right],\nonumber \\
\bTheta&=&\left[
\begin{matrix}
\bLambda &\mathbf{0}_{3\times3}&\mathbf{0}_{3\times1}&\\
\mathbf{0}_{3\times3}&\mathbf{0}_{3\times3}&\mathbf{0}_{3\times1}&\\
\mathbf{0}_{1\times3}&\mathbf{0}_{1\times3}&1&
\end{matrix}
\right], \nonumber \\
\bLambda&=&\left[
\begin{matrix}
0 & 0 & -2 &\\
0 & 1 & 0 & \\
-2 & 0 & 0 &\\
\end{matrix}
\right]. \nonumber
\eeq

Based on the LCA approach, the second issue is resolved by considering the general form $\sum_{i=1}^N \psi_{(\eta,\delta,\lambda)} ([\tilde{\ibX}^\mathrm{T}\tilde{\bal}]_i)$ where $[\cdot]_i$ denotes the $i$th element of the vector.
The problem in \eqref{eq-1.15kk3} or  \eqref{eq-1.15kk4} becomes
\begin{subequations}\label{eq-1.15}
\beq
\min\limits_{\tilde{\bal}} \,\,& \sum_{i=1}^N \psi_{(\eta,\delta,\lambda)} ([\tilde{\ibX}^\mathrm{T}\tilde{\bal}]_i)\\
\mbox{s.t.}\,\, &\tilde{\bal}^\mathrm{ T }\bPhi \tilde{\bal}=1,\\
&\tilde{\bal}^\mathrm{ T }\bTheta \tilde{\bal}=\epsilon.
\eeq
\end{subequations}
When we set $\eta\rightarrow \infty$, $\delta=1$ and $\lambda=1$, (\ref{eq-1.15}a)
is the $l_1$-norm  objective function. 
On the other hand, the proximate $l_0$-norm expression is achieved by setting $\eta\rightarrow \infty$, $\delta=0$ and $\lambda=1$ in (\ref{eq-1.15}a).

To exploit the concept of LCA, we introduce the a dummy vector $\ibz=\tilde{\ibX}^\mathrm{T} \tilde{\bal}$. Then (\ref{eq-1.15}) becomes
\begin{subequations}\label{eq-1.17}
\beq
\min\limits_{\tilde{\bal},\ibz} \,\,& \sum_{i=1}^N
\psi_{(\eta,\delta,\lambda)} (z_i), \\
\mbox{s.t.}\,\, &\ibz=\tilde{\ibX}^\mathrm{T} \tilde{\bal}, \\
&\tilde{\bal}^ \mathrm{ T }\bPhi \tilde{\bal}=1, \\
&\tilde{\bal}^ \mathrm{ T }\bTheta \tilde{\bal}=\epsilon.
\eeq
\end{subequations}

\subsection{LPNN for Ellipse Fitting}
According to~\eqref{eq-1.17} and the concept of LPNN, we first construct the following Lagrangian function:
\begin{eqnarray}
\!\!\!\!L(\tilde{\bal},\ibz,\bzeta,\beta,\gam)\!\!&=&\!\!
\sum_{i=1}^N
\psi_{(\eta,\delta,\lambda)} (z_i)+
\bzeta^ \mathrm{T}(\ibz-\tilde{\ibX}^\mathrm{T} \tilde{\bal})\nonumber \\
\!\!\!\!\!\!& &\!\!
+ \!\beta(\tilde{\bal}^ \mathrm{ T }\bPhi \tilde{\bal}\!-\!1)+\gam(\tilde{\bal}^ \mathrm{ T } \bTheta \tilde{\bal}\!-\!\epsilon).
\label{eq-1.18}
\end{eqnarray}
In~\eqref{eq-1.18}, $\tilde{\bal}$ and $\ibz$
are decision variable vectors while $\bzeta \in \mathbb{R}^N$, $\beta$ and $\gam$ are the Lagrange multipliers.
In the next step, we can use~\eqref{eq-1.18} to deduce the neural dynamics for the robust ellipse fitting problem given by~\eqref{eq-1.17}.
However, our preliminary experimental results find that the neural dynamics may not be stable.
To improve the stability and convexity, several augmented terms are introduced into the objective function~\cite{zhang1992lagrange,paper:lpnn1,paper:lpnn2,lpnn3,lpnn4}, then~\eqref{eq-1.17} becomes:
\begin{subequations}\label{eq-1.19}
\beq
\min\limits_{\tilde{\bal},\ibz} \,\,&  \displaystyle \sum_{i=1}^N
\psi_{(\eta,\delta,\lambda)} (z_i) +\frac{C_0}{2}\left\|\ibz-
\tilde{\ibX}^\mathrm{T} \tilde{\bal}\right\|^2_2 \nonumber \\
&+\displaystyle \frac{C_1}{2}\left(\tilde{\bal}^ \mathrm{ T }\bPhi \tilde{\bal}-1\right)^2+\frac{C_2}{2}\left(\tilde{\bal}^ \mathrm{ T }\bTheta\tilde{\bal}-\epsilon \right)^2, \\
\mbox{s.t.}\,\, &\ibz=\tilde{\ibX}^\mathrm{T} \tilde{\bal}, \\
&\tilde{\bal}^ \mathrm{ T }\bPhi\tilde{\bal}=1,\\
&\tilde{\bal}^ \mathrm{ T }\bTheta\tilde{\bal}=\epsilon.
\eeq
\end{subequations}
In~\eqref{eq-1.19}, $C_0$, $C_1$ and $C_2$ are trade-off factors which are used for adjusting the magnitudes of the augmented terms. When they are large enough, the augmented terms will make the objective function of~\eqref{eq-1.19} be convex. But if they are too large, they may result in the dynamics converge to a local optimal solution. These three extra terms do not influence the objective function value at an equilibrium point. It is because at any equilibrium point, the constraints should be satisfied, i.e.,
$\ibz=\tilde{\bX}^\mathrm{T} \tilde{\bal}$,
$\tilde{\bal}^ \mathrm{ T }\bPhi \tilde{\bal}=1$,
and $\tilde{\bal}^ \mathrm{ T }\bTheta \tilde{\bal}=\epsilon$.
In other words, augmented terms will all equal zero at equilibrium point. Then the Lagrangian for \eqref{eq-1.19} is:
\begin{eqnarray}
L(\tilde{\bal},\ibz,\bzeta,\beta,\gam)&=&\sum_{i=1}^N
\psi_{(\eta,\delta,\lambda)} (z_i) +\bzeta^\mathrm{T}\left (\ibz-\tilde{\ibX}^\mathrm{T} \tilde{\bal}\right) \nonumber \\
&&+\beta\left(\tilde{\bal}^ \mathrm{ T }\bPhi \tilde{\bal}\!-\!1\right)+\gam\left(\tilde{\bal}^ \mathrm{ T } \bTheta \tilde{\bal}\!-\!\epsilon\right) \nonumber \\
&&+\frac{C_0}{2}\left\|\ibz-\tilde{\ibX}^\mathrm{T} \tilde{\bal}\right\|^2_2 \nonumber \\
&&+\frac{C_1}{2}\left(\tilde{\bal}^ \mathrm{ T }\bPhi \tilde{\bal}-1\right)^2 \nonumber \\
& &+\frac{C_2}{2}\left(\tilde{\bal}^ \mathrm{ T }\bTheta\tilde{\bal}-\epsilon\right)^2.
\label{eq-1.20}
\end{eqnarray}
For constructing neural dynamics, we need to calculate the gradient of Lagrangian~\eqref{eq-1.20} with respect to its decision variables and Lagrange variables. To handle the non-differentiable term, we utilize the concept of LCA, introduce an internal state variable $\ibu$ for $\ibz$ and the relationship between $\ibu$ and $\ibz$ is given by (\ref{eq-1.14}).

Now it is ready for us to define the dynamics, $(d u_i)/(dt)$, $ i=1,\cdots,N$, and $(d \tilde{\bal})/(dt)$  for the state variables, and the dynamics,  $(d \bzeta)/(dt)$, $(d \beta)/(dt)$ and $(d \gam) /(dt)$ for the Lagrangian variables.
For the state variables $u_i$'s, we apply the LPNN and combine it with the LCA concepts,
thus, their dynamics are
\begin{equation}
\frac{d u_i}{dt}=
-\frac{\partial L(\tilde{\bal},\ibz,\bzeta,\beta,\gam)}{\partial z_i}.
\label{dynamic1}
\end{equation}
Based on (\ref{eq-1.8}a), the dynamics for the state variables in $\tilde{\bal}$ are given by
\begin{equation}
\frac{d \tilde{\bal}}{dt}=-\frac{\partial L(\tilde{\bal},\ibz,\bzeta,\beta,\gam)}{\partial \tilde{\bal}}.
\label{dynamic2}
\end{equation}
For the Lagrangian variables, their dynamics are obtained from (\ref{eq-1.8}b) as:
\begin{eqnarray}
\frac{d \bzeta}{dt}&=&\frac{\partial L(\tilde{\bal},\ibz,\bzeta,\beta,\gam)}{\partial
\bzeta},
\label{dynamic3} \\
\frac{d \beta}{dt}&=&\frac{\partial L(\tilde{\bal},\ibz,\bzeta,\beta,\gam)}{\partial \beta},
\label{dynamic4} \\
\frac{d \gam}{dt}&=&\frac{\partial L(\tilde{\bal},\ibz,
\bzeta,\beta,\gam)}{\partial \gam}
\label{dynamic5}.
\end{eqnarray}
According to~\eqref{eq-1.20} and~\eqref{internal1x2016}, the dynamics given by \eqref{dynamic1}--\eqref{dynamic5} become
\begin{eqnarray}
\frac{d \ibu}{dt}
&=&-\ibu+\ibz-\bzeta-C_0\left(\ibz-\tilde{\ibX}^\mathrm{T} \tilde{\bal}\right),
\label{dynamic_a1} \\
\frac{d \tilde{\bal}}{dt}
&=&
\tilde{\ibX}\bzeta-2\beta\bPhi \tilde{\bal}-2\gam\bTheta \tilde{\bal}-C_0\tilde{\ibX}\left(\ibz-\tilde{\ibX}^\mathrm{T}\tilde{\bal}\right)\nonumber\\
&&-2C_1\left(\tilde{\bal}^\mathrm{T}\bPhi\tilde{\bal}-1\right)
\bPhi\tilde{\bal} \nonumber \\
&&-2C_2\left(\tilde{\bal}^\mathrm{T}\bTheta\tilde{\bal}-\epsilon\right)\bTheta\tilde{\bal},
\label{dynamic_a2} \\
\frac{d \bzeta}{dt}&=& \ibz-\tilde{\ibX}^\mathrm{T} \tilde{\bal},
\label{dynamic_a3} \\
\frac{d \beta}{dt}&=&\tilde{\bal}^\mathrm{T}\bPhi\tilde{\bal}-1,
\label{dynamic_a4} \\
\frac{d \gam}{dt}&=&
\label{dynamic_a5}\tilde{\bal}^\mathrm{T}\bTheta\tilde{\bal}
-\epsilon.
\end{eqnarray}
It should be noticed that for the relationship given by \eqref{eq-1.14}, if the problem is formulated using the $l_0$-norm objective, we should set $\eta$ as a large number, $\delta=0$ and $\lambda=1$. 
While, if the problem uses the $l_1$-norm objective function, we should set $\eta=\infty$, $\delta=1$ and $\lambda=1$.

\subsection{Properties and Simulation Method}
In the LPNN approach, the circuit complexity depends on the time derivative calculations.
From (\ref{dynamic_a1})--(\ref{dynamic_a5}), the most computationally demanding step is to determine the product of a $N \times 7$ matrix and $7 \times 1$ vector.
Hence  the complexity to obtain the time derivatives is equal to  ${\cal O}(N)$ only.

In simulation, we can update (\ref{dynamic_a1})--(\ref{dynamic_a5}) as rules:
\begin{eqnarray}
\ibu^{(k+1)} & = & \ibu^{(k)} + \mu\frac{d\ibu^{(k)}}{dt}, \\
\tilde{\bal}^{(k+1)} & = & \tilde{\bal}^{(k)}+\mu\frac{d\tilde{\bal}^{(k)}}{dt}, \\
\blam^{(k+1)}  & = & \blam^{(k)} +\mu\frac{d\blam^{(k)} }{dt},\\
\beta^{(k+1)}  & = & \beta^{(k)} +\mu\frac{d\beta^{(k)} }{dt}, \\
\gam^{(k+1)}  & = &\gam^{(k)} +\mu\frac{d\gam^{(k)} }{dt},
\end{eqnarray}
where $^{(k)}$ corresponds to the estimate at the $k$th iteration and $\mu>0$ is the step size which should not be too large to avoid the divergence.
Upon convergence of the iterative procedure, we obtain the estimate of $\tilde{\bal}$, denoted by $\tilde{\bal}^{\ast}$. From $\tilde{\bal}^{\ast}$, the ellipse parameter estimates  $\{a^{\ast},b^{\ast},c^{\ast}_x,c^{\ast}_y,\theta^{\ast}\}$ are then computed from:
\begin{eqnarray}
\!\!\!\!\!\!\!\!\theta^{\ast} & = &\!\!\!\!\frac{1}{2}\tan^{-1}\left(\frac{\tilde{\alpha}^*_2}{\tilde{\alpha}^*_1-\tilde{\alpha}^*_3}\right),\label{eq-1.24}\\
\!\!\!\!\!\!\!\!\left[
\begin{matrix}
c_x^{\ast}\\
c_y^{\ast}
\end{matrix}
\right] &= &
\!\!\!\!\left[
\begin{matrix}
-2\tilde{\alpha}^*_1& -\tilde{\alpha}^*_2 \\
-\tilde{\alpha}^*_2 & -2\tilde{\alpha}^*_3
\end{matrix}
\right]^{-1}
\left[
\begin{matrix}
\tilde{\alpha}^*_4\\
\tilde{\alpha}^*_5
\end{matrix}
\right],\label{eq-1.25}\\
\!\!\!\!\!\!\!\!a^{\ast} & = &\!\!\!\!\sqrt{\frac{\left[
\begin{matrix}
c_x^{\ast}\\
c_y^{\ast}
\end{matrix}
\right]^\mathrm{T}
\left[
\begin{matrix}
\tilde{\alpha}^*_1& \tilde{\alpha}^*_2/2\\
\tilde{\alpha}^*_2/2& \tilde{\alpha}^*_3
\end{matrix}
\right]
\left[
\begin{matrix}
c_x^{\ast}\\
c_y^{\ast}
\end{matrix}
\right]+1}
{\tilde{\alpha}^*_1 \cos^2\theta^{\ast}\!+\!\tilde{\alpha}^*_2 \sin \theta^{\ast} \cos\theta^{\ast}\!+\!\tilde{\alpha}^*_3 \sin^2\theta^{\ast}}},\label{eq-1.26}\\
\!\!\!\!\!\!\!\! b^{\ast}&=&\!\!\!\!\sqrt{\frac{\left[
\begin{matrix}
c_x^{\ast}\\
c_y^{\ast}
\end{matrix}
\right]^\mathrm{T}
\left[
\begin{matrix}
\tilde{\alpha}^*_1& \tilde{\alpha}^*_2/2\\
\tilde{\alpha}^*_2/2& \tilde{\alpha}^*_3
\end{matrix}
\right]
\left[
\begin{matrix}
c_x^{\ast}\\
c_y^{\ast}
\end{matrix}
\right]+1}
{\tilde{\alpha}^*_1 \sin^2\theta^{\ast}\!-\!\tilde{\alpha}^*_2 \sin\theta^{\ast}\cos\theta^{\ast}\!+\!\tilde{\alpha}^*_3 \cos^2\theta^{\ast}}}.\label{eq-1.27}
\end{eqnarray}

\begin{figure}[!htb]
\centering
\subfloat[]{
\label{fig dynamics 1}
\includegraphics[width=1.6in]{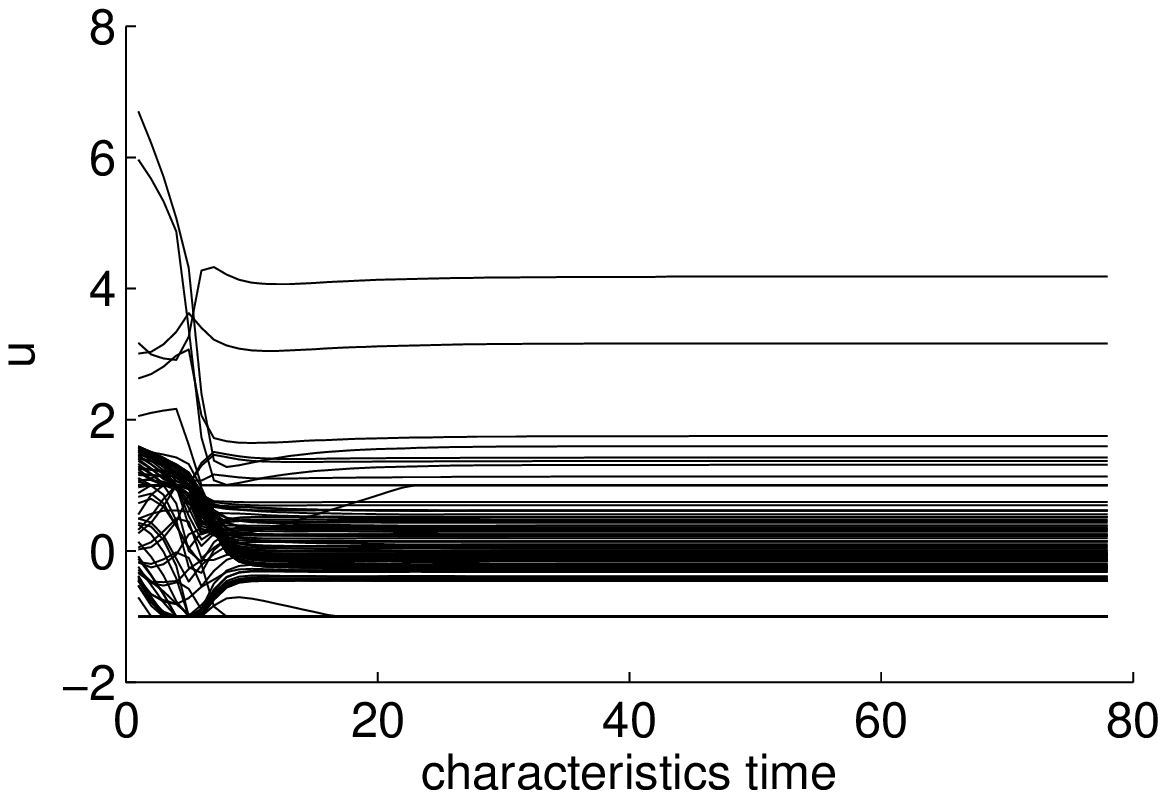}}
\subfloat[]{
\label{fig dynamics 2}
\includegraphics[width=1.6in]{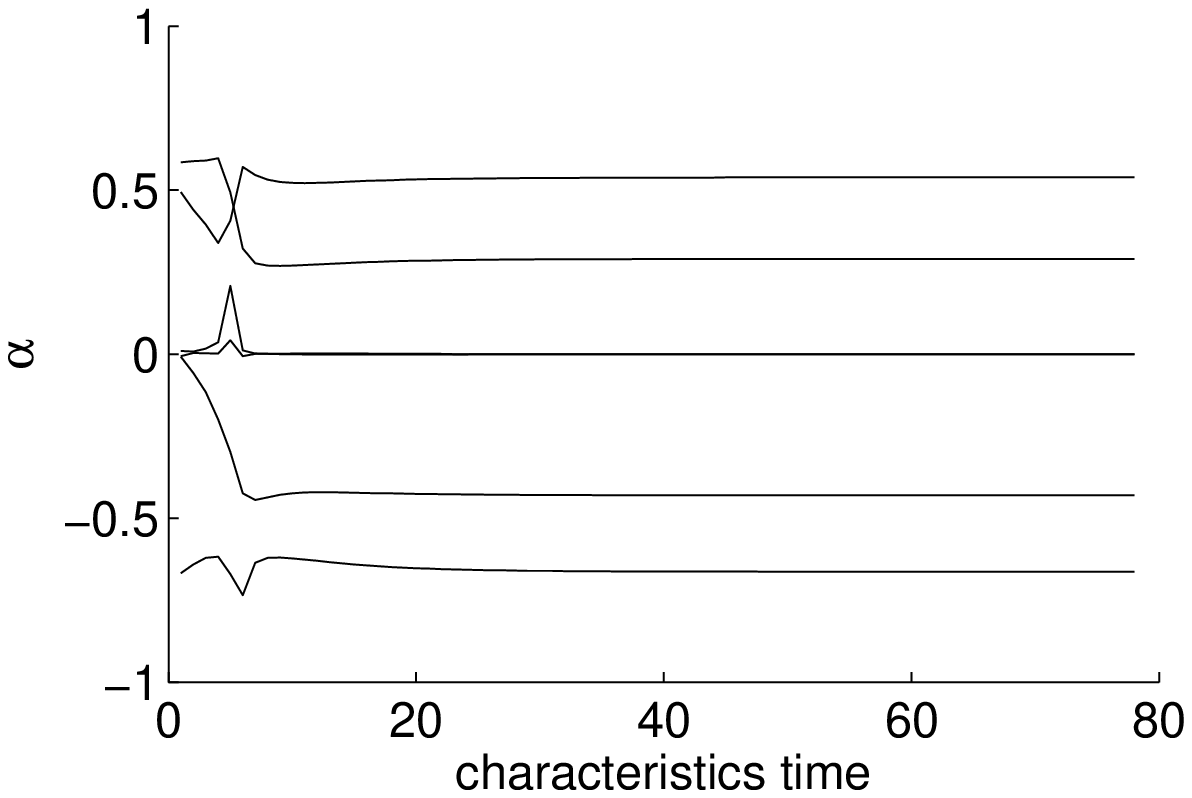}}
\hfil
\subfloat[]{
\label{fig dynamics 3}
\includegraphics[width=1.6in]{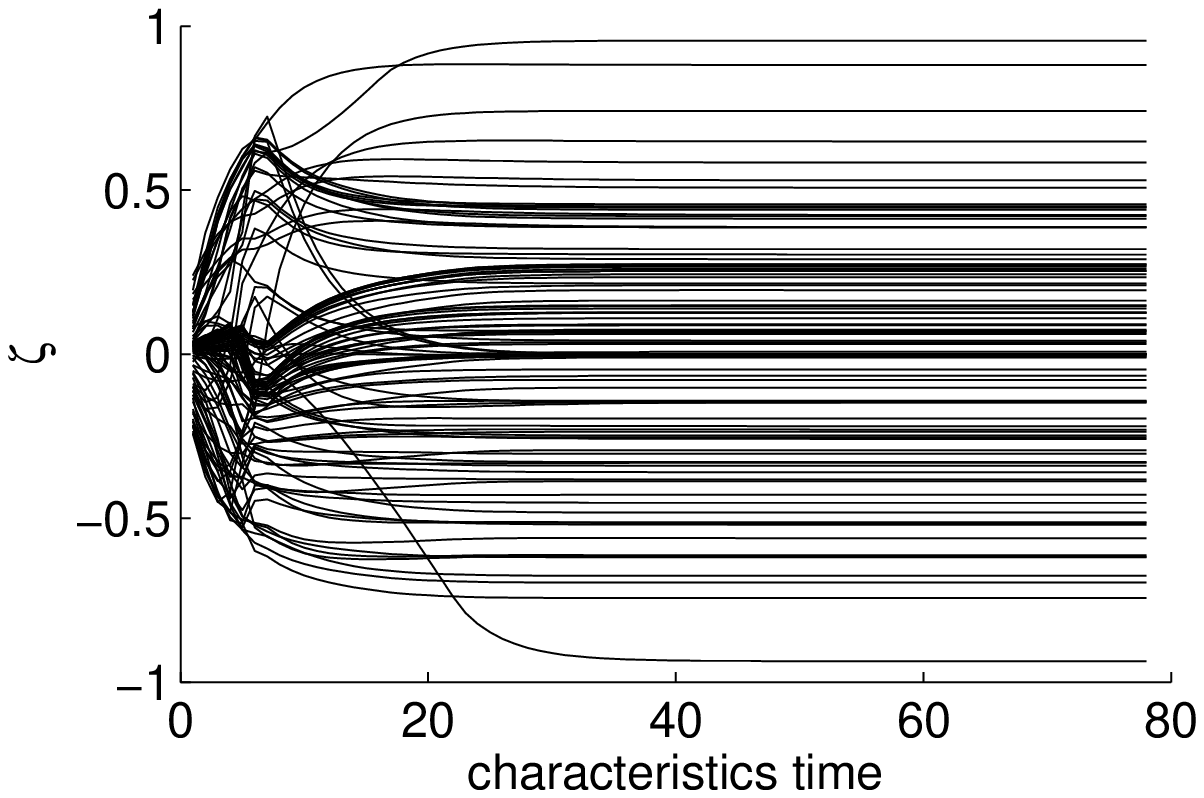}}
\subfloat[]{
\label{fig dynamics 4}
\includegraphics[width=1.6in]{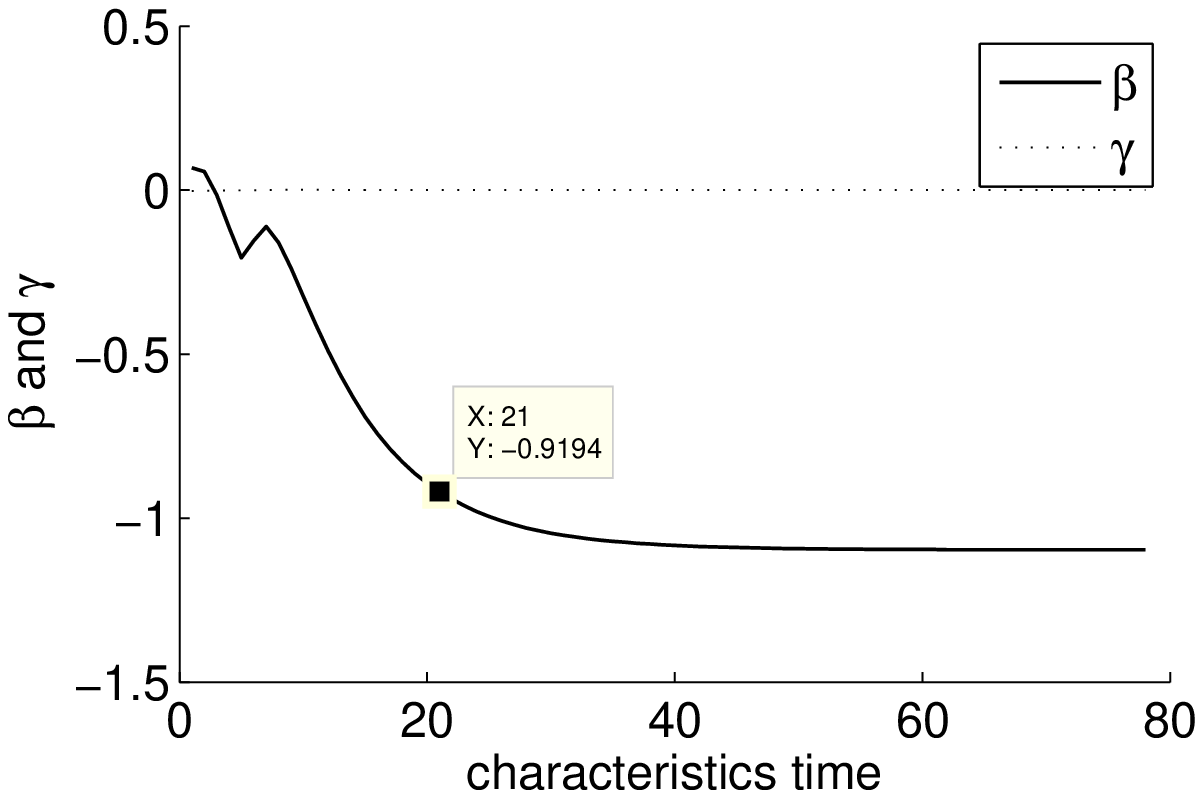}}
\caption{Dynamics of estimated parameters when the noise level is
equal to $0.8\sqrt{2}$. (a) $\bu$; (b) $\tilde{\bal}$; (c) $\bzeta$; (d) $\beta$ and $\gamma$}.
\label{dyn}
\end{figure}

Fig.~\ref{dyn} shows the dynamics of the estimated parameters in a typical experiment.
The settings are described in Section~V.B. It is seen that the network can settle down within 40 characteristic times.

\section{Stability of Proposed Algorithm}\label{section4}
For analog neural networks, the stability of its dynamics is a crucial property that needs to be investigated.
For the ellipse fitting model shown in \eqref{eq-1.19}, its global stability is hard to be proved.
In this section, we mainly discuss its local stability which means that a minimum point should be stable. Otherwise, the network can never converges to the minimum.

Let $\{\tilde{\bal}^{*},\ibu^{*},\bzeta^{*},\beta^{*},\gam^{*}\}$ be a minimum point of the dynamics given by \eqref{dynamic_a1}-\eqref{dynamic_a5}, $\tilde{\bal}^{*},\ibu^{*}$ is the corresponding state variable vector.
There are two sufficient conditions for local stability in the LPNN approach.
The first one is that the Hessian matrix of the Lagrangian \eqref{eq-1.20} at $\{\tilde{\bal}^{*},\ibu^{*},\bzeta^{*},\beta^{*},\gam^{*}\}$ should be positive definite.
It has been achieved by introducing the augmented terms.
Because according to ~\cite{zhang1992lagrange,paper:lpnn1,paper:lpnn2,lpnn3,lpnn4}, as long as the augmented terms are large enough, at an equilibrium point, the Hessian is positive definite under mild conditions.

The second condition is that at the minimum point, the gradient vectors of the constraints with respect to the state variables should be linearly independent.
In \eqref{eq-1.19}, we have $N+2$ constraints given by
\begin{eqnarray}
h_1(\tilde{\bal},\ibz)&=& \tilde{\bal}^\mathrm{ T } \bPhi \tilde{\bal}-1 \\
h_2(\tilde{\bal},\ibz)&=& \tilde{\bal}^\mathrm{T } \bTheta \tilde{\bal}-\epsilon \\
h_{i+2}(\tilde{\bal},\ibz)&=& z_i- \tilde{\bal}^\mathrm{T} \tilde{\ibx}_i, \quad i=1,\cdots,N.
\end{eqnarray}
The gradient vectors with respect to $\{\tilde{\bal}^{*},\ibu^{*}\}$ are given by
\begin{eqnarray}
\!\!\!\!\!\!\!\!\!\!\!\!& &\left\{
\left[
  \begin{array}{c}
    \displaystyle \frac{\partial h_1(\tilde{\bal}^*,\ibz^*) }{\partial \tilde{\bal}} \\
    \displaystyle \frac{\partial h_1(\tilde{\bal}^*,\ibz^*) }{\partial \ibu} \\
  \end{array}
\right],\displaystyle \cdots,
\left[
  \begin{array}{c}
    \displaystyle \frac{\partial h_{N+2} (\tilde{\bal}^*,\ibz^*) }{\partial \tilde{\bal}} \\
    \displaystyle \frac{\partial h_{N+2} (\tilde{\bal}^*,\ibz^*) }{\partial \ibu} \\
  \end{array}
\right]
\right\} \nonumber \\
\!\!\!\!\!\!\!\!\!\!\!\!& &=\!\!{\small
\left\{
\left[\!
  \begin{array}{c}
    2A \\
    2B \\
    2C \\
    2D \\
    2E \\
    2F \\
    0 \\
    0 \\
    0 \\
    \vdots \\
    0 \\
  \end{array}
\!\right]\!,\!
\left[\!
  \begin{array}{c}
    -2C \\
     B \\
    -2A \\
    0 \\
    0 \\
    0 \\
    G \\
    0 \\
    0 \\
    \vdots \\
    0 \\
  \end{array}
\!\right]\!,
\!\left[\!
  \begin{array}{c}
    -x_1^2 \\
     -x_1 y_1 \\
    -y_1^2 \\
    -x_1 \\
    -y_1 \\
    -1 \\
    0 \\
    g_1 \\
    0 \\
    \vdots \\
    0 \\
  \end{array}
\!\right]\!,\! \cdots\!,\!
\left[
 \! \!\begin{array}{c}
    -x_N^2 \\
     -x_N y_N \\
    -y_N^2 \\
    -x_N \\
    -y_N \\
    -1 \\
    0 \\
    0 \\
    0 \\
    \vdots \\
    g_N \\
  \end{array}
\!\!\right]
\right\}} \label{gradVector}
\end{eqnarray}
where
\beq
g_i &=& \frac{\partial h_{i+2}(\tilde{\bal},\ibz)}{\partial z_i}\frac{\partial z_i}{\partial u_i}
=\frac{1}{1 + \exp{(-\eta (|u_i| - \lambda))}} \nonumber \\
&&+ \frac{\eta(|u_i|-\delta\lambda)\exp{(-\eta (|u_i| - \lambda))}}{(1 + \exp{(-\eta (|u_i| - \lambda))})^2}. \nonumber
\eeq
For the case with $l_1$-norm objective function, $\eta\rightarrow \infty$, $\delta=1$.
If we assume $z_i\neq 0$, in other words, all data points are influenced by noise, thus we have $g_i=1$ for $\forall i=1,\dots,N$. When the proximate $l_0$-norm objective function is used, we let $\eta$ be a large positive number, $\delta=0$. Without any assumption, we can deduce that, for $\forall i=1,\dots,N$, $g_i$ is a positive constant.

In \eqref{gradVector}, there are $N+2$ gradient vectors where each has $N+7$ elements.
Firstly, it is easy to note that the last $N$ vectors are linear independent with each other.
Besides, they are all linear independent with the first two vectors.
Secondly, the first two vectors are linear independent with each other.
Because to make sure the fitting result is an ellipse, $B^2-4AC<0$, i.e., for $B^2-4AC+G^2=\epsilon$, ($\epsilon$ is a very small negative value, which can be considered as $0$ here), we can deduce that $G^2>0$, in other words $G\neq0$. And for satisfying $(\tilde{\bal}^*)^\mathrm{T} \bPhi{\tilde{\bal}^*}=1$, the first vector cannot be $\mathbf{0}$.
Therefore, $\{\tilde{\bal}^{*},\ibu^{*},\bzeta^{*},\beta^{*},\gam^{*}\}$ is an asymptotically stable point of the neural network. For any points nearby, they must converge to this minimum point.
\begin{figure}[h]
\centering
\centerline{\includegraphics[height=2in]{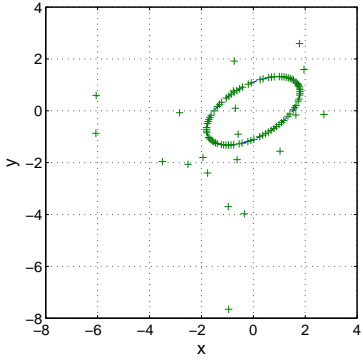}}
\caption{Ellipse data with 20 scattering points contaminated by Laplacian noise.}
\label{Laplacenoise}
\end{figure}

\section{Numerical Examples}\label{section5}
In this section, we conduct simulations and experiments to evaluate the performance of the proposed LPNN approach.
We evaluate our algorithm with the $l_2$-norm, $l_1$-norm and proximate $l_0$-norm.
For the $l_2$-norm, we apply the LPNN to solve:
\begin{subequations}\label{eq:l2}
\beq
\min\limits_{\bal,\ibz} \,\,&
\|\ibz\|^2_2, \\
\mbox{s.t.}\,\, &\ibz=\ibX^\mathrm{T} \tilde{\bal}, \\
&\tilde{\bal}^ \mathrm{ T }\bPhi\tilde{\bal}=1,\\
&\tilde{\bal}^ \mathrm{ T }\bTheta\tilde{\bal}=\epsilon.
\eeq
\end{subequations}
It is expected that \eqref{eq:l2} is just an alternative implementation of the CLS estimator in \cite{gander1994least} with an additional constraint to make sure the fitting result is an ellipse.
In the $l_1$-norm version, we set $\eta\rightarrow \infty$, $\delta=1$ and $\lambda=1$. That means the threshold is given by
\begin{equation}
z_i = T_{1}(u_i) = \left\{ \begin{array}{lcl}
0, &  |u_i| \leq 1, \\
u_i - \mbox{sign} (u_i),  & |u_i| > 1.
\end{array}\right.
\end{equation}
For LPNN with the proximate $l_0$-norm, we set $\eta=10,000$, $\delta=0$ and $\lambda=1$, and the threshold is:
\begin{equation}
z_i=T_{(10000,0,1)}(u_i)=\mbox{sign} (u_i)\frac{ |u_i|}{1 + e^{-10000(|u_i|-1)}}.
\end{equation}

Then, we discuss the parameter settings and initialization. $C_0, C_1, C_2$ are three tuning parameters, and we use trial-and-error method to select them.
We try 6 $C_0$ values: $C_0=\{1,2,3,4,5,6\}$ and 6 $C_1$, $C_2$ values: $C_1=C_2=\{2,4,6,8,10,12\}$, and finally choose $C_0=5, C_1=10, C_2=10$.
In the discrete simulation, the step size $\mu$ is selected as $0.0001$.
We also need to initialize the state variables $\tilde{\bal}$ and $\ibu$, and the Lagrangian variables $\bzeta, \beta$ and $\gam$.
The $\tilde{\bal}$ is not initialized with the CLS method because its solution may not correspond to an ellipse.
Instead, we compute the initial estimate of $\tilde{\bal}$ by assuming that the data points are sampled from a circle.
That is, the circle center is given by the midpoint of the data set while the radius is a small positive random value. Once the circle is constructed, it is easy to initialize $\tilde{\bal}$.
We can also get initial estimates of $\bu$ by $\bu=\tilde{\ibX}^\mathrm{T} {\tilde{\bal}}$.
The initial values of the Lagrangian variables $\lam, \beta$ and $\gam$ are small random values.

Several state-of-the-art ellipse fitting algorithms are implemented for performance comparison. They are the direct least squares fitting (DLSF)~\cite{fitzgibbon_direct_1999}, SBM \cite{liang_robust_2013}, and RCLS \cite{liang_robust_2015}. Note that for the DLSF algorithm, it solves a generalized eigenvalue problem to fit an ellipse.
The SBM method~\cite{liang_robust_2013} introduced two regularized terms and determines ellipse parameters by solving a second-order cone programming (SOCP) problem.
The RCLS algorithm combines the maximum correntropy criterion with the CLS method.

\begin{figure}[ht]
\begin{tabular}{@{\extracolsep{1mm}}c@{\extracolsep{1mm}}c}
\mbox{\epsfig{figure=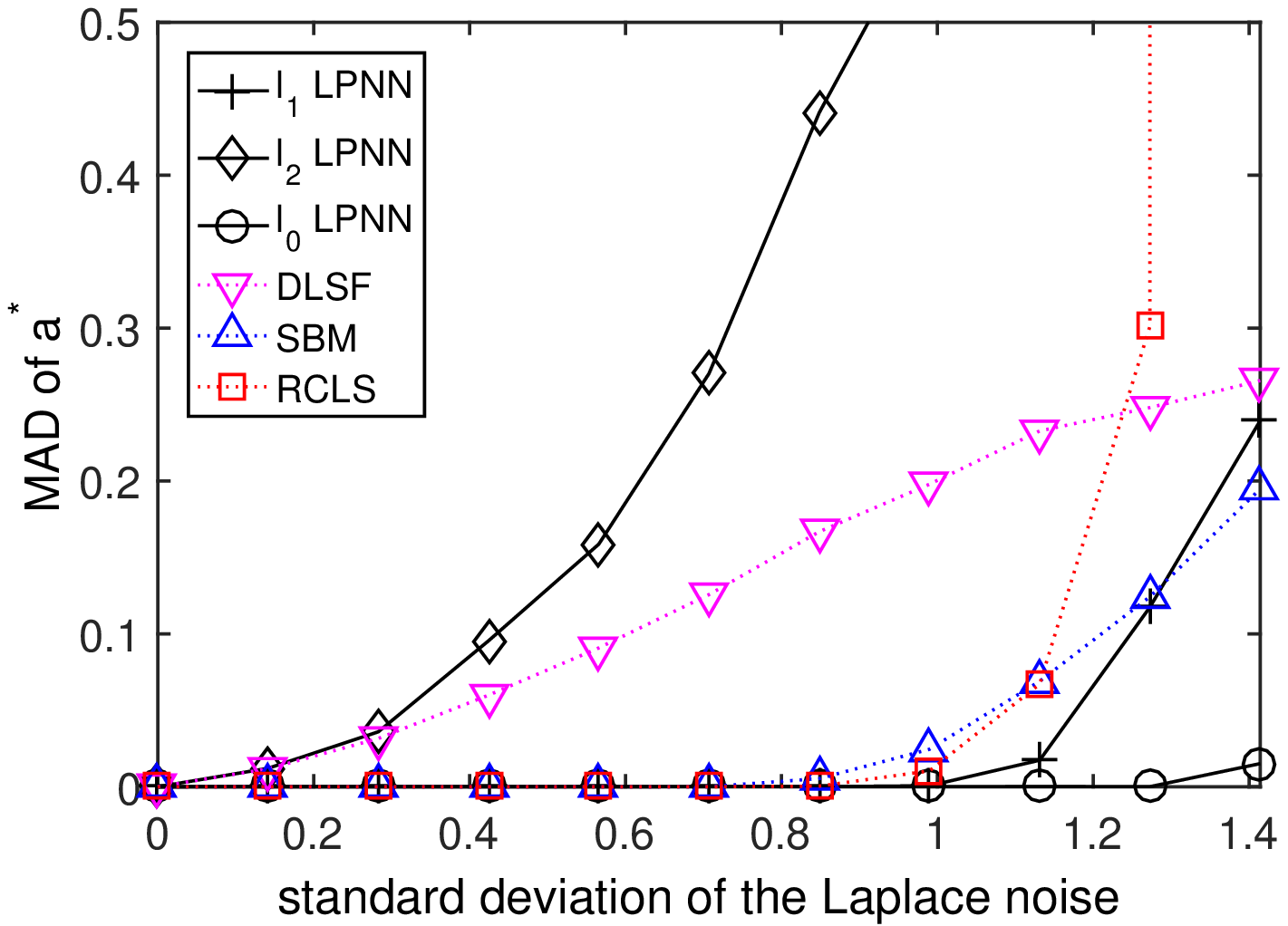,width=1.65in}} &
\mbox{\epsfig{figure=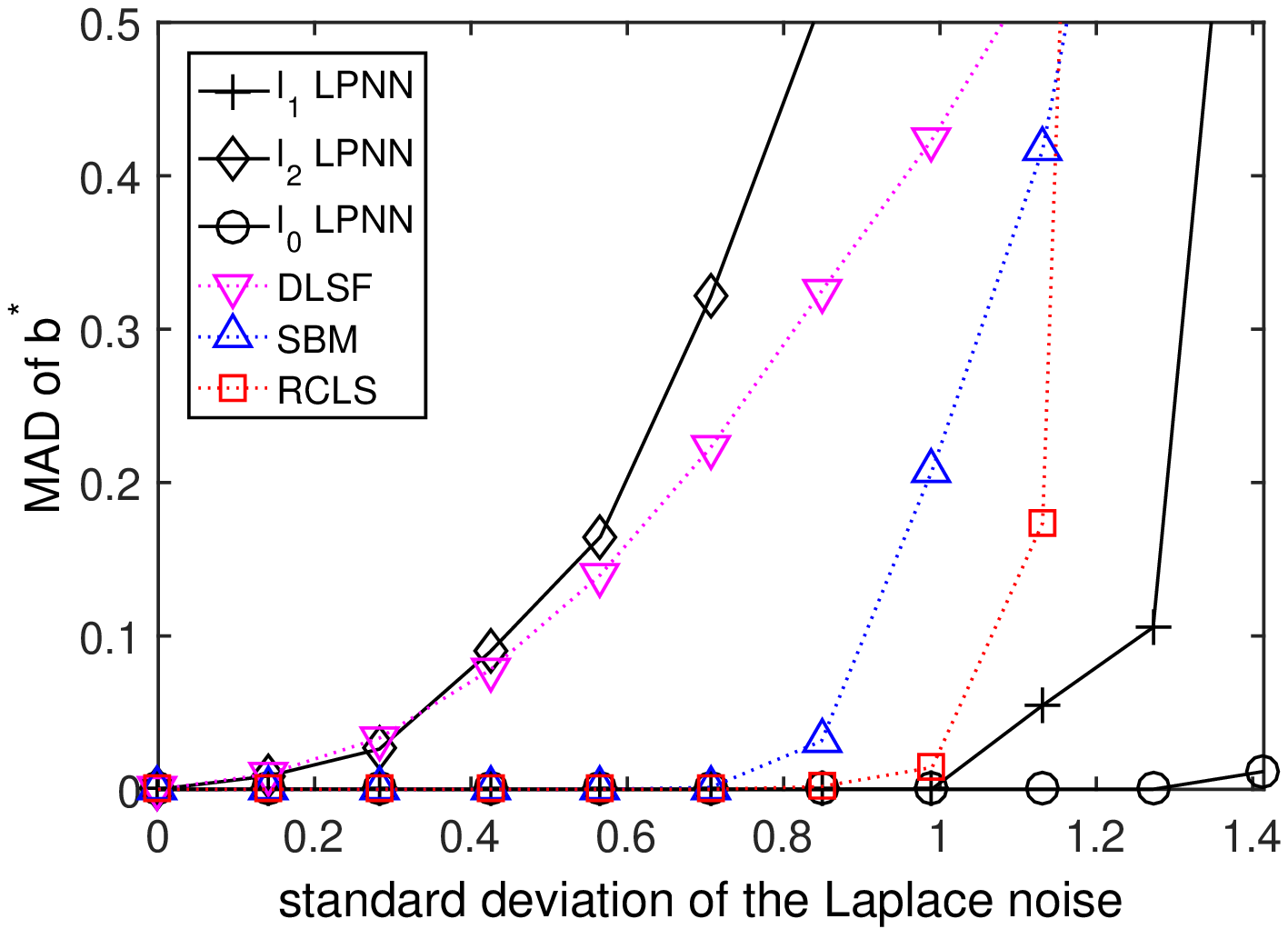,width=1.65in}}  \\
(a) MAD of $a^*$ & (b) MAD of $b^*$  \\
\mbox{\epsfig{figure=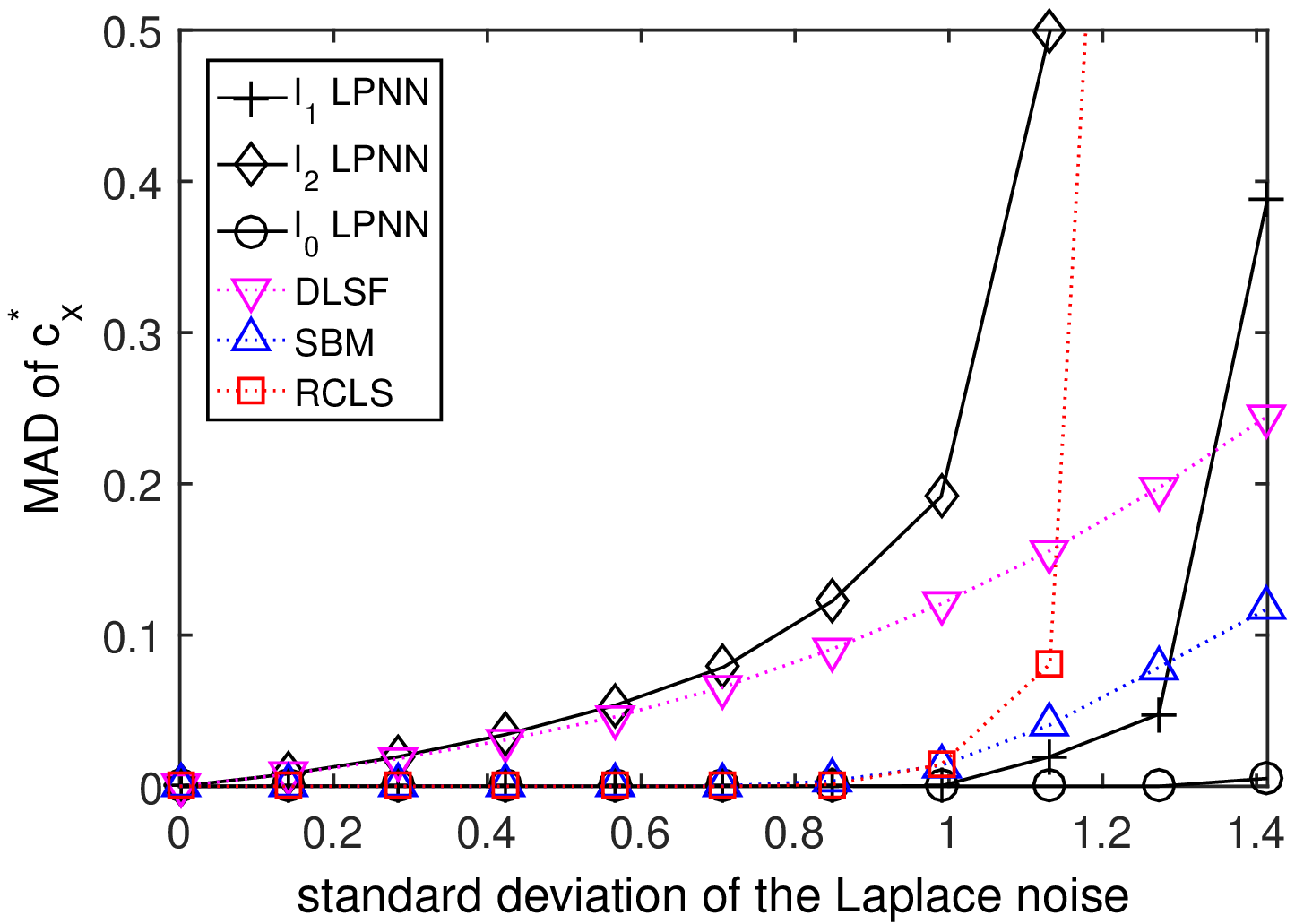,width=1.65in}} &
\mbox{\epsfig{figure=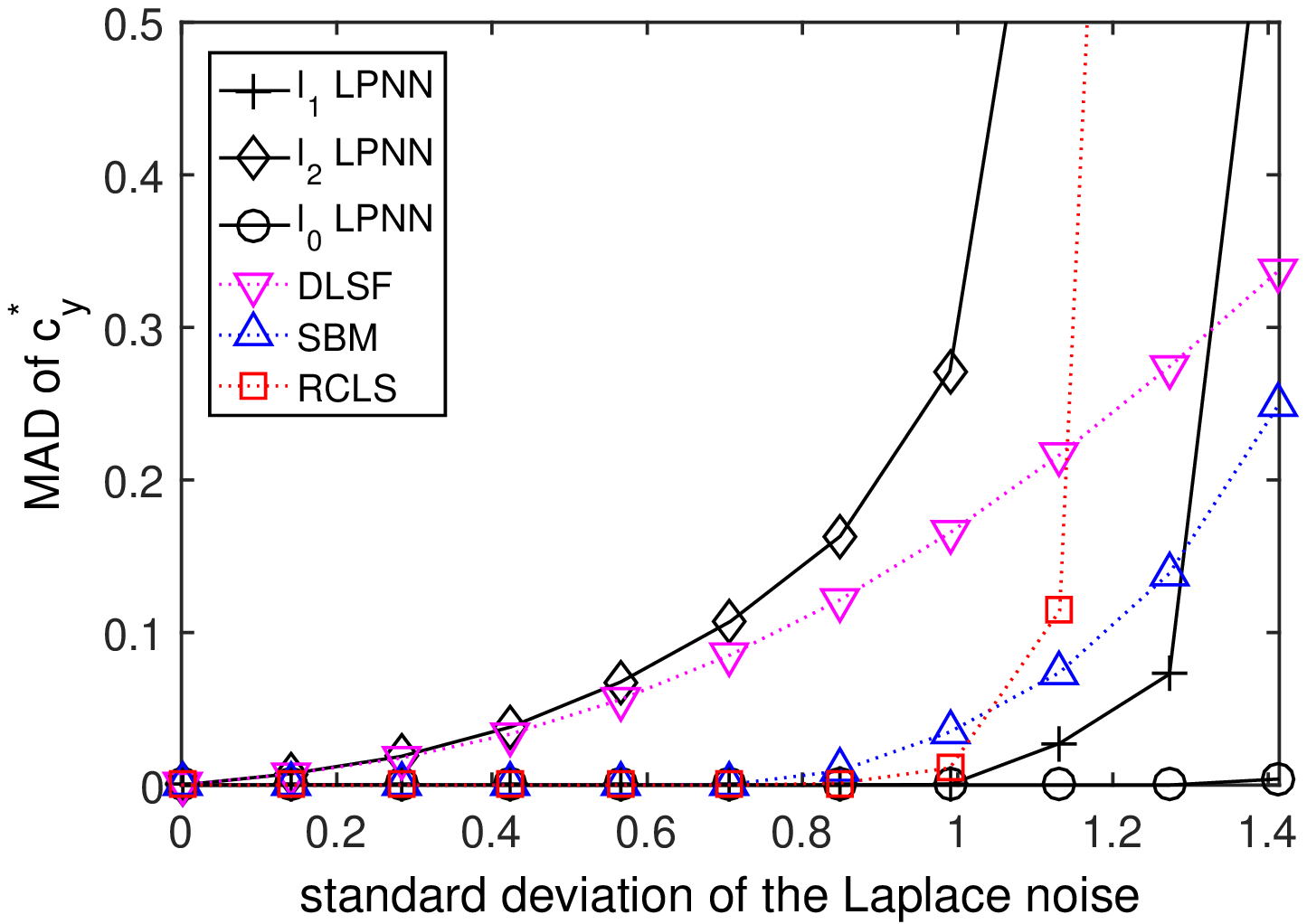,width=1.65in}} \\
(c) MAD of $c^*_x$ & (d) MAD of $c^*_y$ \\
\mbox{\epsfig{figure=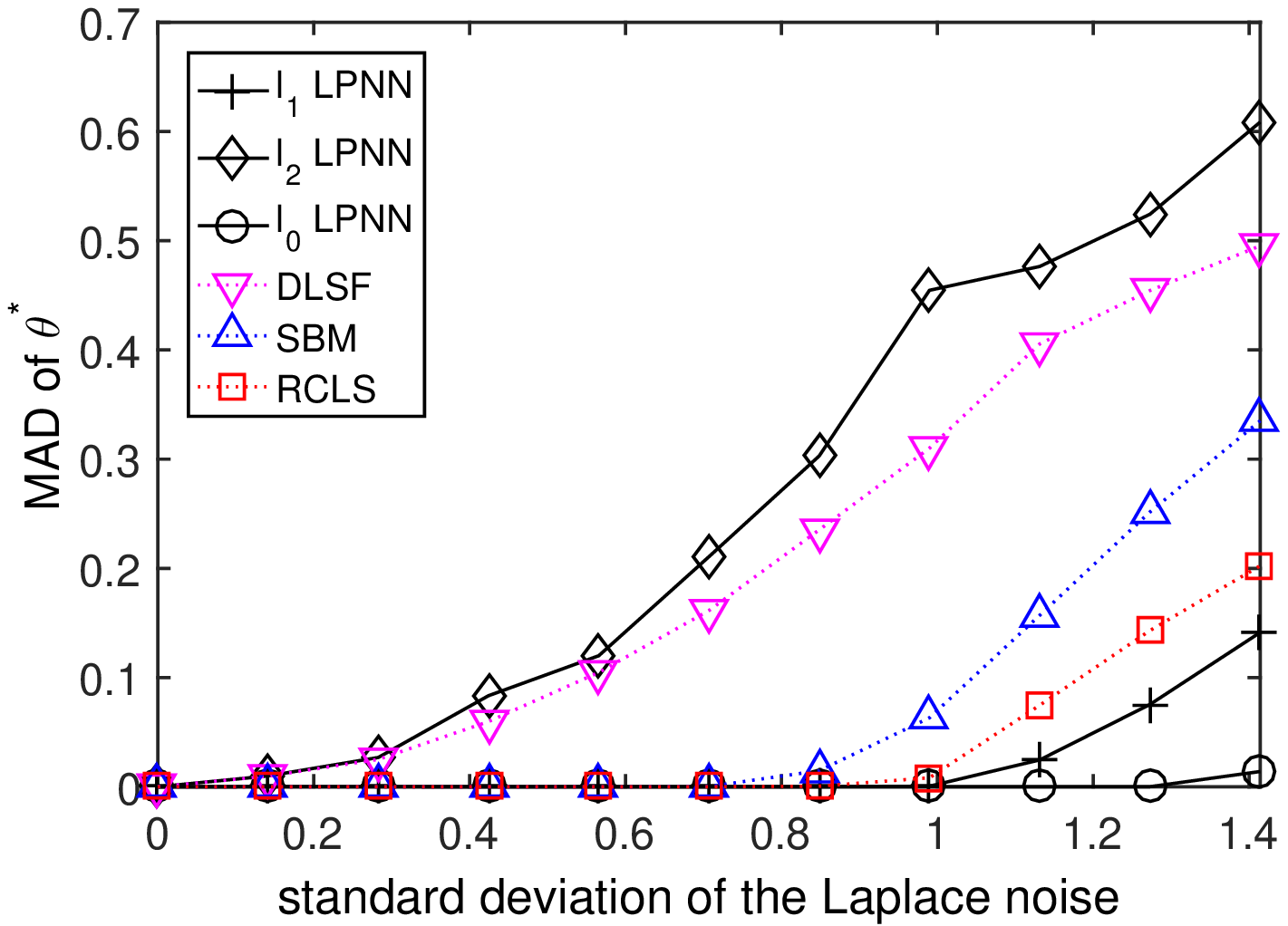,width=1.65in}} & \\
(e) MAD of $\theta^*$ &
\end{tabular}
\caption{MAD results of different algorithms. The Laplacian noise level is varied from 0 to $\sqrt{2}$.}
\label{Comparing_Laplace}
\end{figure}

\subsection{Experiment 1: Ellipse Fitting in Laplacian Noise}
In this experiment, we test the performance of our proposed approach in different Laplacian noise levels. Firstly, we generate an ellipse with $100$ data points, which is shown in Fig.~\ref{Laplacenoise}.
The true elliptical parameters are {$c_x=0$, $c_y=0$, $a=2$, $b=1$, $\theta=30^{\circ}$}.
We then add small Gaussian noise with variance $10^{-8}$ to these points, randomly choose $20$ points from the data set and add zero-mean Laplacian noise into them, which is also illustrated in Fig.~\ref{Laplacenoise}.
The standard deviation of the Laplacian noise is varied from 0 to $\sqrt{2}$. We repeat the experiment $100$ times at each noise level, and compute the mean absolute deviation (MAD) of the estimated parameters ($c_x^*$, $c_y^*$, $a^*$, $b^*$, $\theta^*$). The results are shown in Fig.~\ref{Comparing_Laplace}. It is seen that the $l_2$-norm LPNN and DLSF algorithms are very sensitive to outliers. The SBM and RCLS methods can effectively decrease the impact of outliers. However, both of them start to break down when the Laplacian noise level is greater than $0.9899$. For $l_1$-norm LPNN, we can increase the threshold point to $1.1314$. Furthermore, the $l_0$-norm LPNN still works very well up to the noise level of $\sqrt{2}$.

\begin{figure}
\begin{tabular}{@{\extracolsep{2mm}}c@{\extracolsep{1mm}}c@{\extracolsep{1mm}}c}
\mbox{\epsfig{figure=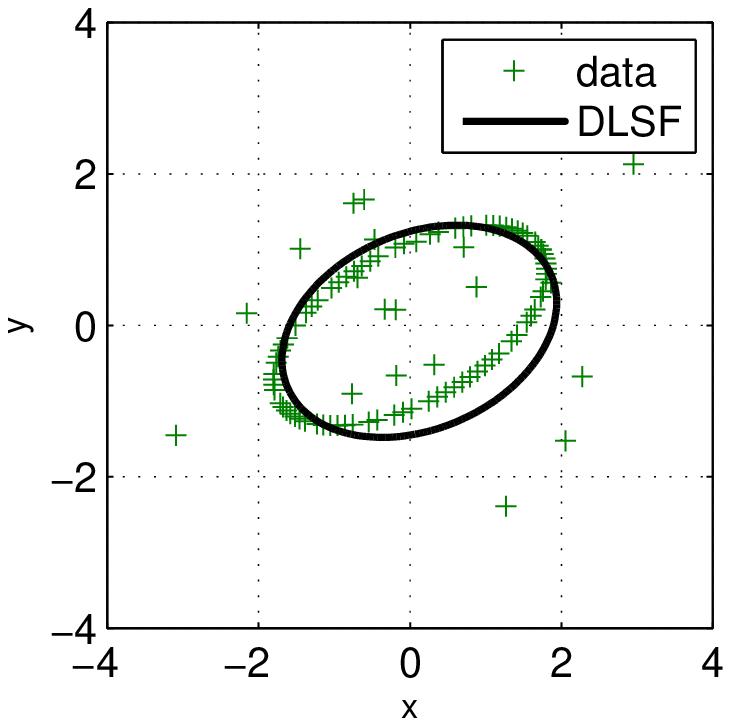,width=1.1in}} &
\mbox{\epsfig{figure=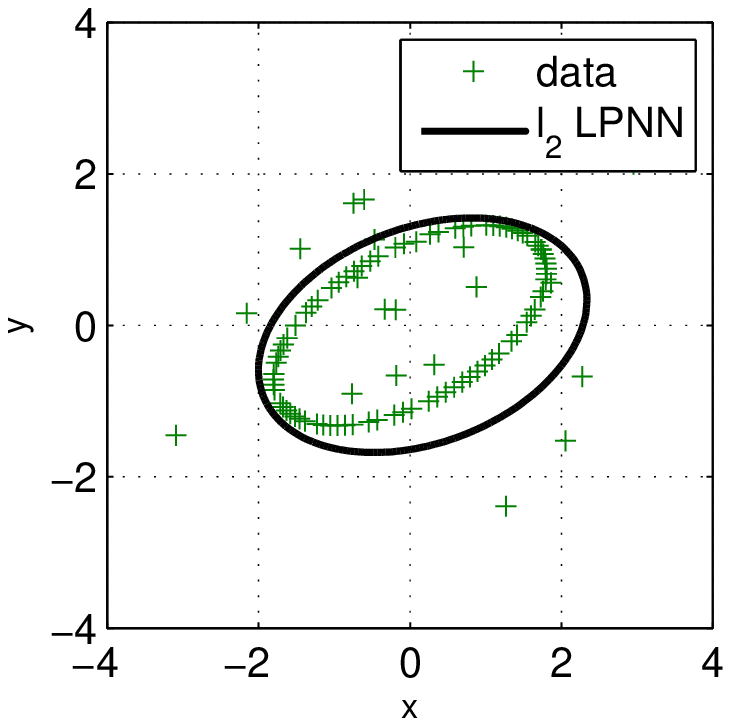,width=1.1in}} &
\mbox{\epsfig{figure=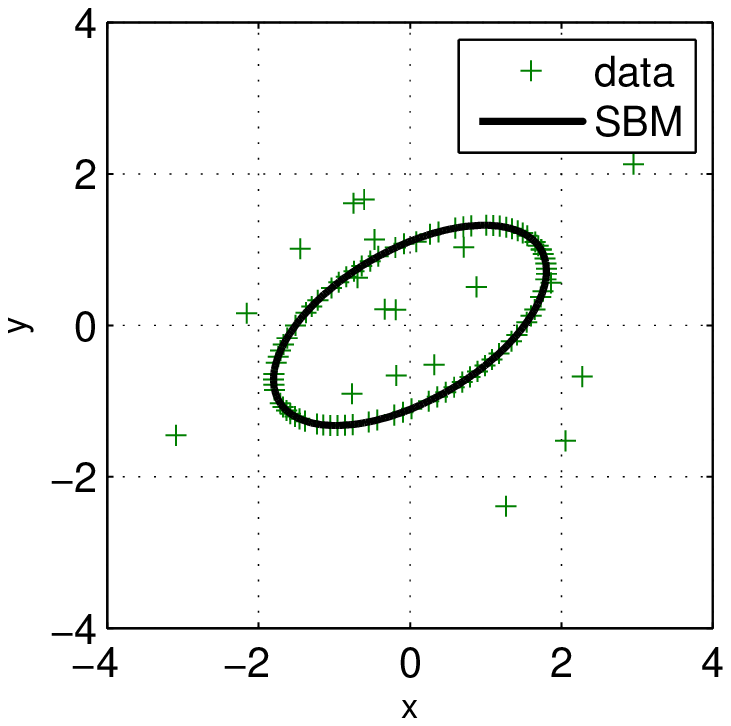,width=1.1in}} \\
(a) DLSF & (b) $l_2$ LPNN & (c) SBM \\
\mbox{\epsfig{figure=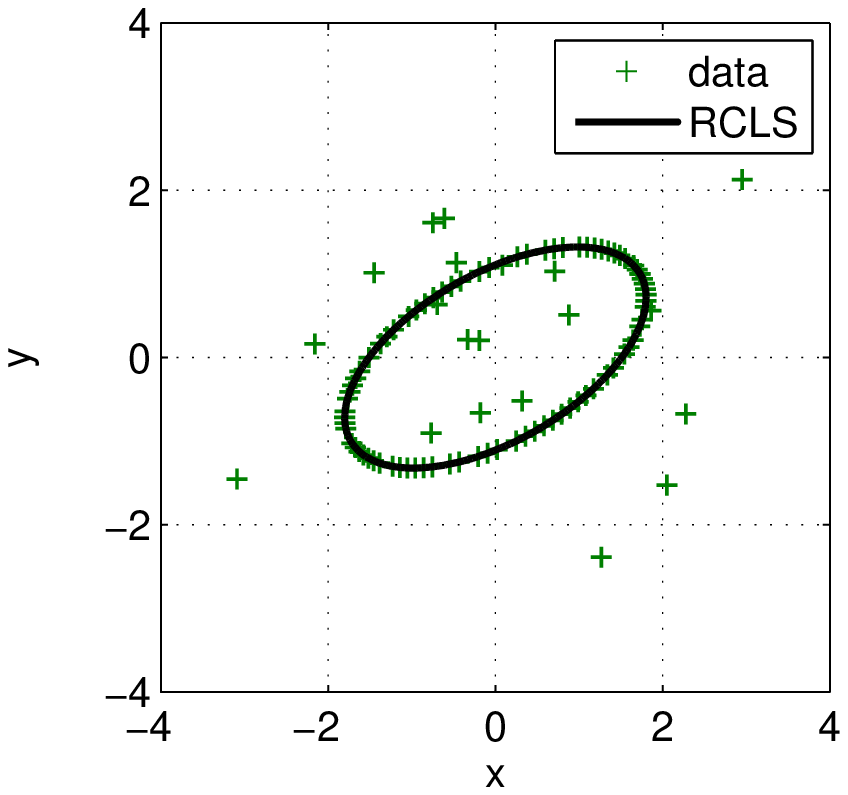,width=1.05in}} &
\mbox{\epsfig{figure=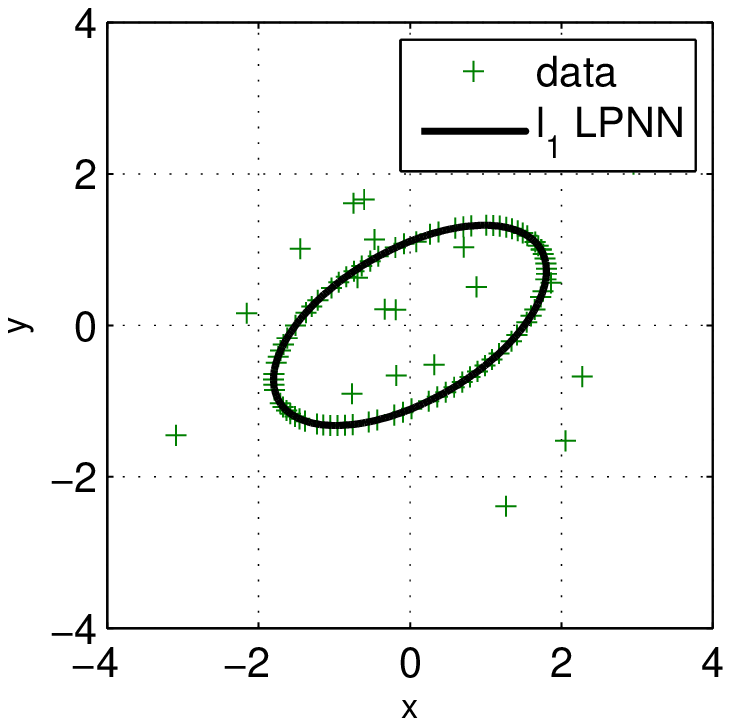,width=1.1in}} &
\mbox{\epsfig{figure=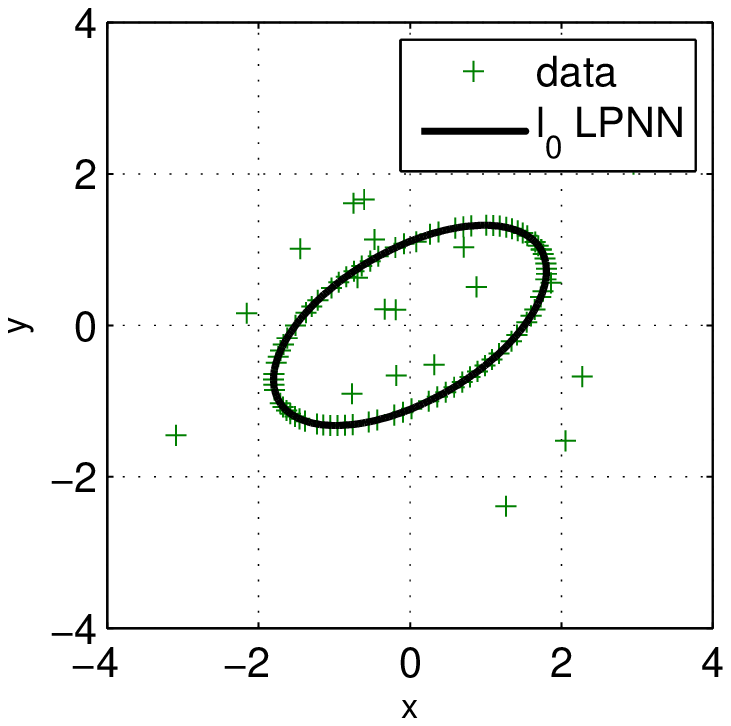,width=1.1in}} \\
 (d) RCLS & (e) $l_1$ LPNN & (f) $l_0$ LPNN
\end{tabular}
\caption{Fitting result of a typical run at Laplacian noise of $0.7\sqrt{2}$ (around $0.9899$).}
\label{lap_result_70}
\end{figure}

\begin{figure}
\begin{tabular}{@{\extracolsep{-2mm}}c@{\extracolsep{1mm}}c@{\extracolsep{1mm}}c}
\mbox{\epsfig{figure=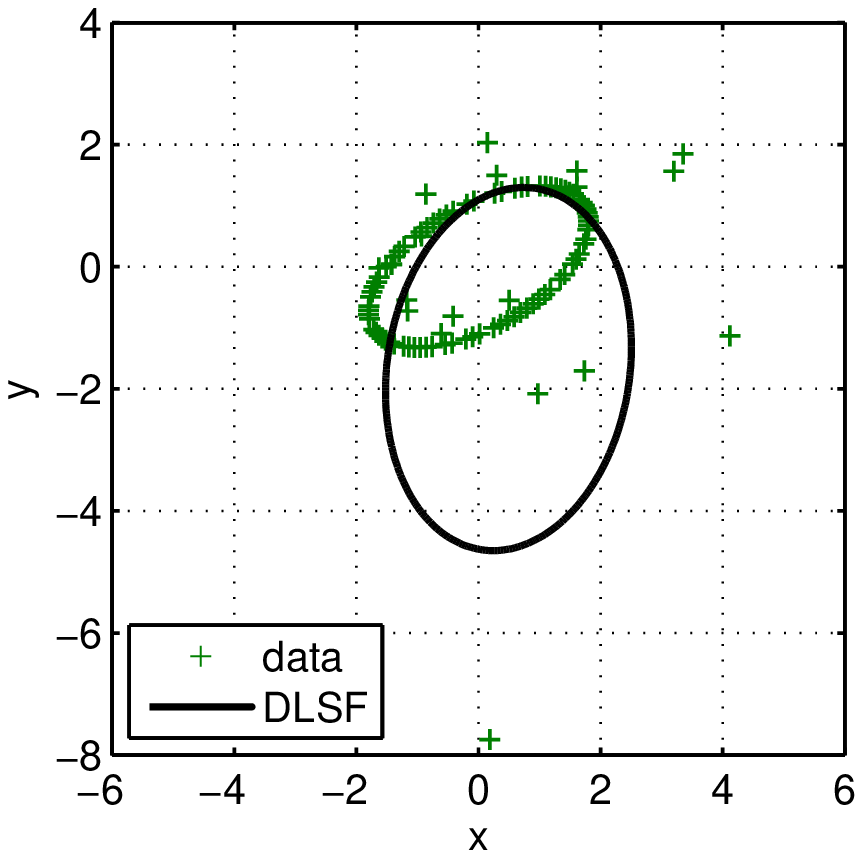,width=1.05in}} &
\mbox{\epsfig{figure=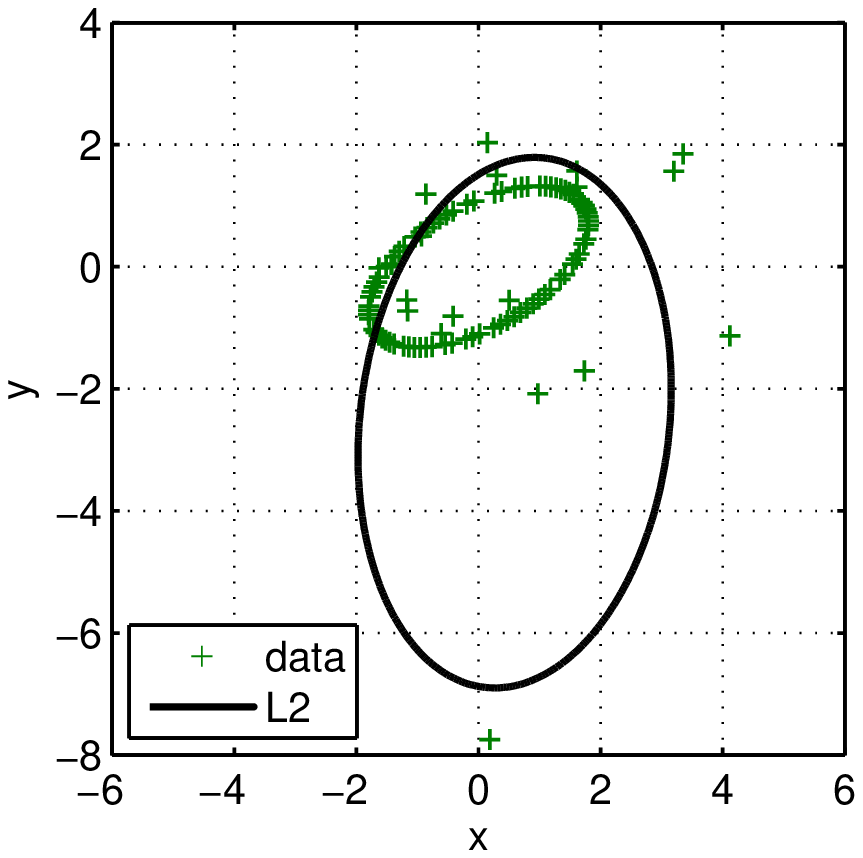,width=1.05in}} &
\mbox{\epsfig{figure=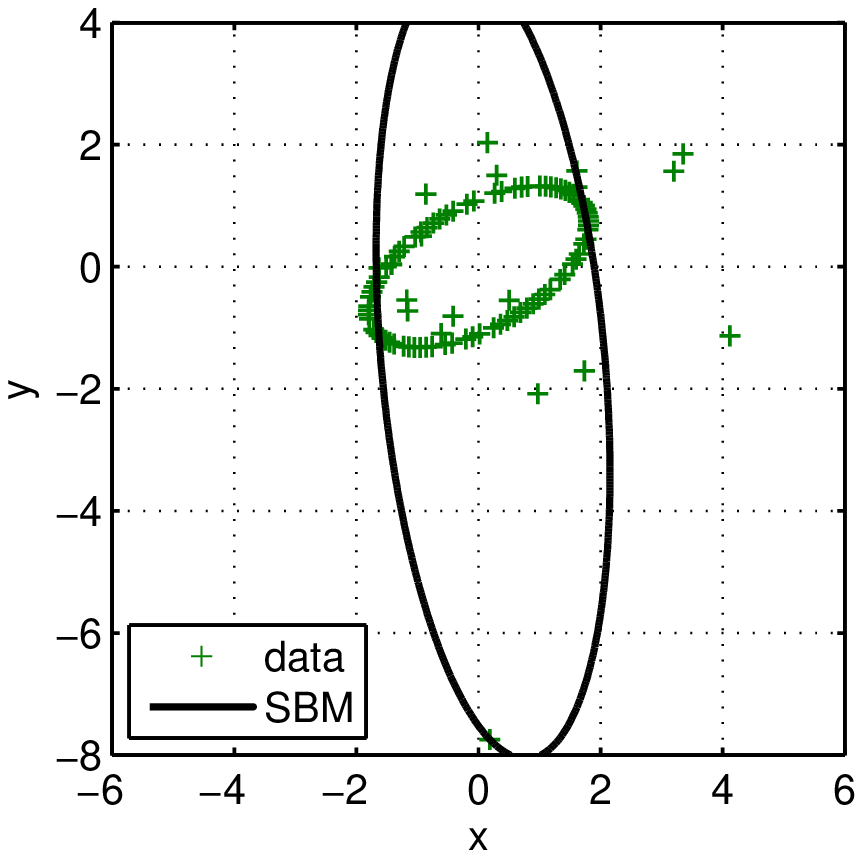,width=1.05in}} \\
(a) DLSF & (b) $l_2$ LPNN & (c) SBM \\
\mbox{\epsfig{figure=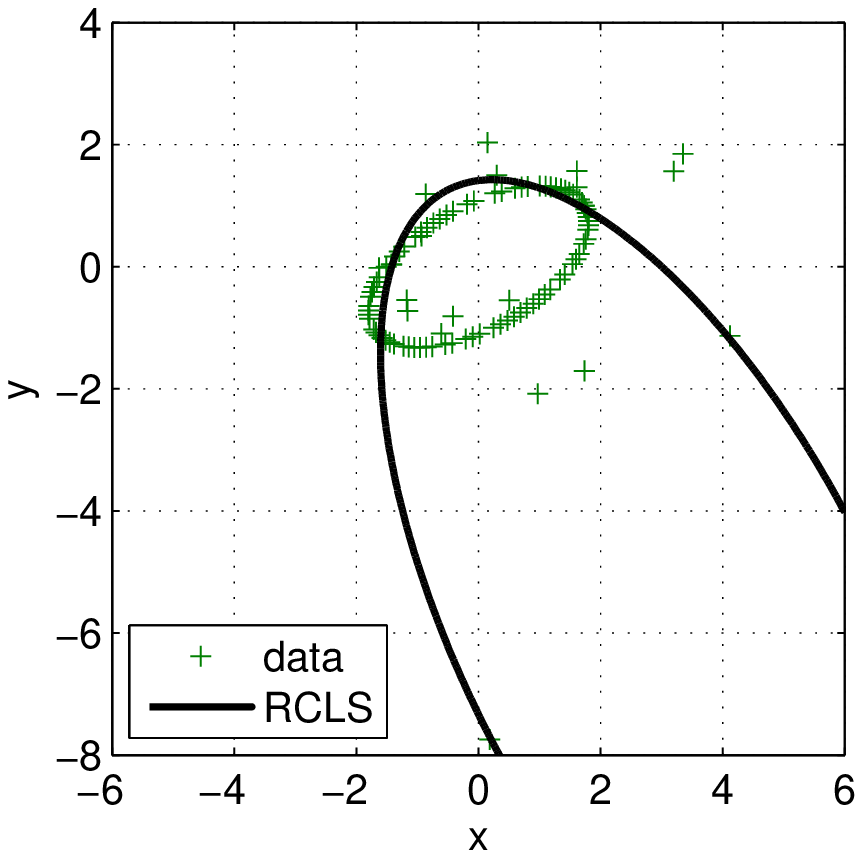,width=1.05in}} &
\mbox{\epsfig{figure=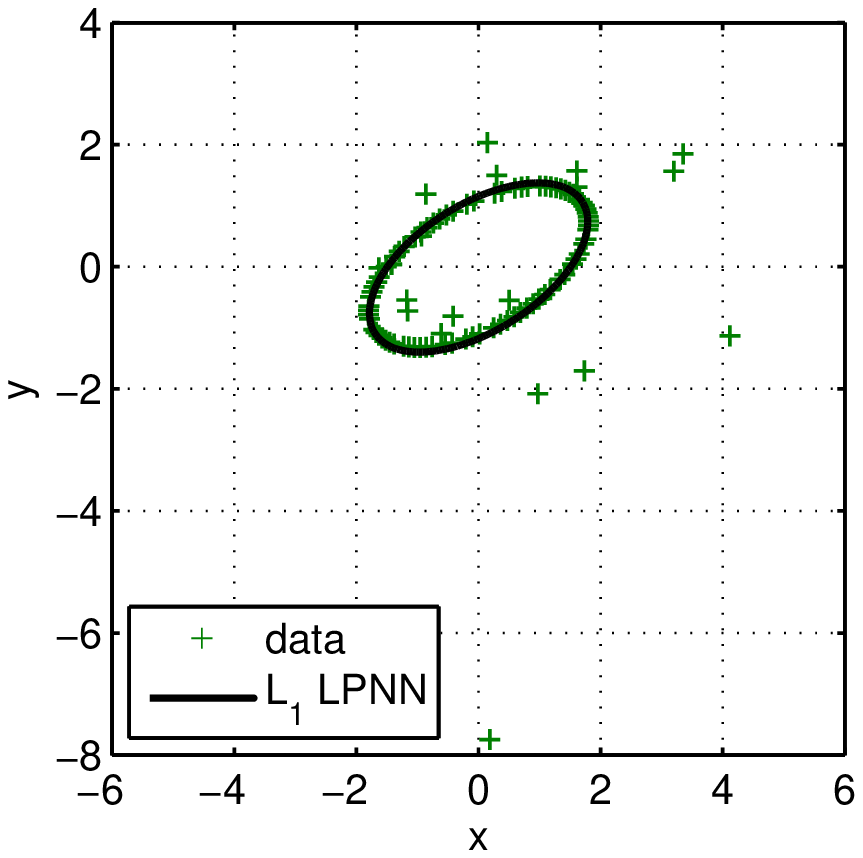,width=1.05in}} &
\mbox{\epsfig{figure=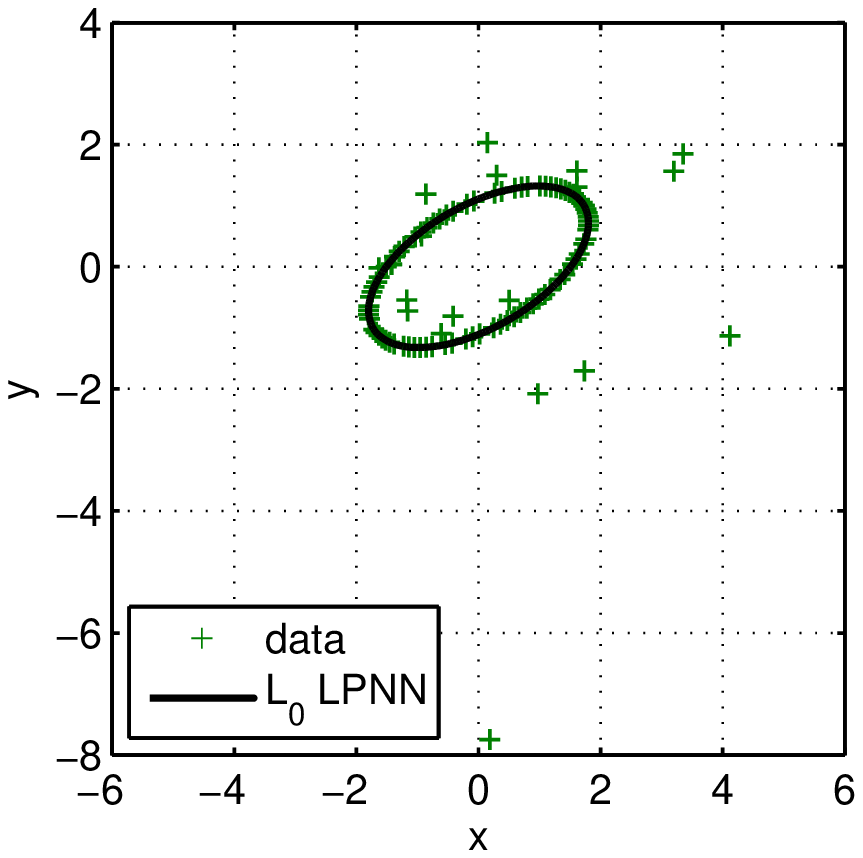,width=1.05in}} \\
 (d) RCLS & (e) $l_1$ LPNN & (f) $l_0$ LPNN
\end{tabular}
\caption{Fitting result of a typical run at Laplacian noise of $0.9 \sqrt{2}$ (around $1.2728$).}
\label{lap_result_90}
\end{figure}

\begin{figure}
\begin{tabular}{@{\extracolsep{-2mm}}c@{\extracolsep{1mm}}c@{\extracolsep{1mm}}c}
\mbox{\epsfig{figure=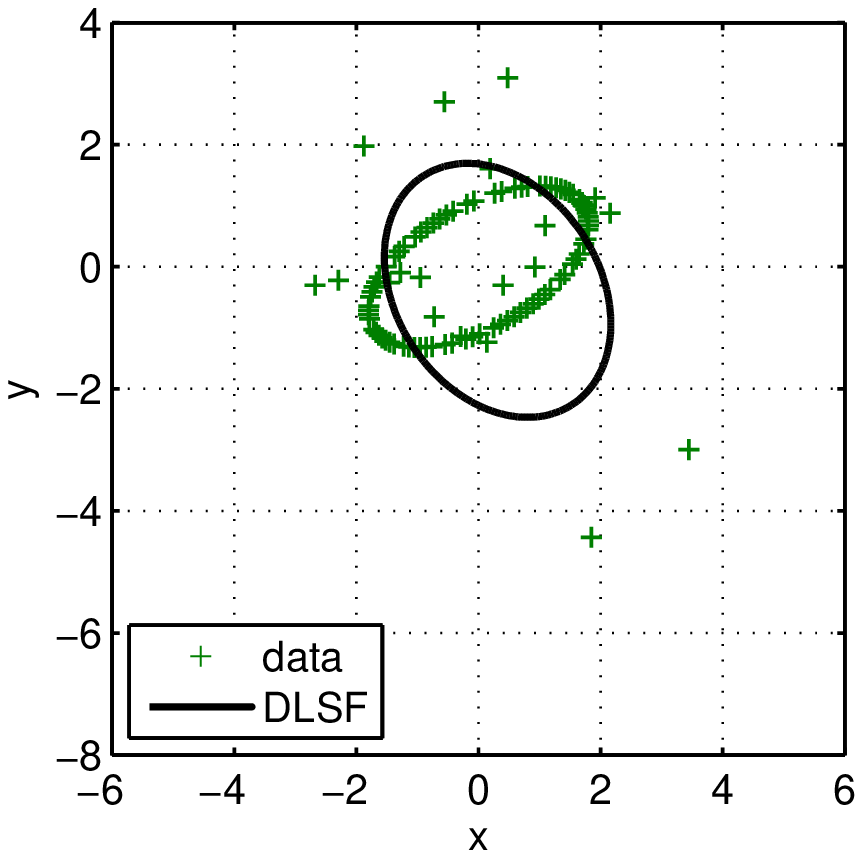,width=1.05in}} &
\mbox{\epsfig{figure=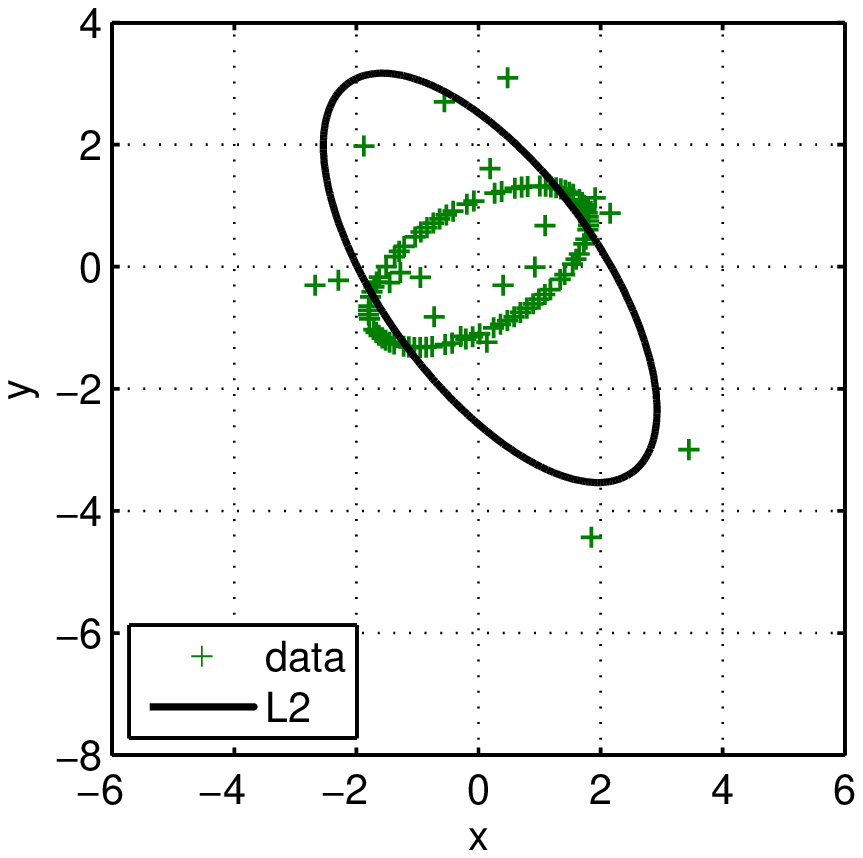,width=1.05in}} &
\mbox{\epsfig{figure=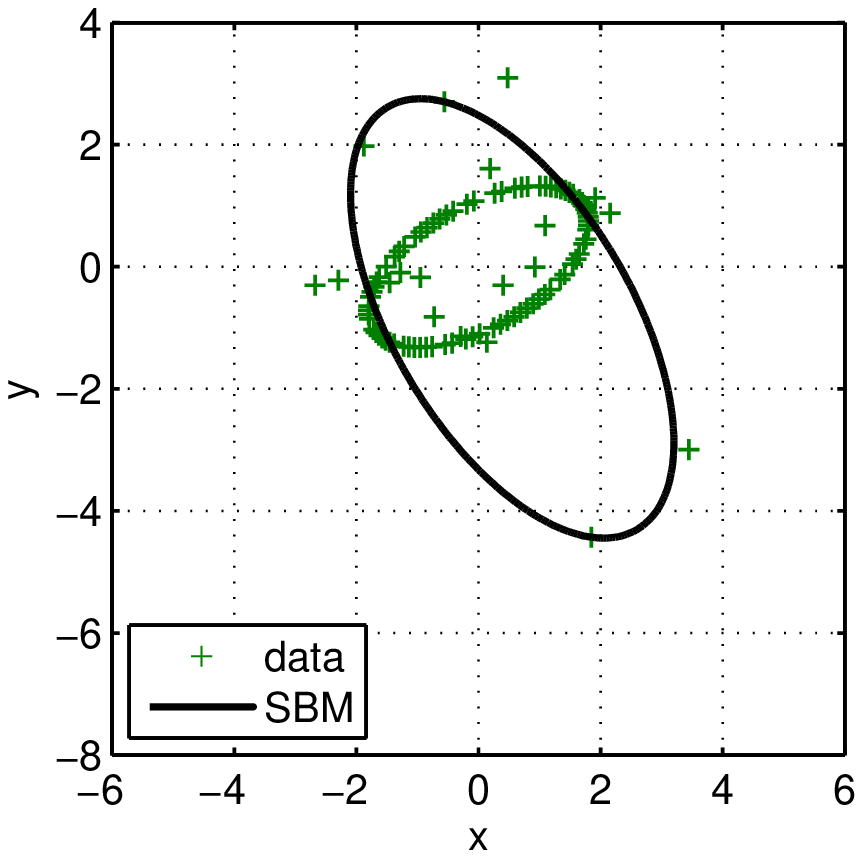,width=1.05in}} \\
(a) DLSF & (b) $l_2$ LPNN & (c) SBM \\
\mbox{\epsfig{figure=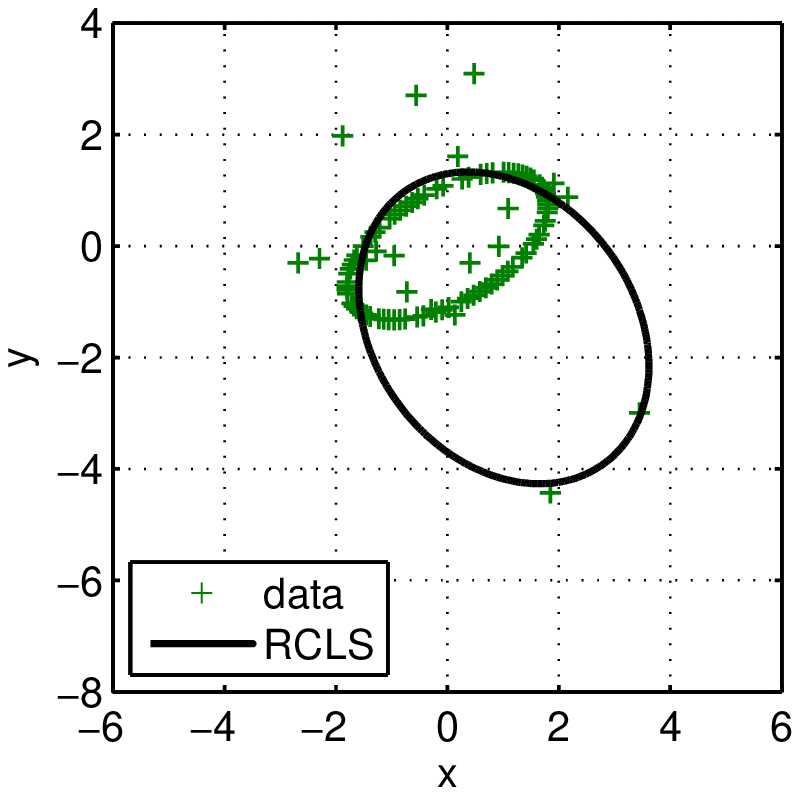,width=1.05in}} &
\mbox{\epsfig{figure=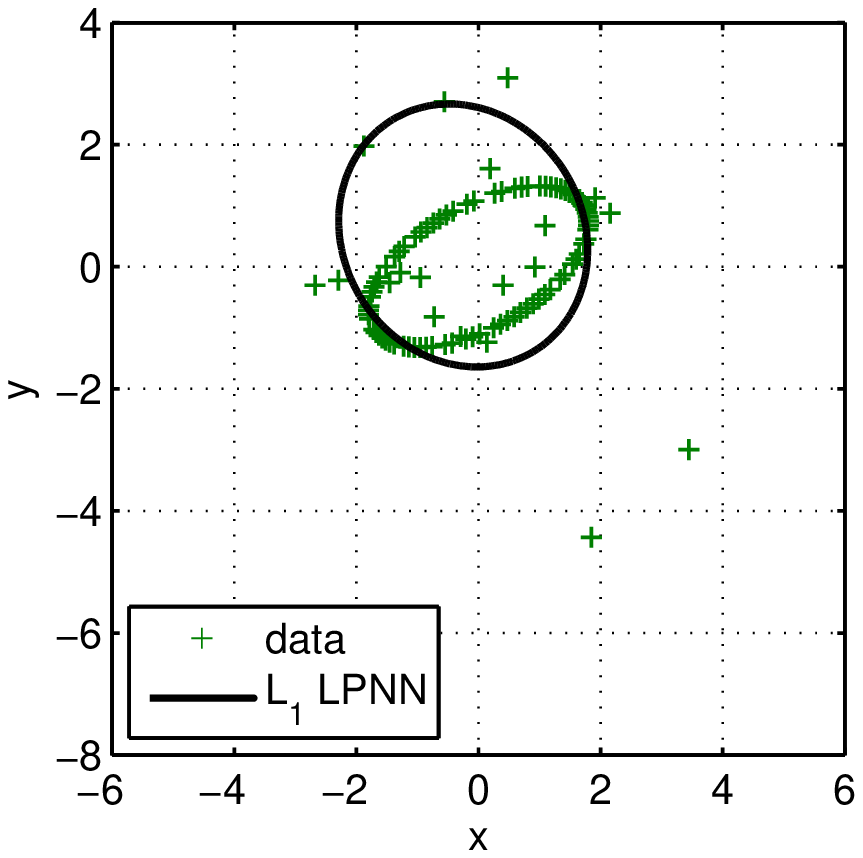,width=1.05in}} &
\mbox{\epsfig{figure=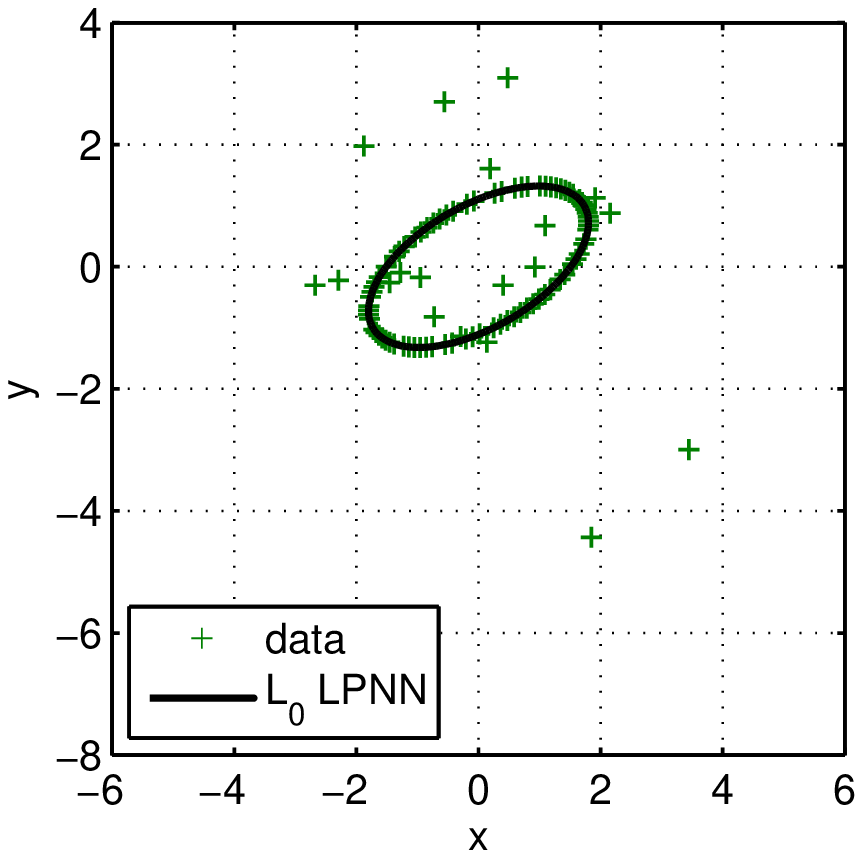,width=1.05in}} \\
 (d) RCLS & (e) $l_1$ LPNN & (f) $l_0$ LPNN
\end{tabular}
\caption{Fitting result of a typical run at Laplacian noise of $\sqrt{2}$ (around $1.4142$).}
\label{lap_result_100}
\end{figure}

\begin{figure}[ht]
\begin{tabular}{@{\extracolsep{1mm}}c@{\extracolsep{1mm}}c}
\mbox{\epsfig{figure=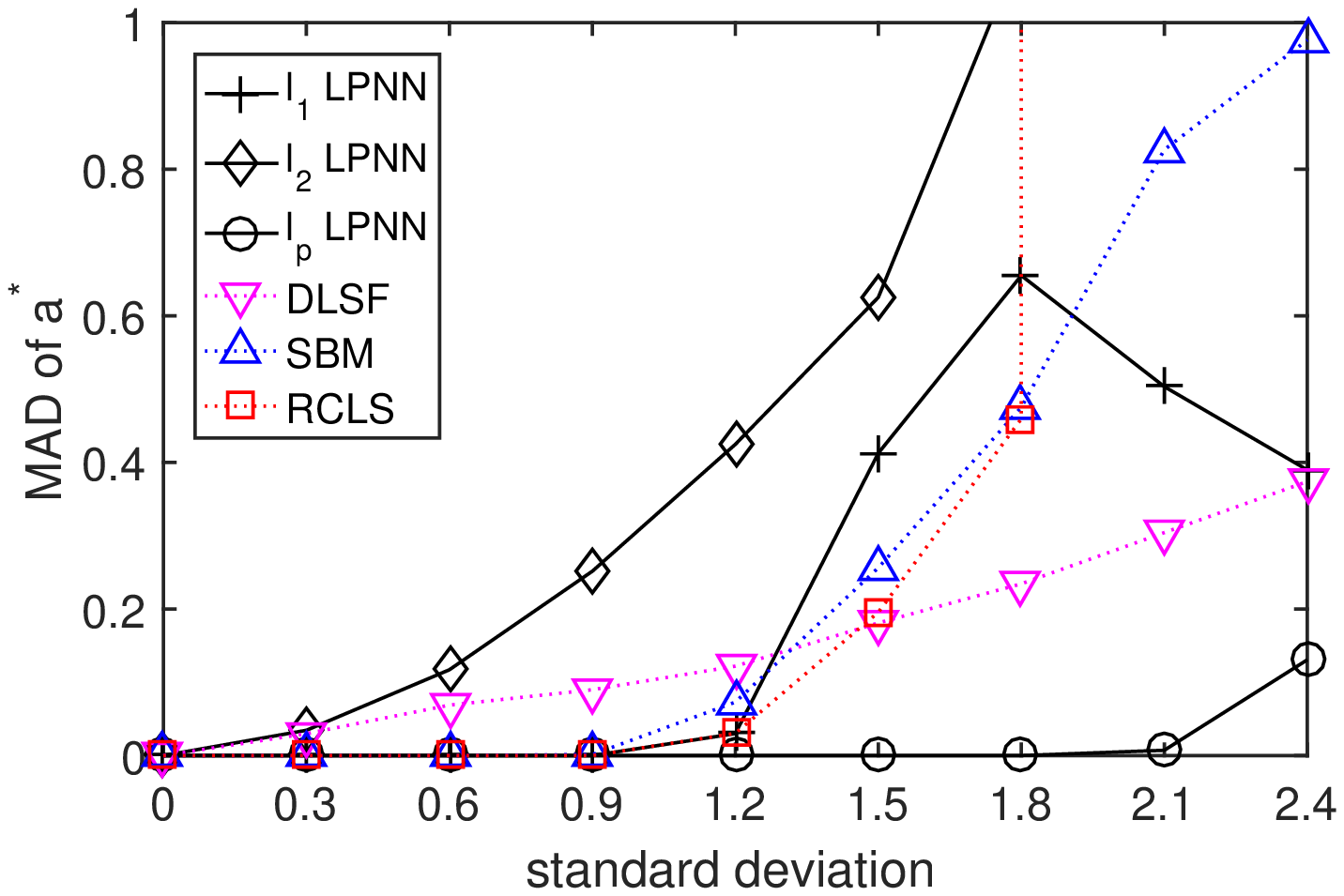,width=1.65in}} &
\mbox{\epsfig{figure=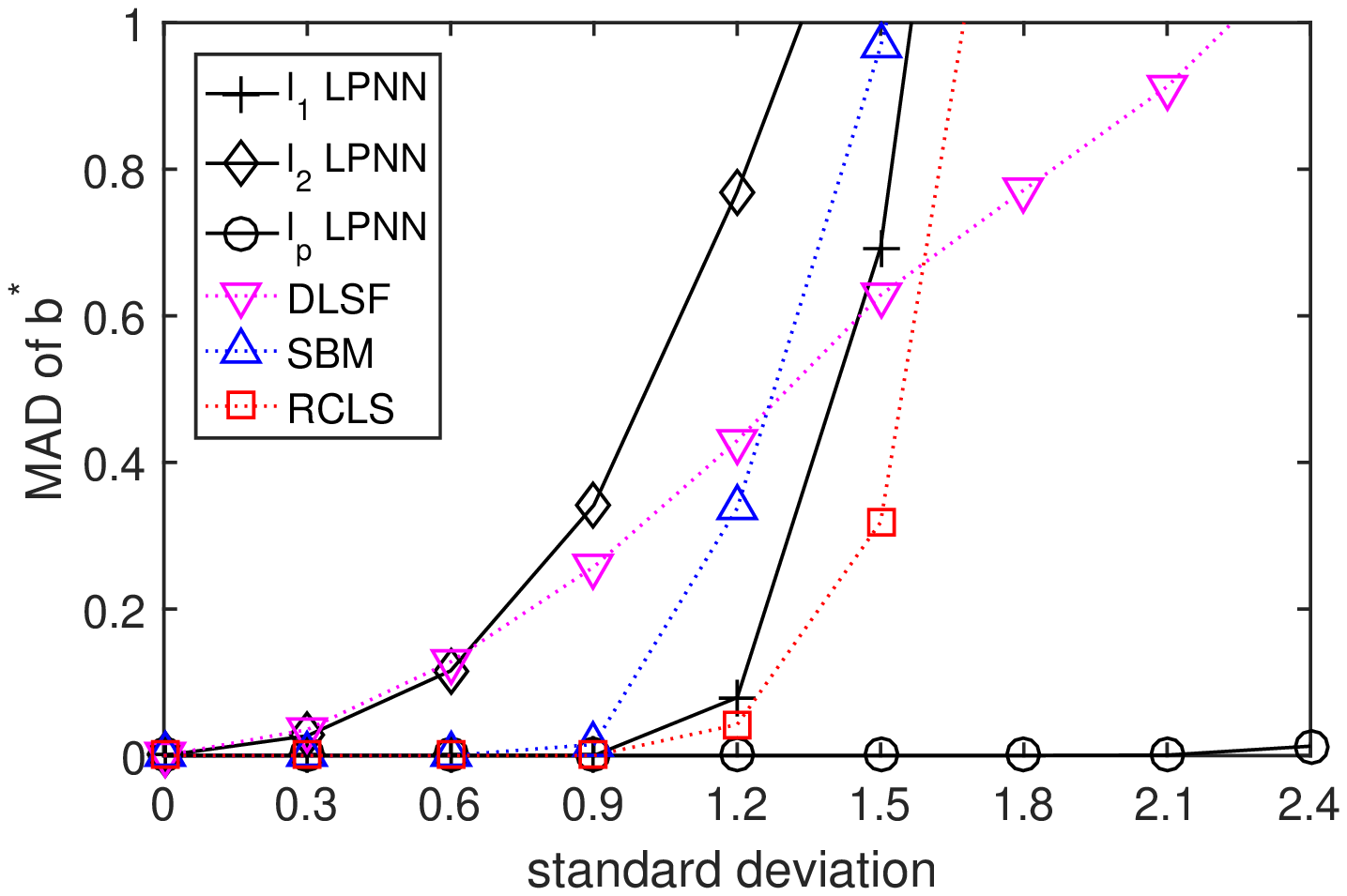,width=1.65in}}  \\
(a) MAD of $a^*$ & (b) MAD of $b^*$ \\
\mbox{\epsfig{figure=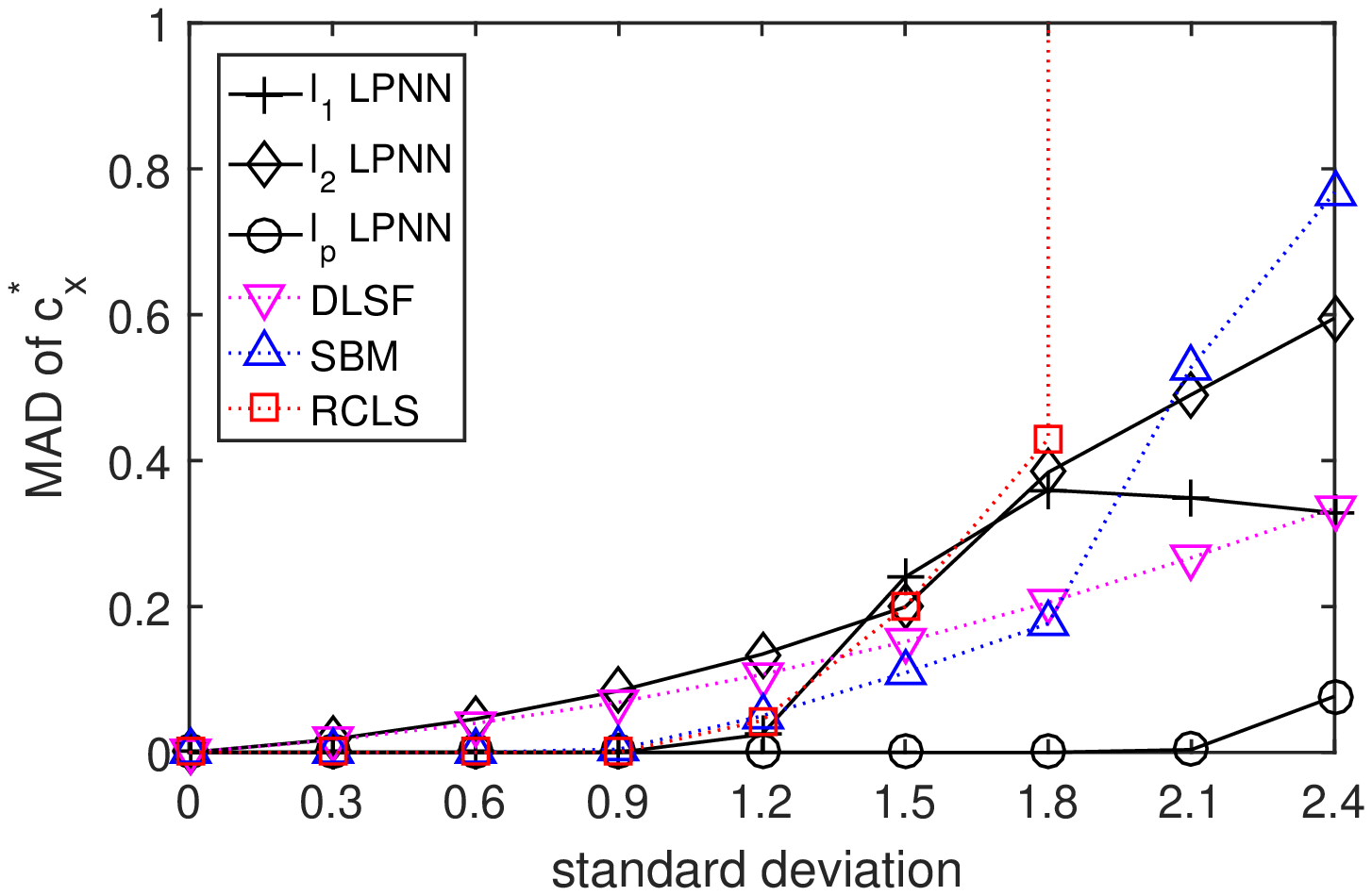,width=1.65in}} &
\mbox{\epsfig{figure=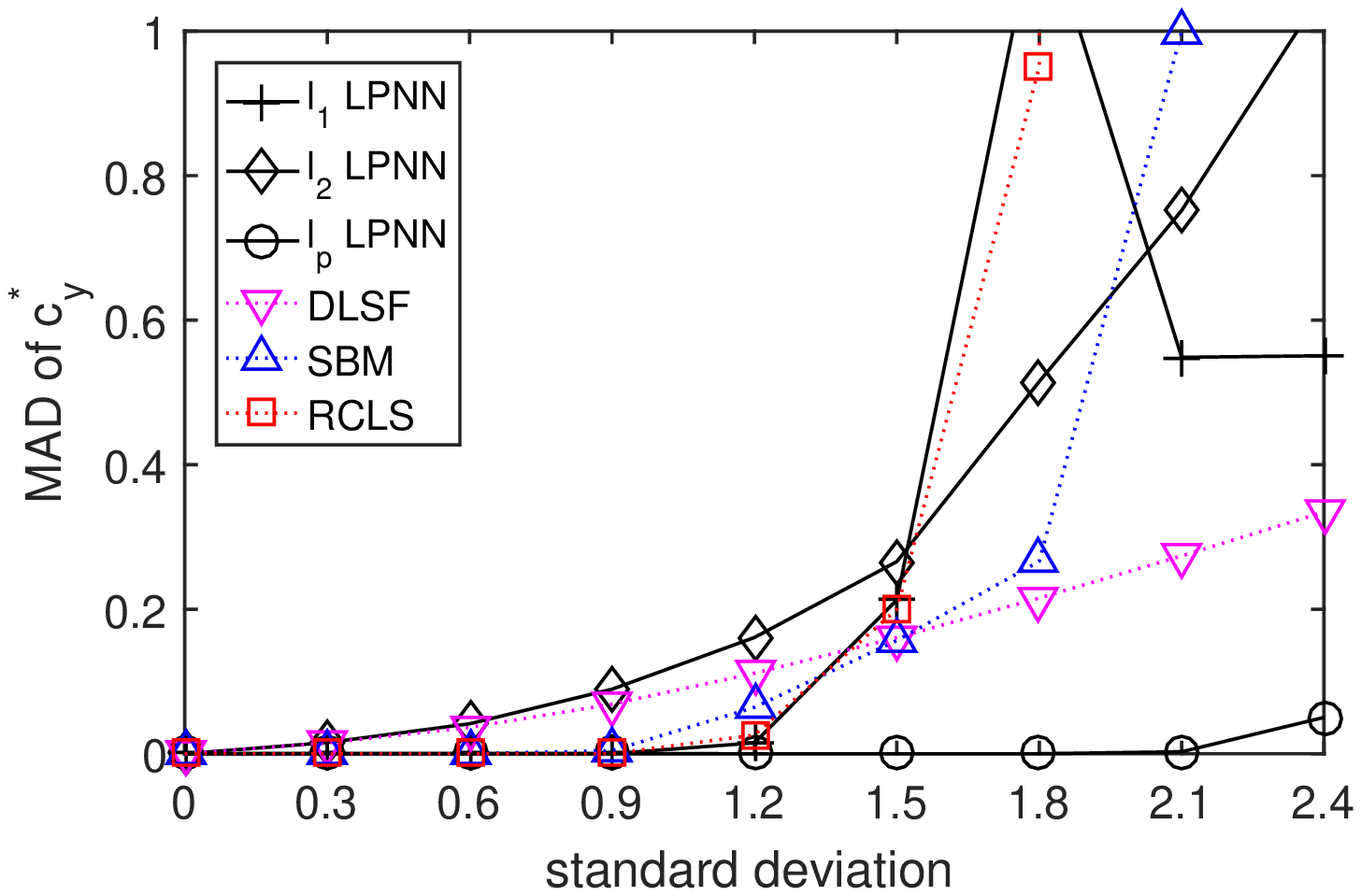,width=1.65in}} \\
(c) MAD of $c^*_x$ & (d) MAD of $c^*_y$\\
\mbox{\epsfig{figure=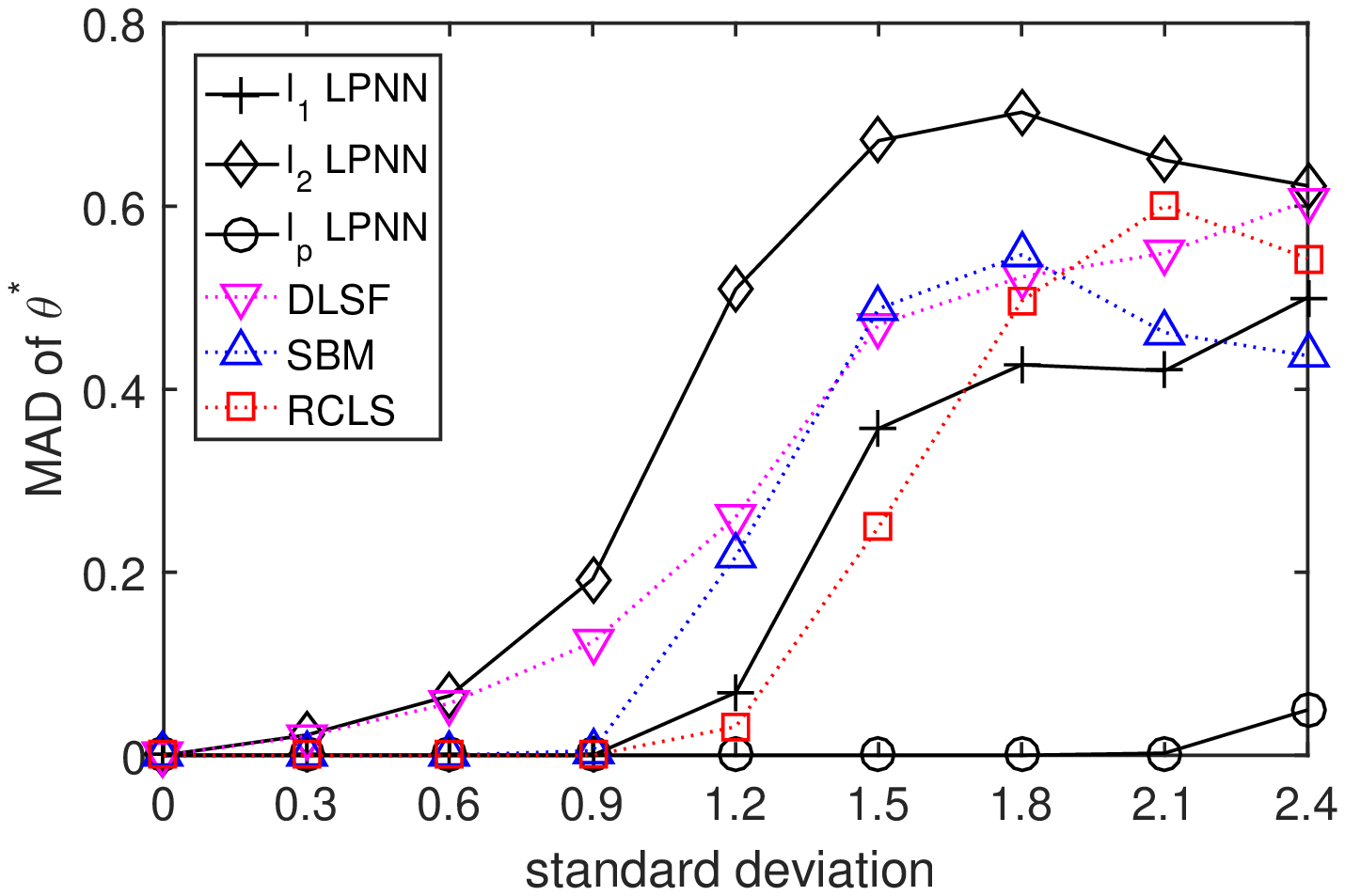,width=1.65in}}\\
(e) MAD of $\theta^*$ &
\end{tabular}
\caption{The MAD results of different algorithms. The uniform noise level is varied from $0$ to $2.4$.
We repeat the experiment $100$ times at each noise level.}
\label{Comparing_uniform}
\end{figure}

Fig.~\ref{lap_result_70} shows the fitting result of a typical run when the noise level is equal to $0.9899$.
It can be seen that the $l_2$-norm LPNN and DLSF methods do not offer reliable result.
While the remaining algorithms can provide satisfactory fitting.
Fig.~\ref{lap_result_90} plots the fitting result of a typical run at the noise level of $1.2728$.
We observe that only the $l_1$-norm and $l_0$-norm LPNN algorithms can achieve accurate ellipse fitting.
When we increase the noise level to $1.4142$, only the $l_0$-norm LPNN algorithm works well, which is shown in Fig.~\ref{lap_result_100}.

\subsection{Experiment 2: Ellipse Fitting in Uniform Noise}
\begin{figure}[h]
\begin{tabular}{@{\extracolsep{-2mm}}c@{\extracolsep{1mm}}c@{\extracolsep{1mm}}c}
\mbox{\epsfig{figure=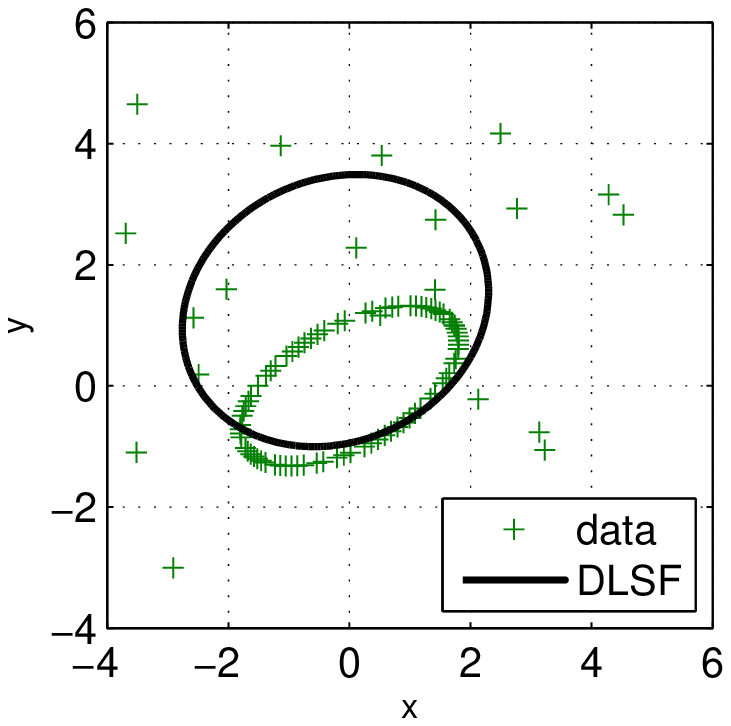,width=1.1in}} &
\mbox{\epsfig{figure=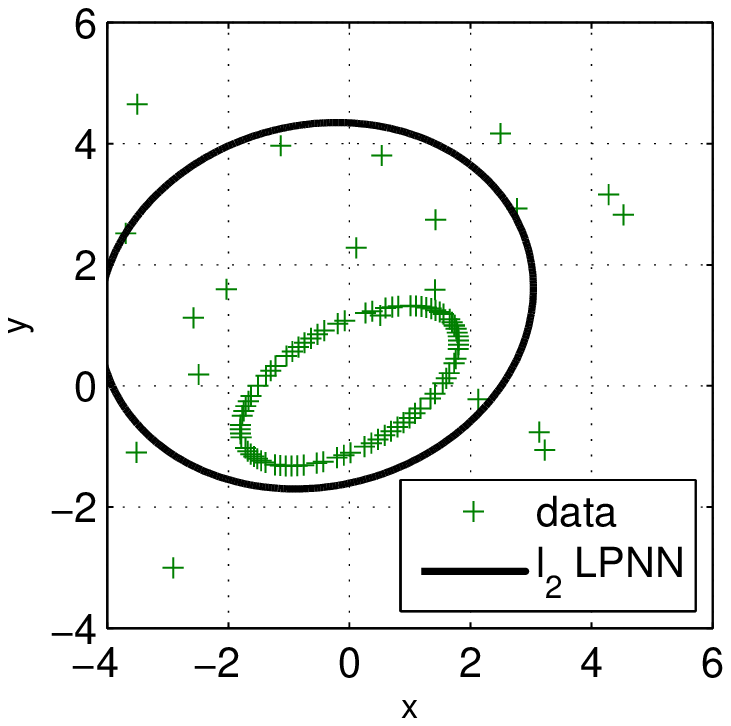,width=1.1in}} &
\mbox{\epsfig{figure=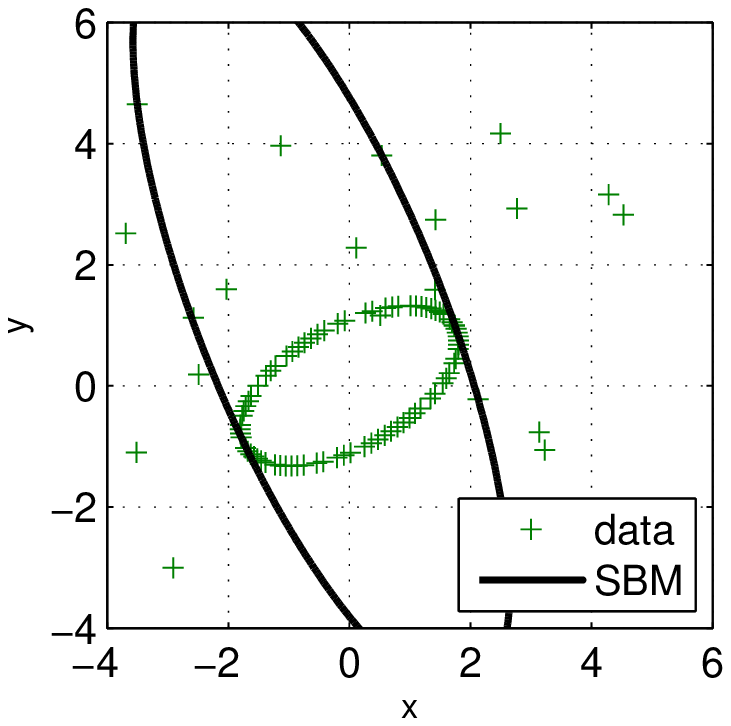,width=1.1in}} \\
(a) DLSF & (b) $l_2$ LPNN & (c) SBM \\
\mbox{\epsfig{figure=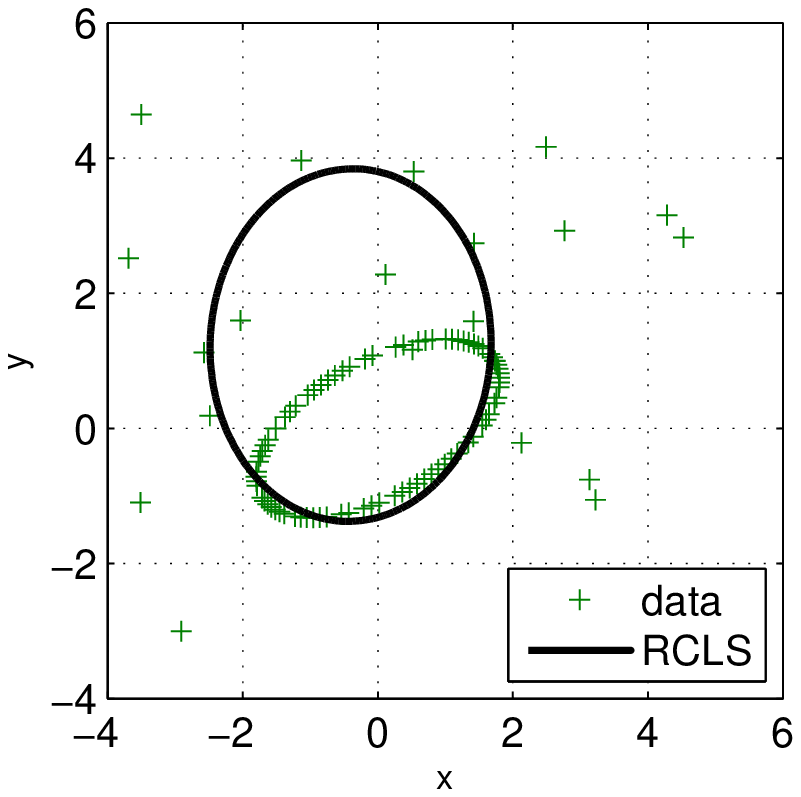,width=1.0in}} &
\mbox{\epsfig{figure=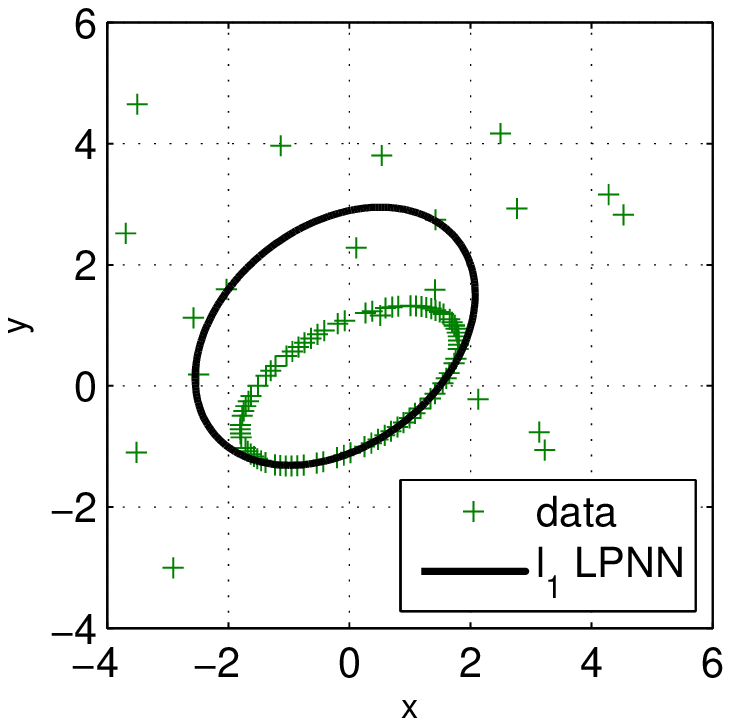,width=1.1in}} &
\mbox{\epsfig{figure=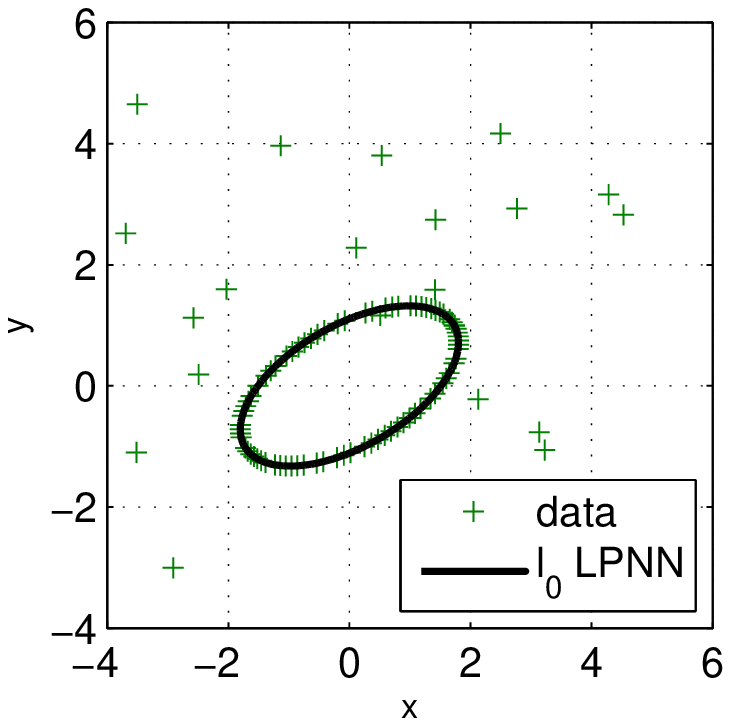,width=1.1in}} \\
 (d) RCLS & (e) $l_1$ LPNN & (f) $l_0$ LPNN
\end{tabular}
\caption{The fitting result of a typical run for
the uniform noise level equal to $2.4$.}
\label{uni_result_900}
\end{figure}

In the second experiment, we test the performance of different algorithms under uniform noise. The experimental setting is the same as Experiment~1, except that the Laplacian noise is replaced by the uniform noise. The noise standard deviation is now varied from $0$ to $2.4$.
We repeat the experiment 100 times at each noise level, to compute the MAD of the estimated parameters.
The results are shown in
Fig.~\ref{Comparing_uniform}. It is observed that the $l_2$-norm LPNN and DLSF algorithms are very sensitive to outliers.
The SBM, RCLS and $l_1$-norm LPNN methods start to break down when the uniform noise level is around $0.9$ to $1.2$.
The $l_0$-norm LPNN still works very well up to the noise level of $2.4$.
Fig.~\ref{uni_result_900} shows the fitting result of a typical run at the noise level of  $2.4$.
It can be seen that only the $l_0$-norm LPNN method produces satisfactory fitting result.

\subsection{Experiment 3: Ellipse Fitting with Different Number of Noisy Data Points}
\begin{figure}[ht]
\begin{tabular}{@{\extracolsep{1mm}}c@{\extracolsep{1mm}}c}
\mbox{\epsfig{figure=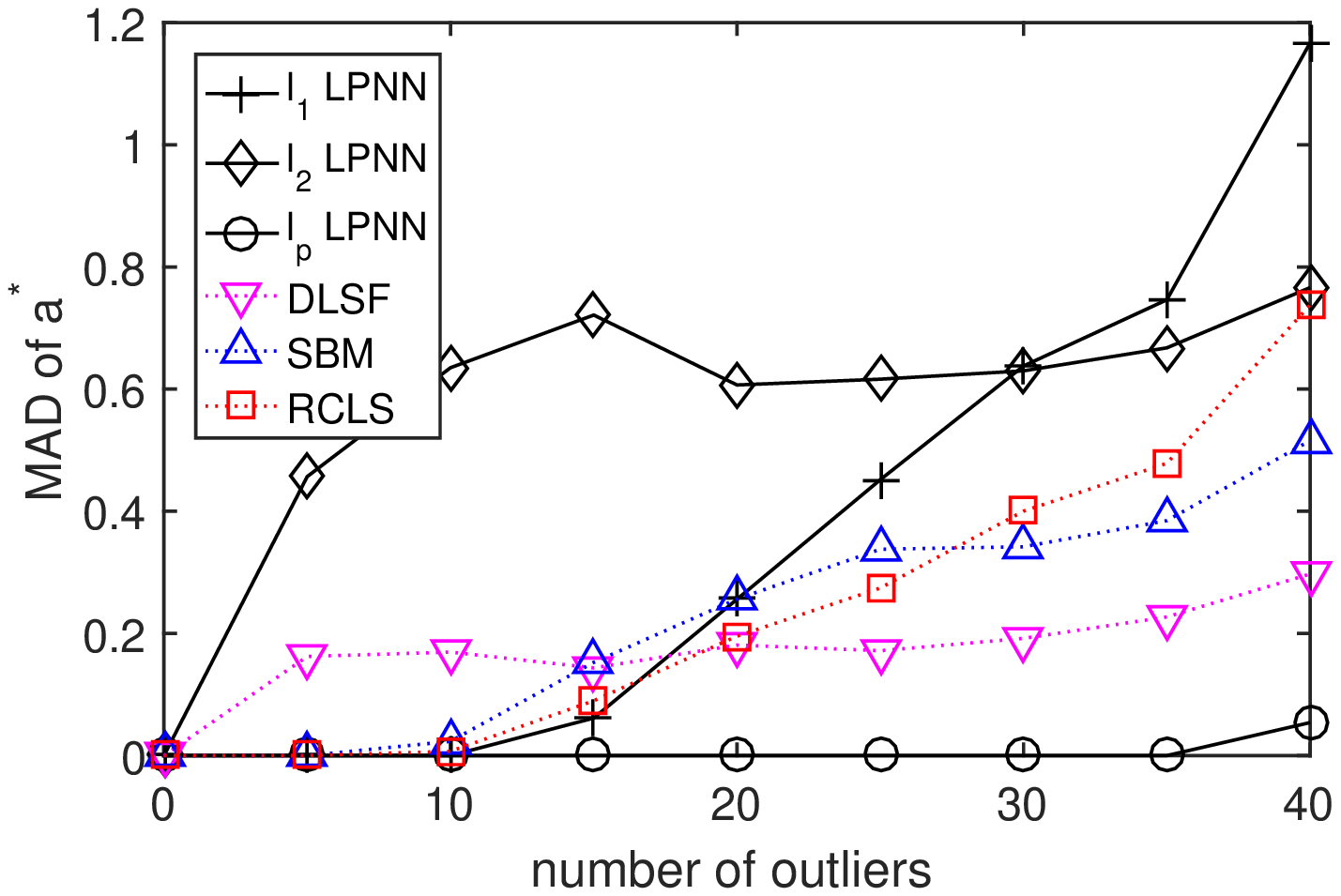,width=1.65in}} &
\mbox{\epsfig{figure=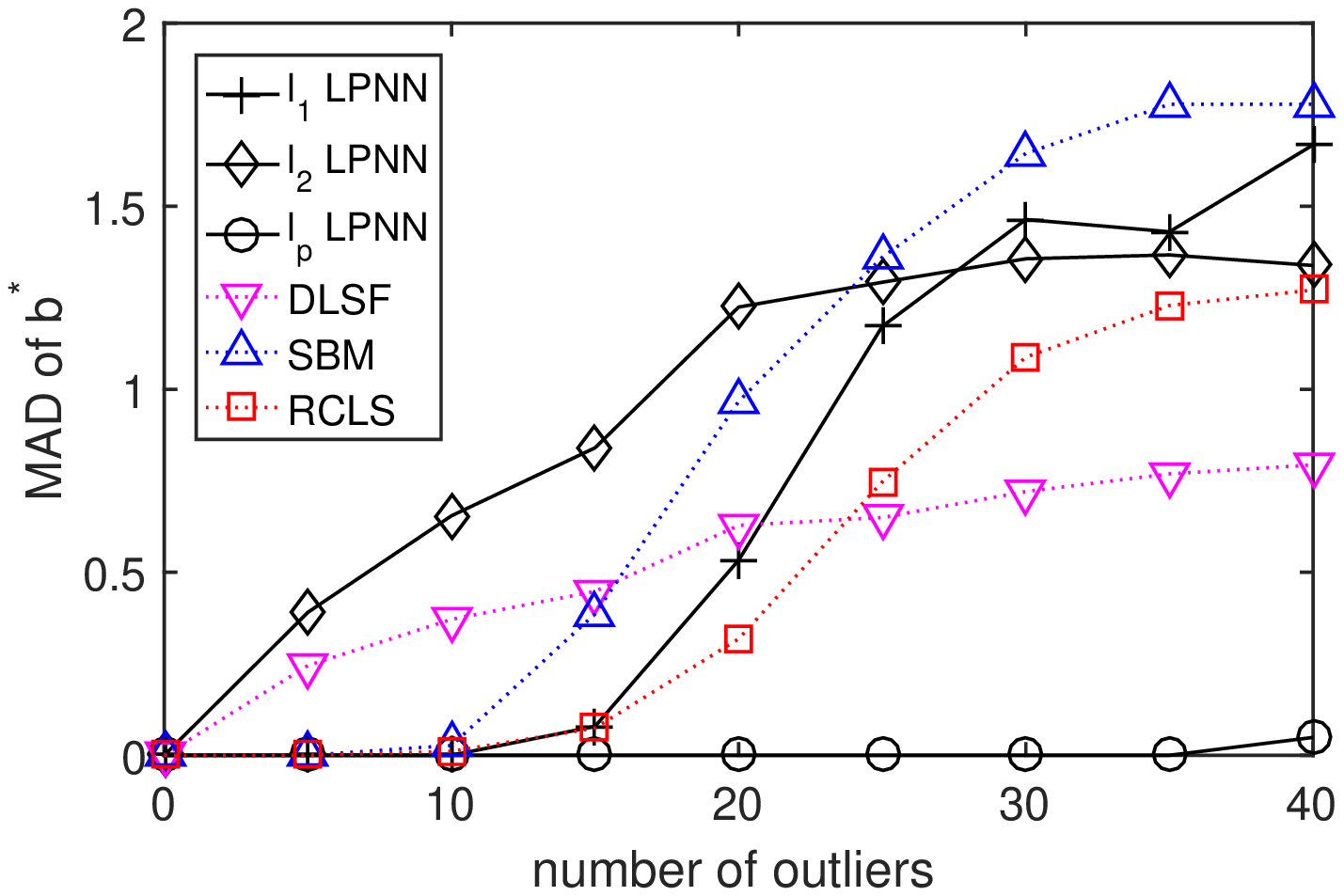,width=1.65in}}  \\
(a) MAD of $a^*$ & (b) MAD of $b^*$ \\
\mbox{\epsfig{figure=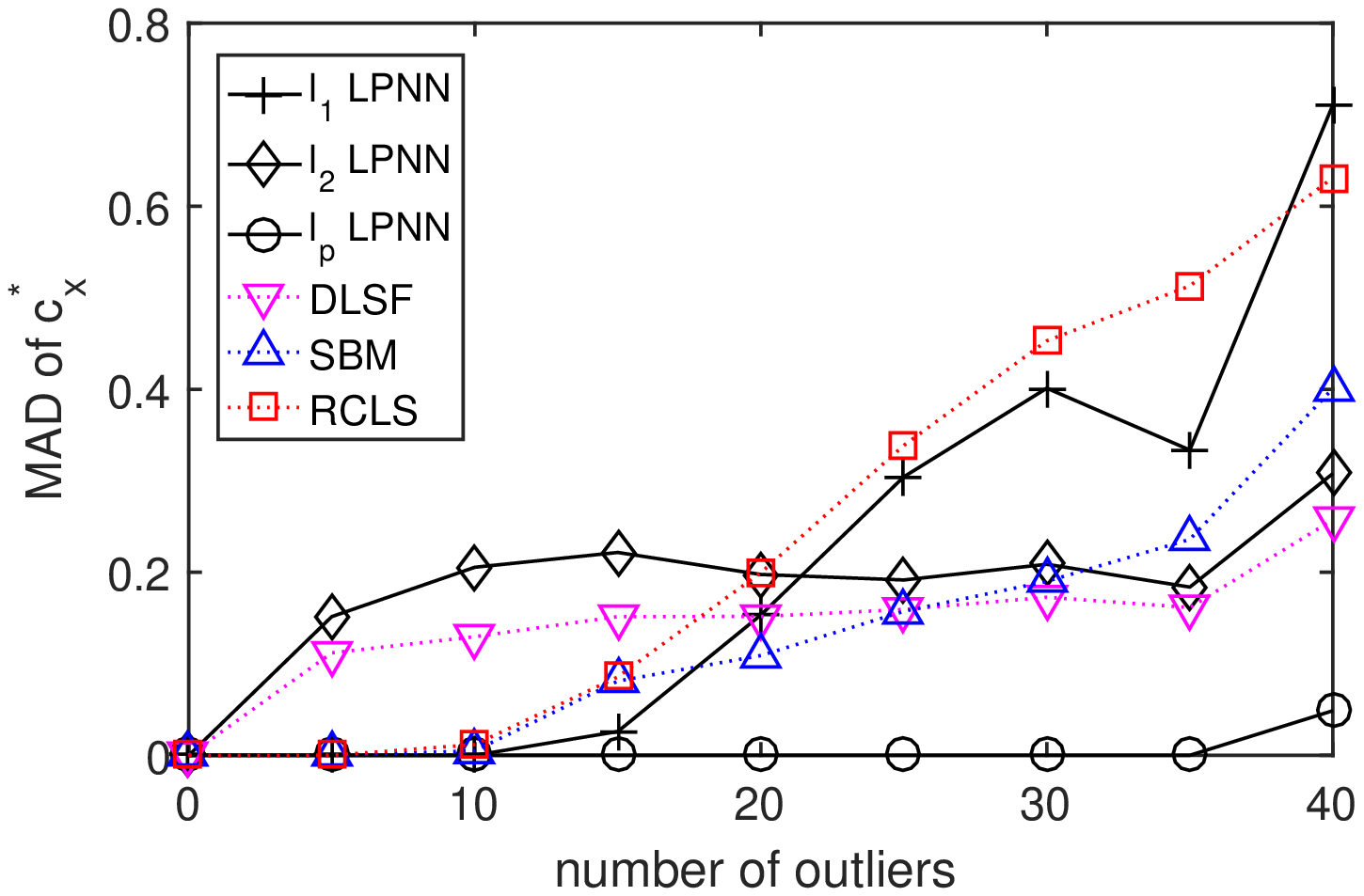,width=1.65in}} &
\mbox{\epsfig{figure=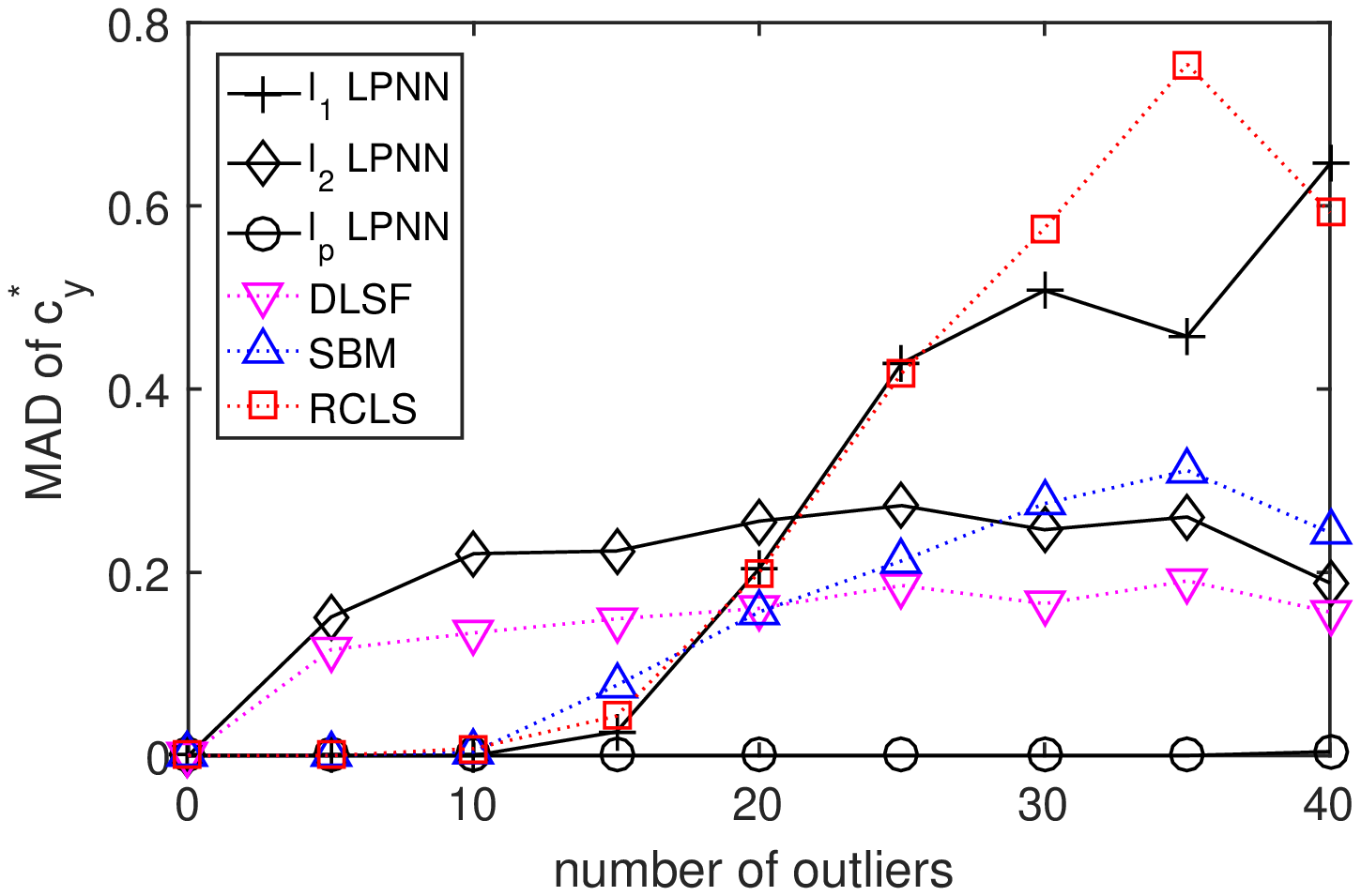,width=1.65in}} \\
(c) MAD of $c^*_x$ & (d) MAD of $c^*_y$\\
\mbox{\epsfig{figure=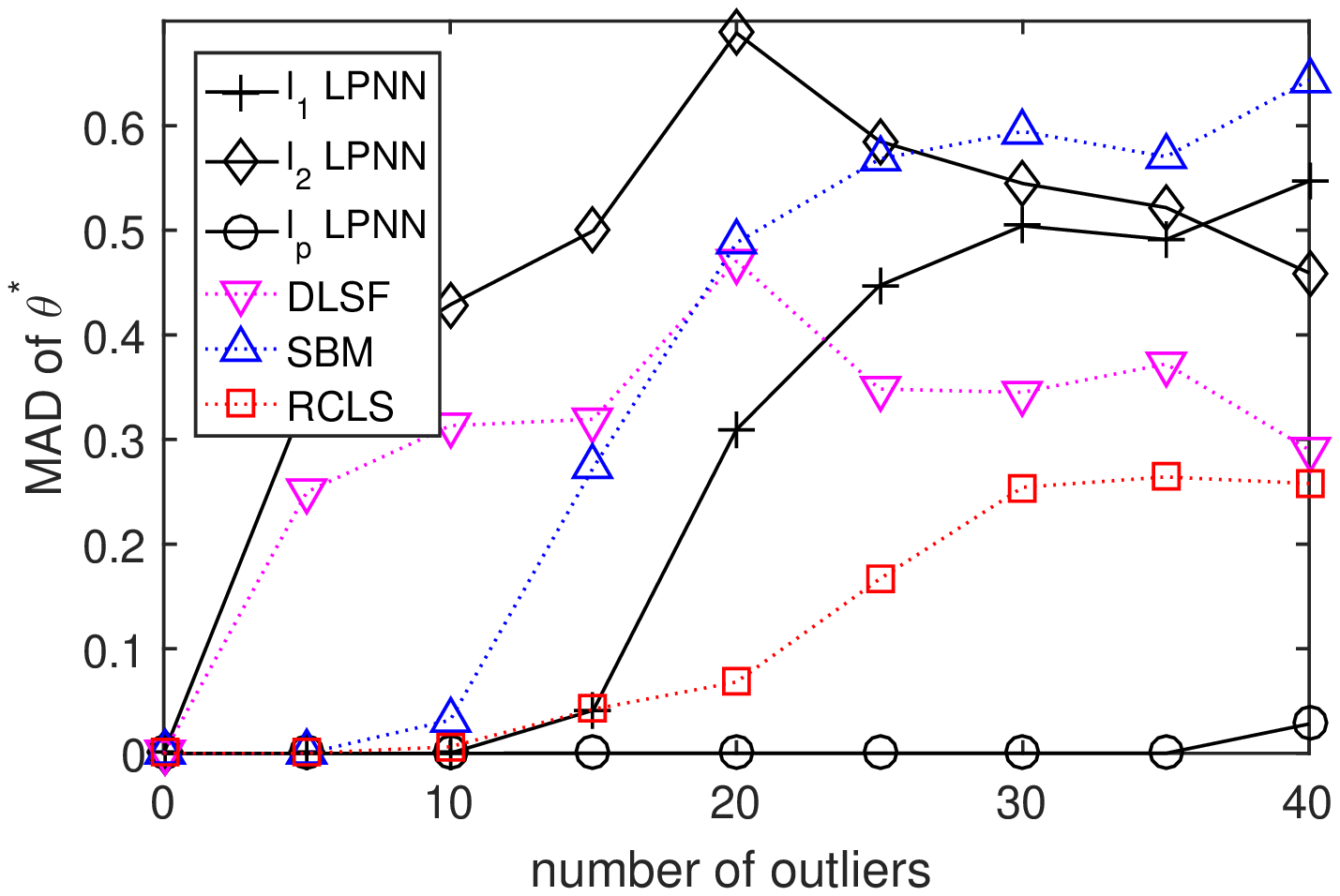,width=1.65in}}\\
(e) MAD of $\theta^*$ &
\end{tabular}
\caption{The MAD results of different algorithms. The uniform noise level is fixed at $1.5$, but number of noisy points changes from 0 to 40. We repeat the experiment 100 times at each different number of noise.}
\label{Comparing_uniform_var_numb}
\end{figure}

In the third experiment, we fix the standard deviation of the uniform noise at $1.5$, but change the number of noise from $0$ to $40$. Other settings are same as the Experiment~2. We repeat the experiment 100 times at each different number of noisy points. The results are shown in Fig.~\ref{Comparing_uniform_var_numb}. We can see that the $l_2$-norm LPNN and DLSF algorithms are also very sensitive to the quantity of outliers. The SBM, RCLS and $l_1$-norm LPNN methods cannot work when the number of uniform noise is larger than $10$. The $l_0$-norm LPNN can give satisfactory result until the number of noise is $40$.

\subsection{Experiment 4: Real Data with Pepper Noise}
In the fourth experiment, we test the performance of different algorithms with real data.

\begin{figure}[h]
\begin{tabular}{c@{\extracolsep{2mm}}c}
\mbox{\epsfig{figure=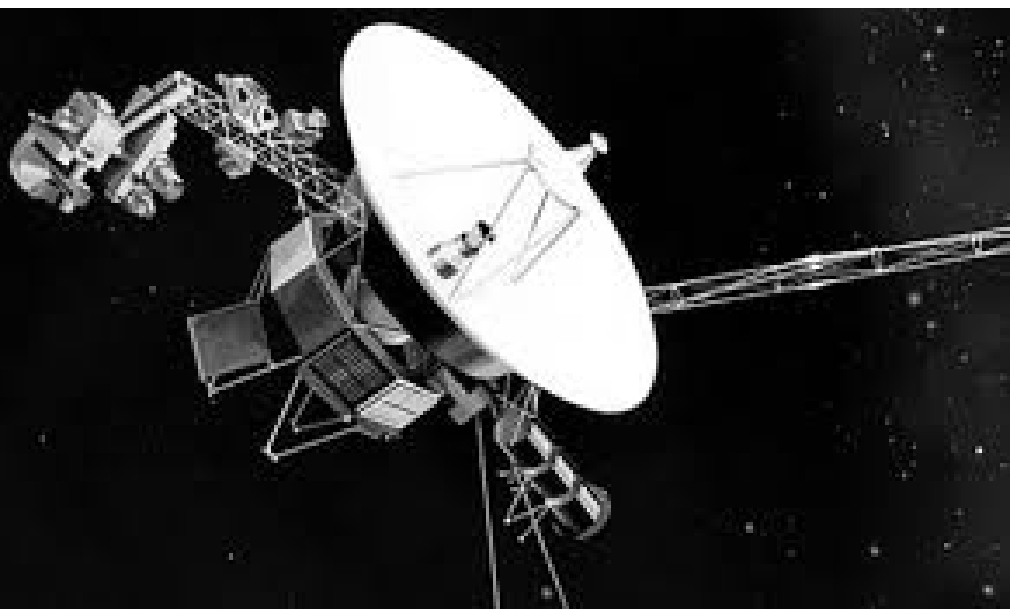,width=1.5in}} &
\mbox{\epsfig{figure=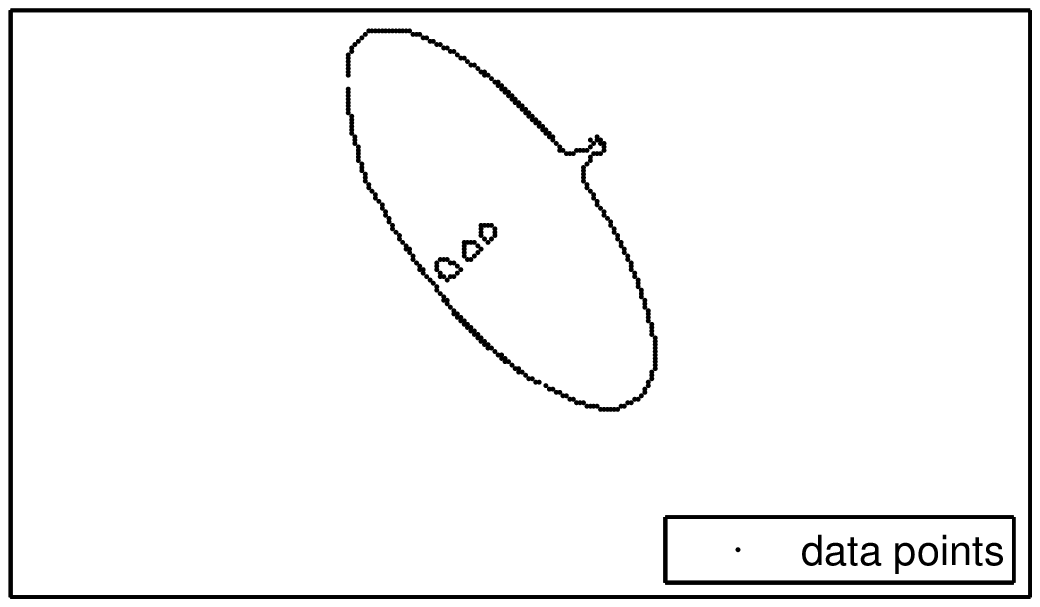,width=1.56in}}  \\
(a)  & (b) \\
\mbox{\epsfig{figure=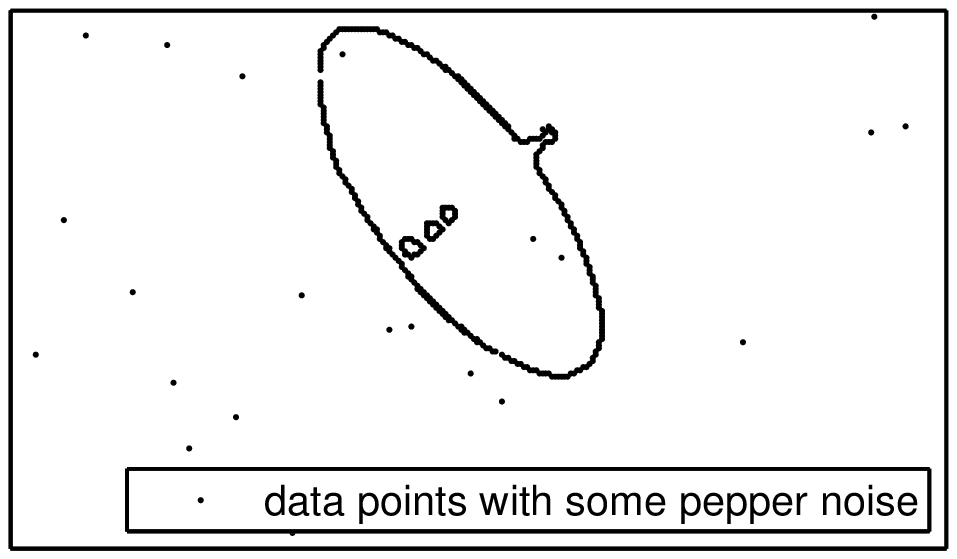,width=1.56in}} &
\mbox{\epsfig{figure=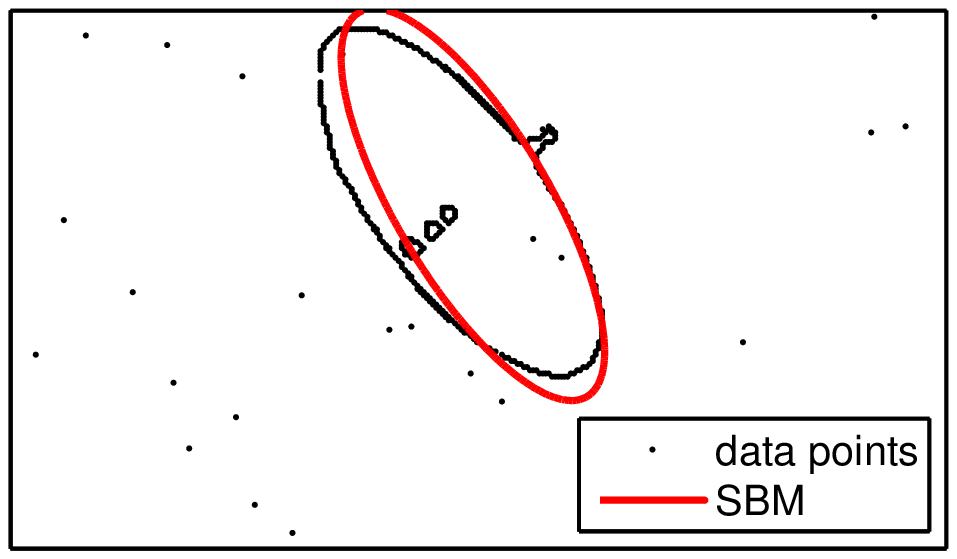,width=1.56in}} \\
(c) & (d)  \\
\mbox{\epsfig{figure=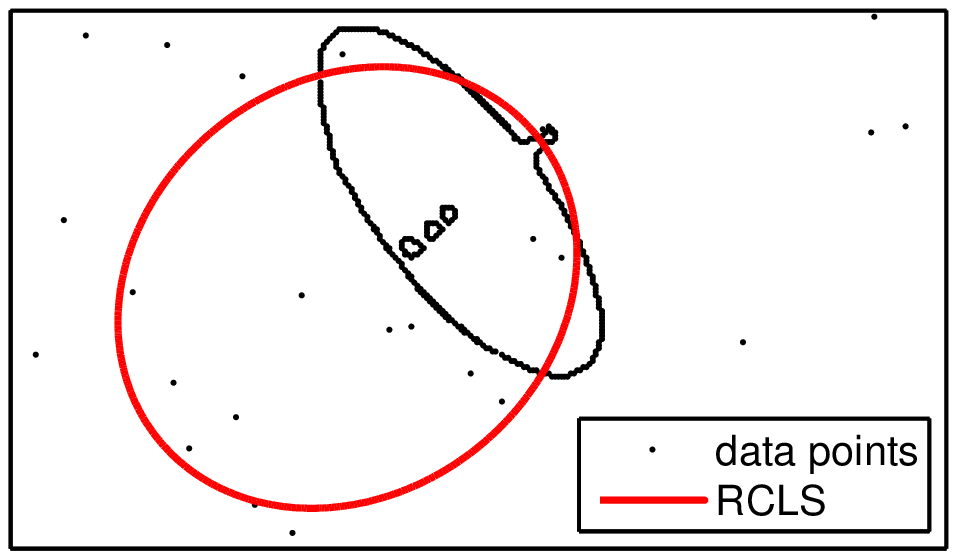,width=1.56in}} &
\mbox{\epsfig{figure=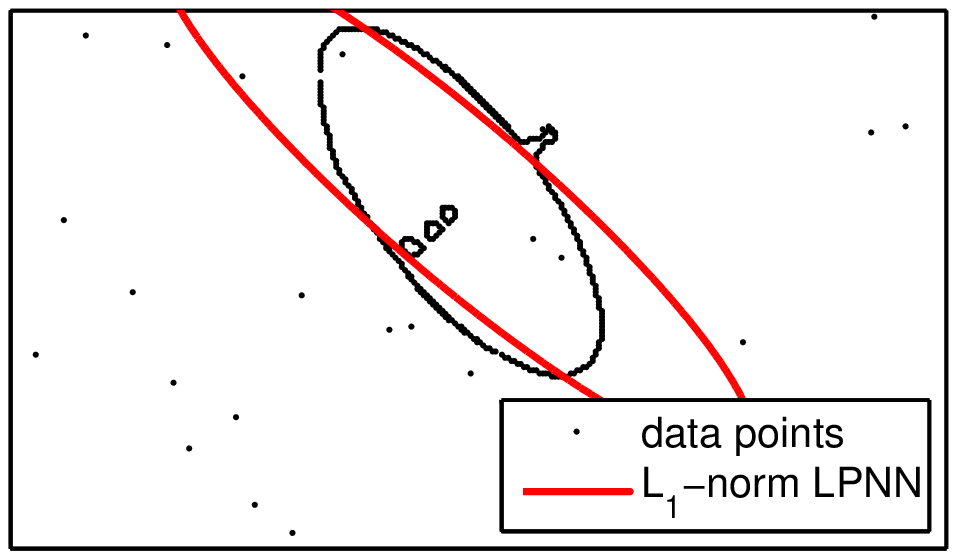,width=1.56in}} \\
(e) & (f) \\
\mbox{\epsfig{figure=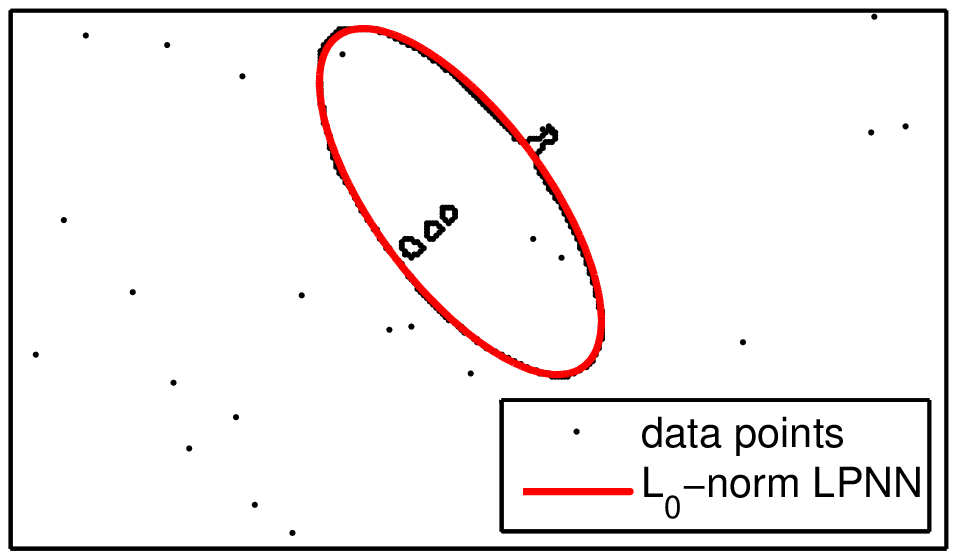,width=1.56in}} & \\
(g)
\end{tabular}
\caption{Fitting results of a space probe image.
(a) Actual image. (b) Data points after edge extraction. (c) Observations with pepper noise.
(d) Fitting result of SBM. (e) Fitting result of RCLS. (f) Fitting result of $l_1$-norm LPNN.
(g) Fitting result of $l_0$-norm LPNN.}
\label{fig:image}
\end{figure}

Fig.~\ref{fig:image} (a) shows a real image of space probe \cite{liang_robust_2015} and here the task is to fit the circumference of the antenna. After edge detection, Fig.~\ref{fig:image} (b) is obtained. For the extracted image, we randomly add some pepper noise whose density is 0.001. The resultant observed data are given in Fig.~\ref{fig:image} (c). Fig.~\ref{fig:image} (d)-Fig.~\ref{fig:image} (g)
show the fitting results of the SBM, RCLS, $l_1$-norm LPNN, and $l_0$-norm LPNN.
It can be seen that the SBM, RCLS, $l_1$-norm LPNN methods do not work very well.
On the other hand, only the $l_0$-norm LPNN scheme best fits the
circumference of the antenna.


\begin{figure}[h]
\begin{tabular}{c@{\extracolsep{6mm}}c}
\mbox{\epsfig{figure=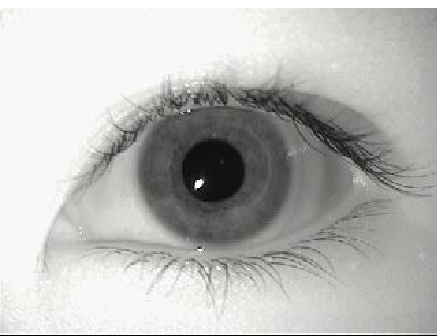,width=1.5in}}    &
\mbox{\epsfig{figure=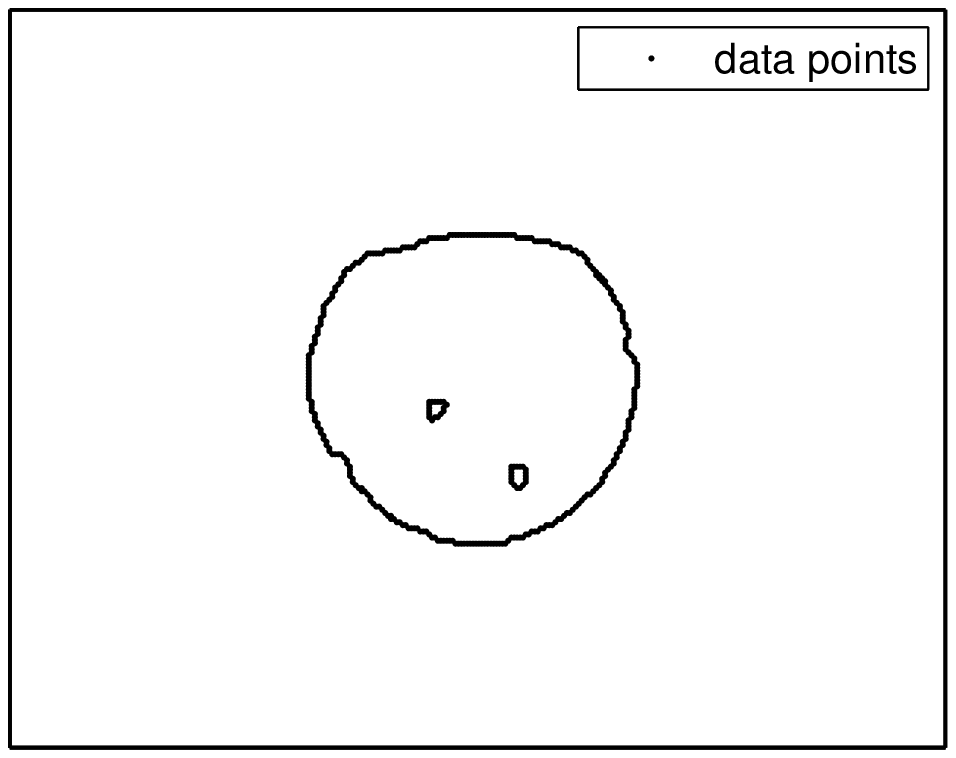,width=1.5in}}  \\
(a)     & (b) \\
\mbox{\epsfig{figure=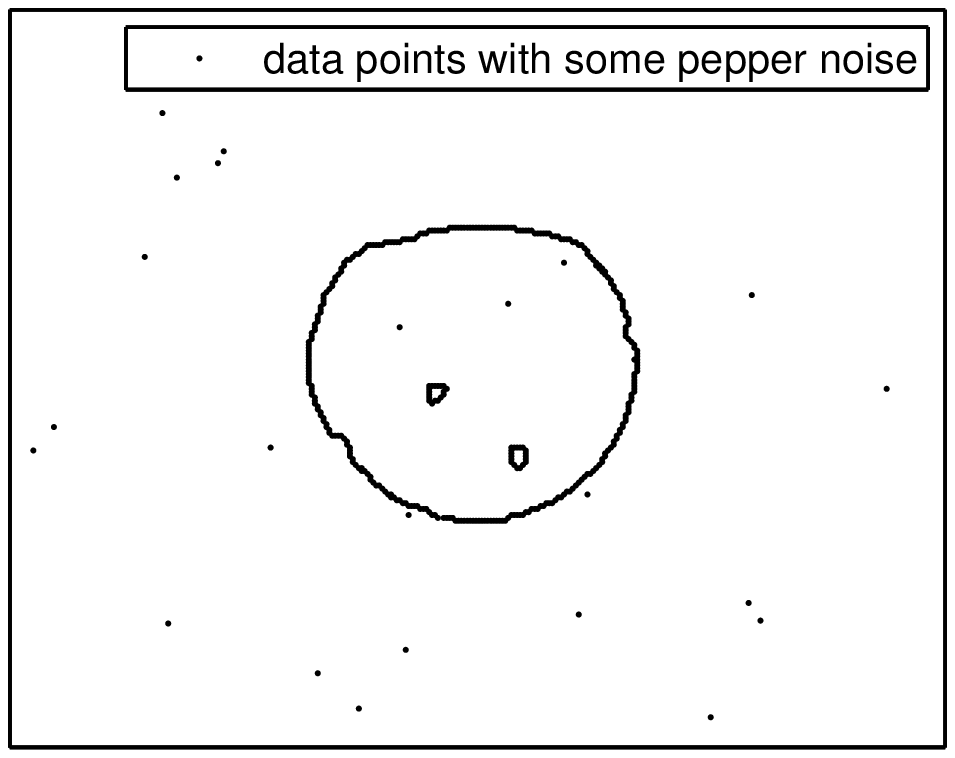,width=1.5in}}    &
\mbox{\epsfig{figure=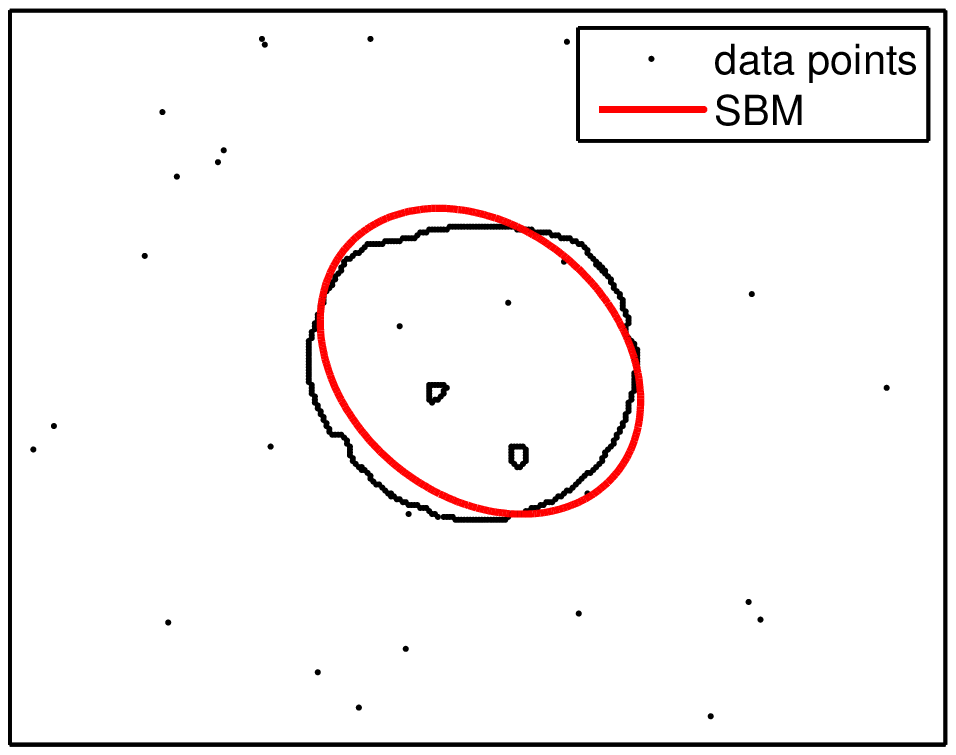,width=1.5in}} \\
(c)    & (d)  \\
\mbox{\epsfig{figure=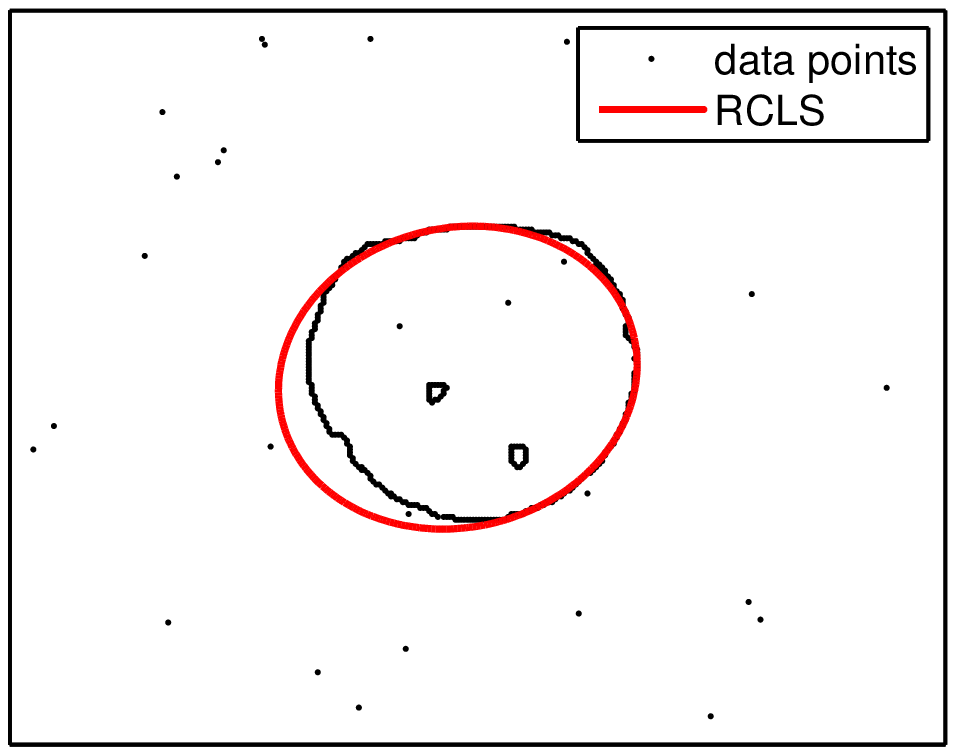,width=1.5in}}    &
\mbox{\epsfig{figure=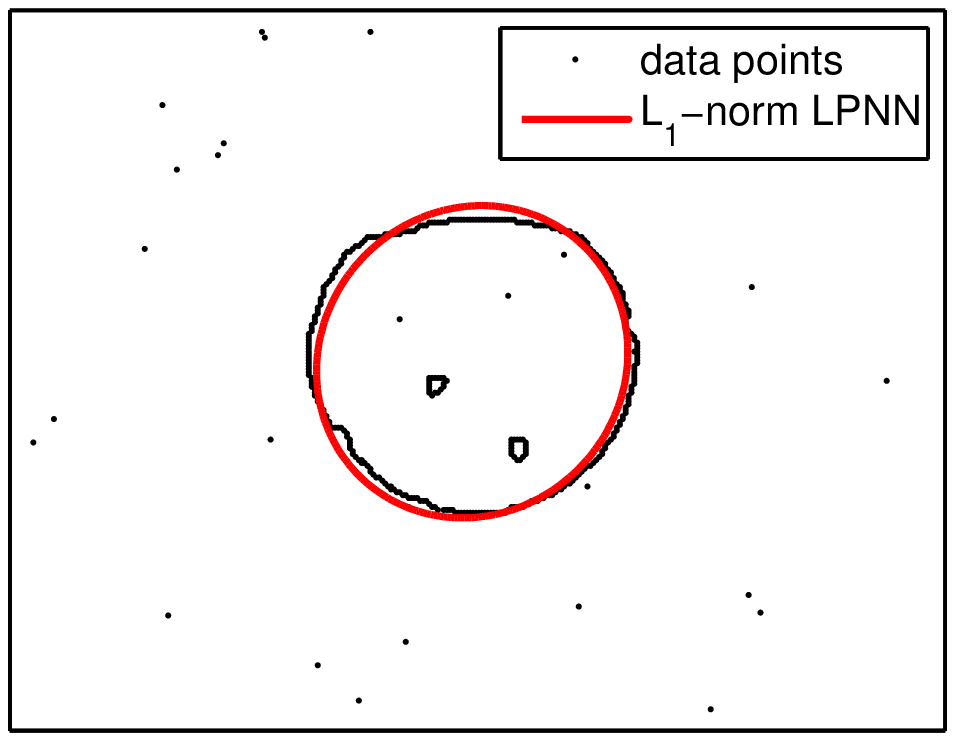,width=1.5in}} \\
(e)    & (f) \\
\mbox{\epsfig{figure=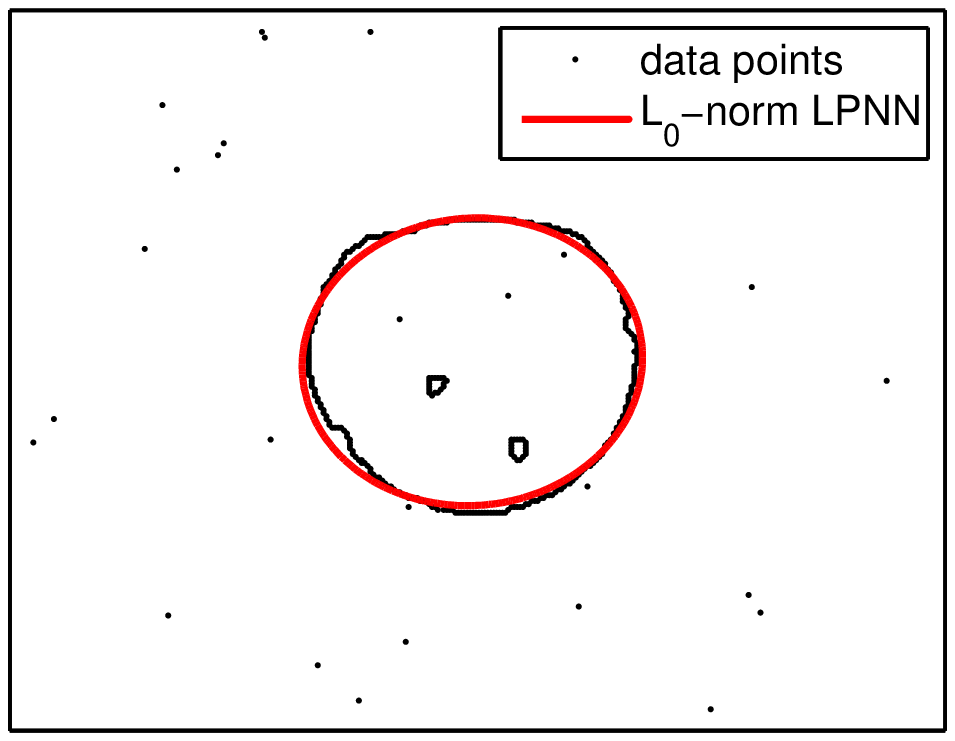,width=1.5in}}    & \\
(g)
\end{tabular}
\caption{Fitting results of a human eye image.
(a) Actual image. (b) Data points after edge extraction. (c) Observations with pepper noise.
(d) Fitting result of SBM. (e) Fitting result of RCLS. (f) Fitting result of $l_1$-norm LPNN.
(g) Fitting result of $l_0$-norm LPNN.}
\label{fig:eye}
\end{figure}

Furthermore, Fig.~\ref{fig:eye} (a) shows a human eye image [16] and this kind of images is frequently used in iris recognition where a key step is to find out the correct pupil region. In this test, our target is to fit the pupil region of the eye. After edge extraction, Fig.~\ref{fig:eye} (b) is obtained. Same as the process mentioned before, we add pepper noise whose density is 0.001. The observations are provided in Fig.~\ref{fig:eye} (c). Finally, we apply different robust ellipse fitting algorithms to the data and the results are given by Fig.~\ref{fig:eye} (d)-Fig.~\ref{fig:eye} (g). We can see that the RCLS and SBM both are influenced by the pepper noises, but $l_1$-norm LPNN, and $l_0$-norm LPNN give out satisfied results.


\begin{figure}[ht]
\begin{tabular}{c@{\extracolsep{-4mm}}c}
\mbox{\epsfig{figure=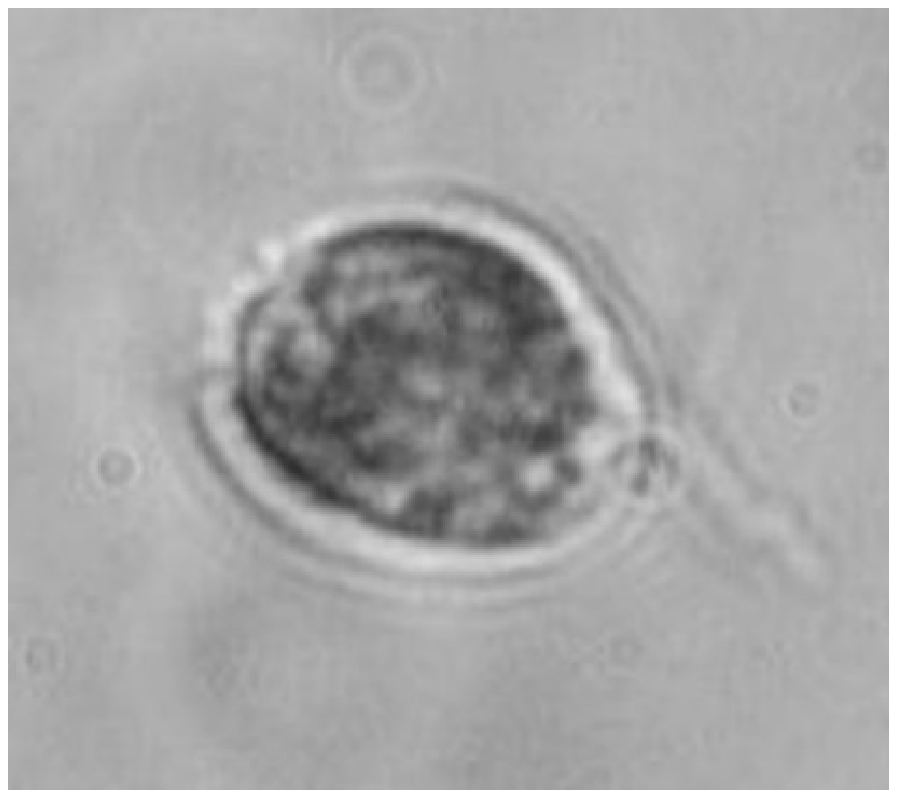,width=2.0in}}    &
\mbox{\epsfig{figure=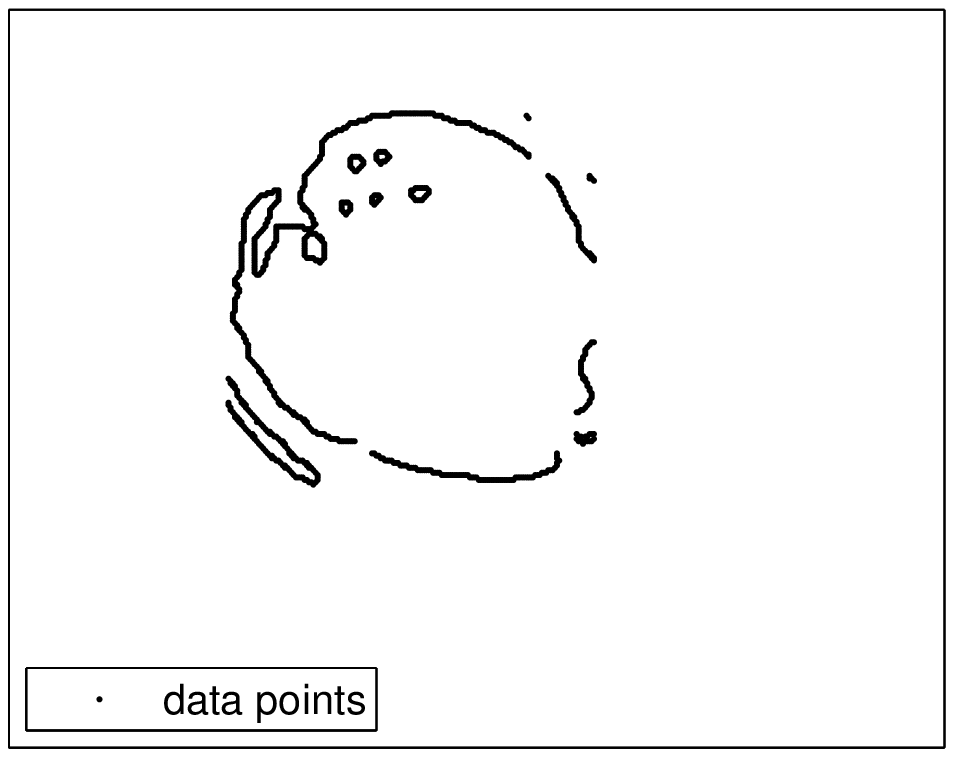,width=1.4in}}  \\
(a)     & (b) \\
\mbox{\epsfig{figure=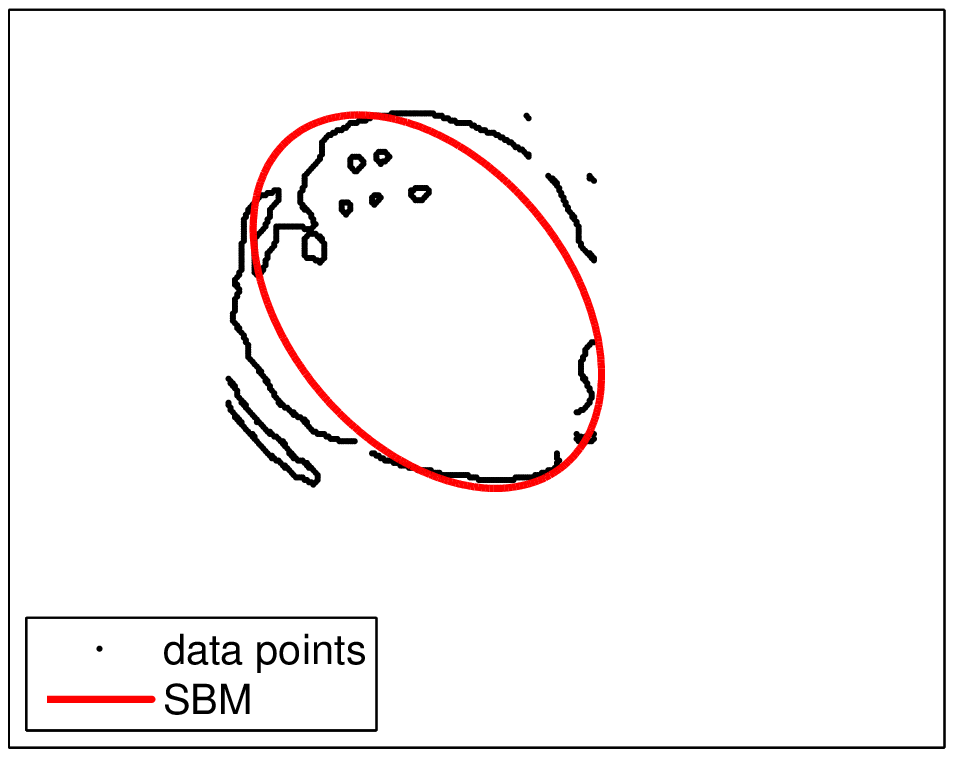,width=1.4in}}    &
\mbox{\epsfig{figure=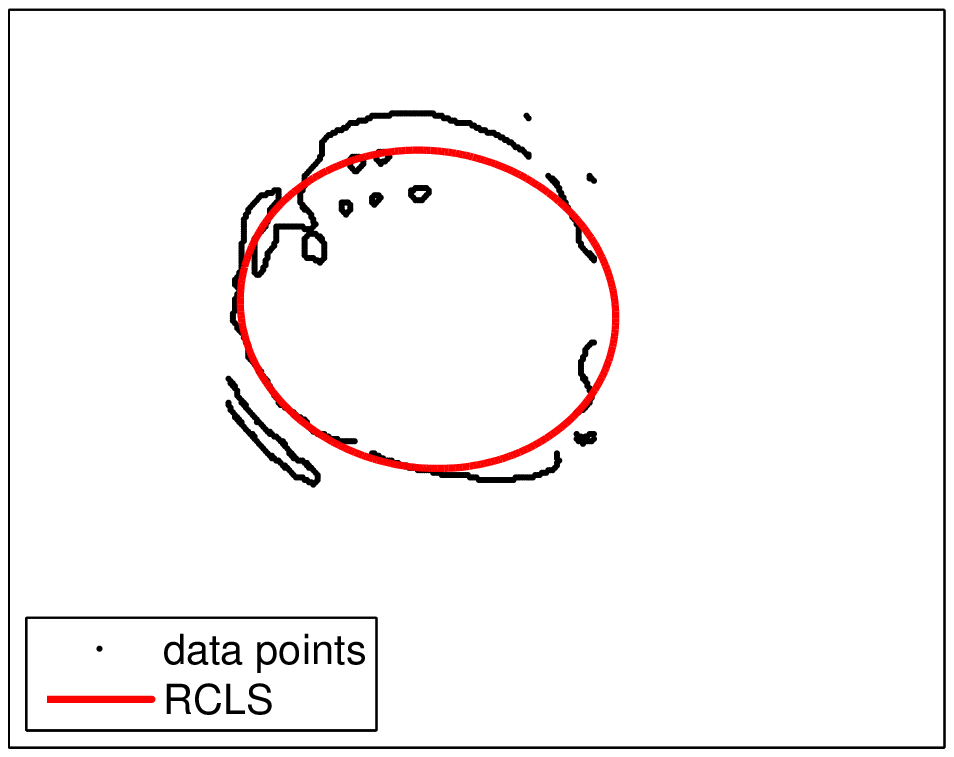,width=1.4in}} \\
(c)    & (d)  \\
\mbox{\epsfig{figure=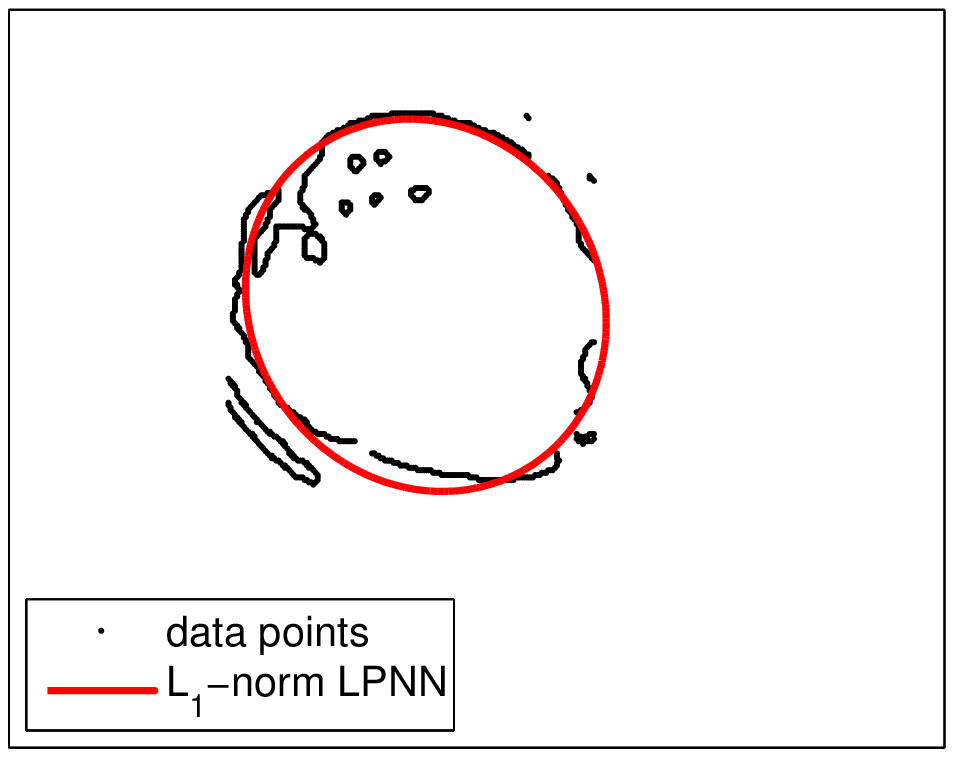,width=1.4in}}    &
\mbox{\epsfig{figure=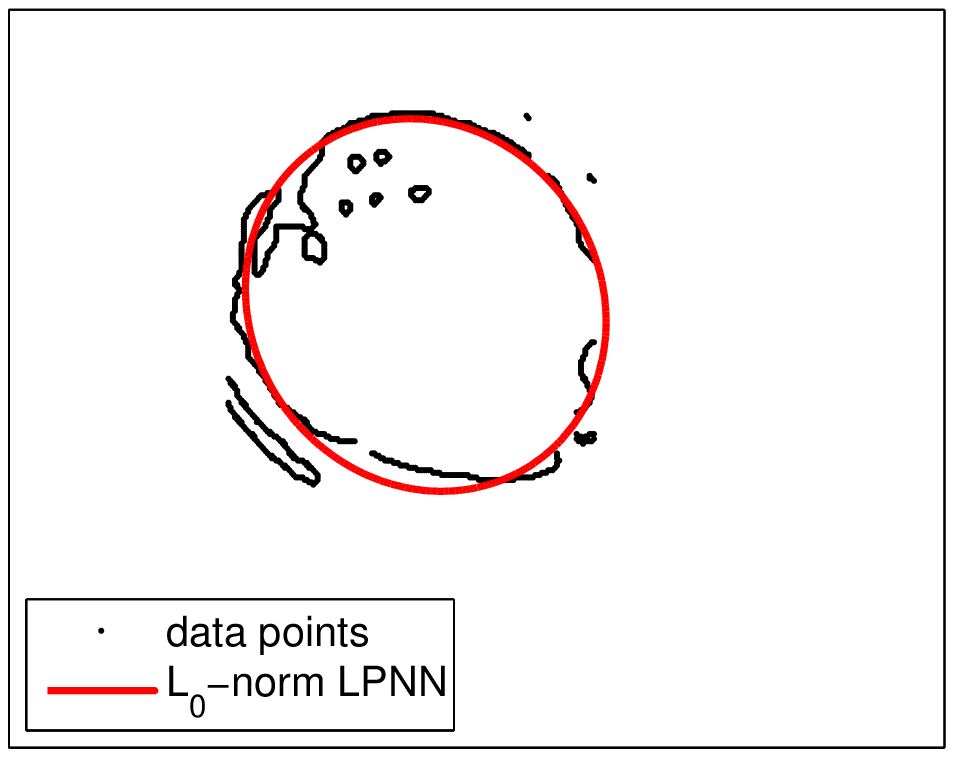,width=1.4in}} \\
(e)    & (f)
\end{tabular}
\caption{Fitting results of a plankton image.
(a)Actual image. (b) Data points after edge extraction. (c) Fitting result of SBM. (d) Fitting result of RCLS. (e) Fitting result of $l_1$-nrom LPNN. (f) Fitting result of $l_0$-norm LPNN.}
\label{fig:planktont}
\end{figure}

Finally, we consider a biological image of a plankton \cite{munoz2014multicriteria} shown in Fig.~\ref{fig:planktont} (a) and our aim is to fit its contour with an ellipse. Apparently, the shape of the plankton is not a regular ellipse. In Fig.~\ref{fig:planktont} (b), we see that after edge extraction, the points on the left edge are very irregular, including a lot of outliers. Hence, we do not add further disturbances in this test. We directly use SBM, RCLS, $l_1$-norm LPNN, $l_0$-norm LPNN to fit these data, the results are shown in Fig.~\ref{fig:planktont} (c) - Fig.~\ref{fig:planktont} (f). We can see that the $l_0$-norm and $l_1$-norm LPNN schemes outperform the existing methods.

\section{Conclusion}\label{section6}
Many applications require fitting 2-D noisy data points with an ellipse. To reduce the influence of outliers, this paper proposes a robust ellipse fitting approach
based on the concept of LPNN.
Inspired by the properties of $l_1$-norm and $l_0$-norm, we redesign the objective function of the original ellipse fitting problem to make it robust against impulsive noise and outliers.
Since the conventional LPNN is able to handle differentiable objective functions only,
we introduce the LCA concept into the LPNN framework.
It is demonstrated that our proposed algorithms can effectively reduce the influence of outliers.
Especially, the proposed $l_0$-norm LPNN method is better than other robust ellipse fitting algorithms.

\bibliographystyle{IEEEtran}
\bibliography{IEEEabrv,my_reference}

\end{document}